\documentclass[journal]{IEEEtran}

\usepackage{cite} 

\usepackage{amsmath}
\usepackage{enumitem}
\usepackage{amsthm}
\usepackage{amssymb}
\usepackage{graphicx}
\usepackage{latexsym}
\usepackage{subfigure}
\usepackage{color}
\usepackage{epsfig}
\usepackage{dsfont}
\usepackage{fancyhdr}
\usepackage{pst-all}
\usepackage[scanall]{psfrag}
\usepackage{multirow}
\usepackage{booktabs}
\usepackage{textcomp}
\usepackage{siunitx}
\usepackage{textcomp}
\usepackage{courier}

\usepackage{amsfonts,amssymb}
\usepackage[german,english]{babel}
\usepackage{url}
\usepackage[colorlinks=true,
            linkcolor=black,     
            citecolor=black,    
             urlcolor=blue,
            ]{hyperref}

\renewenvironment{proof}{\vspace{.1cm}\noindent{\sc Proof.}\hspace{0.10cm}\,\,}{\hfill$\blacksquare$ } 
\newtheorem{theorem}            {Theorem}[section] 
\newtheorem{definition}         [theorem]{Definition} 
\newtheorem{assumption}         {Assumption}[section] 
 
\newtheorem{lemma}              [theorem]{Lemma} 
\newtheorem{proposition}		[theorem]{Proposition}

\newtheorem{remark}	      [theorem]{Remark}

\newcommand{\rline}{{\mathbb R}}

\newcommand{\sbm}[1]{\left[\begin{smallmatrix} #1
	\end{smallmatrix}\right]}

\newcommand{\rfb}[1]{\mbox{\rm
		(\eref{#1})}\ifx\undefined\stillediting\else:\fbox{$#1$}\fi}

\newcommand{\bluff}{{\hbox{\raise 15pt \hbox{\hskip 0.5pt}}}}

\newfont{\roma}{cmr10 scaled 1200}

\usepackage{bm}

\usepackage{algorithm}
\usepackage{algorithmicx}
\usepackage{algpseudocode}
\usepackage{amsmath}
\usepackage{amssymb}
 
\floatname{algorithm}{Algorithm}

\usepackage{mathtools}
\usepackage{cite} 

\begin{document}
\title{Collision-free Source Seeking and Flocking Control of Multi-agents with Connectivity Preservation }

\author{Tinghua Li and Bayu Jayawardhana
\thanks{Tinghua Li and Bayu Jayawardhana are with the Engineering and Technology Institute Groningen (ENTEG), Faculty of Science and Engineering, University of Groningen, 9747 AG Groningen, The Netherlands (e-mail: tinghua.li@outlook.com;
b.jayawardhana@rug.nl).}}
\maketitle

\begin{abstract}
In this article, we present a distributed source-seeking and flocking control method for networked multi-agent systems with non-holonomic constraints. Based solely on identical on-board sensor systems, which measure the source local field, the group objective is attained by appointing a leader agent to seek the source while the remaining follower agents safely form a cohesive flocking with their neighbors using a distributed flocking control law in a connectivity-preserved undirected network. To guarantee safe separation and group motion for all agents and to solve the conflicts with the ``cohesion" flocking rule of Reynolds, the distributed control algorithm is solved individually through feasible CBF-based optimization problem with complex constraints, which guarantees the inter-agent collision avoidance and connectivity preservation. Stability analysis of the closed-loop system is presented and the efficacy of the methods is shown in simulation results. 
\end{abstract}
\IEEEpeerreviewmaketitle

\section{Introduction}
The coordination control of networked multi-agent systems pertains to the development of distributed control protocols, with limited local interactions among neighboring agents, such that the coordinated objectives can be realized (e.g., flocking, rendezvous, formation and consensus \cite{Jia,Deghat}). As a family of distributed control problems, flocking is a typical form of collective motion behavior that can be found in nature and has been studied in various disciplines including computer science \cite{Reynolds}, physics \cite{Frasca,Vicsek}, biology \cite{Emlen,Okubo} and robotics\cite{Jia,Saulnier}. Given the typical flocking rules (cohesion, alignment and separation), flocking is considered to be a motion synchronization or consensus problem where all agents reach an agreement \cite{Yu} and achieve a group control objective \cite{Olfati-1} in a dynamically changing environment \cite{Ji, Olfati-2, Vicsek, Jadbabaie}. 

In practice, the agents face a trade-off between achieving flocking and keeping a safe distance from each other. The potential function method is widely implemented with the structures of leader-follower \cite{Ji-1, Olfati-1} and virtual leader\cite{Gu-1}, where the safe separation among agents relies on the repulsive force, while the attractive force contributes to the flocking cohesion \cite{Gu,Tanner-i}. The dynamic programming method is presented in \cite{Ibuki} to allow the computation of control input with the constraints of pose synchronization and collision avoidance. Recently, learning-based methods have been investigated to solve these flocking challenges where a control policy is learned via reinforcement learning \cite{Morihiro} and Q-learning \cite{Hung}. In addition to this trade-off, the networked multi-agent systems must deal with a dynamic communication graph, since each agent can only obtain local information within its limited sensing range. The graph connectivity is therefore required to be maintained to realize cohesion, and to avoid agent splitting and fragmentation \cite{Ji, Olfati-2, Jadbabaie, Zavlanos}. Various algebraic connectivity algorithms have been proposed in the literature where, in general, the global connectivity is maintained by maximizing the second smallest eigenvalue of the graph Laplacian \cite{kim2005maximizing,de2006decentralized,sabattini2013distributed}.

Though the distributed flocking control problem is well-studied for point-mass agents described by single-integrator \cite{Olfati-2, Jadbabaie, Zavlanos, Gu-1, Ji-1} or by double-integrator \cite{Tanner-i, Ibuki, Gu, Morihiro, Hung, Tanner-iii}, their control algorithms can not be directly implemented on unicycle agents where both the agent's orientation and velocity need to be controlled under non-holonomic constraints. This restriction poses difficulties in the application of control barrier function (CBF)-based method to multi-unicycle systems, where the angular velocity does not affect the time derivative of the CBF, resulting in a mixed relative degree problem.

In this paper, we propose a constraint-driven distributed control framework for safe source-seeking and flocking-cohesion of the multi-unicycle system in an environment where an unknown source with maximum signal strength exists. Our proposed methods use only local measurements to achieve all the aforementioned coordination tasks (source-seeking, flocking-cohesion, collision avoidance, and connectivity preservation). Our main contributions are summarized as follows.
\begin{enumerate}
    \item \textit{Source-signal-guided flocking}: The source signal-based flocking objective is to reach a consensus agreement where each follower maintains a desired relative source signal gradient difference with respect to its neighbors' average. The flocking control relies on the individual source signal gradient measurement on each agent instead of the relative distance (or position) between each pair of connected agents as commonly used in literature. This is motivated by the bacterial chemotactic motion that has been adopted for the control of molecular and microrobots \cite{Izumi}. In this case, the need for range sensor is eliminated, and the signal–gradient–based flocking is considered as a physical-level variant of the distance-based flocking mechanism, while maintaining comparable efficiency in achieving cohesive group motion within the cluttered, obstacle-rich environments.
    We assume that each agent has a local reference frame located in its mass center and shares the same orientation of North and East, such that the locally measured source signal gradient vector can be converted into this frame. Only connected agents exchange the local source measurement with each other, which implies that the state of the leader will not be shared with all followers and vice versa (i.e., it is not an all-to-all structure). 
    \item \textit{Flocking-cohesion control of unicycle agents with nonholonomic constraints}: Two types of flocking-cohesion forms and the corresponding control laws are proposed for the multi-unicycle system, the group flocking behavior can be achieved with any randomly selected initial states in the connected undirected graph. 
    \item \textit{Inter-agent control barrier function with uniform relative degree}: We propose an inter-agent CBF construction for each connected pair of unicycle agents with extended state space, ensuring a uniform relative degree with respect to distinct control inputs (of acceleration and angular velocity). Compared to the general solutions that only constrain angular velocity for safe motion, our design imposes safety constraints on both control inputs and avoids the problem of confined control performance.
    \item \textit{Distributed CBF-QPs for inter-agent safety and connectivity}: A distributed CBF-based quadratic programming (QP) framework with complex constraints is proposed for the multi-agent system, guaranteeing inter-agent safety (collision-avoidance) and connectivity preservation. In particular, the piecewise feasibility of the QP is ensured by updating weight parameters for constraints.
\end{enumerate}

The structure of the paper is organized as follows. The problem formulation and notations are stated in Section \ref{sec:problem}. We review the background of the source-seeking control algorithm and analyze the flocking control design in Section \ref{sec:contr_flocking}. The inter-agent safety and connectivity preservation are discussed in Section \ref{set:connectivity}. The efficacy of the proposed algorithm with extensive simulation results is presented in Section \ref{sec:simulation}. Finally, Section \ref{sec:conclusion} concludes the summary and discussion of the work.

\section{Preliminaries and Problem Formulation}\label{sec:problem}
In this section, we provide the relevant background and formulate our safe source-seeking (leader) and flocking-cohesion (followers) problem. We first present the unicycle model and the flocking topology of the multi-agent system, then introduce the control barrier function and its application for maintaining the safety of system. 
\begin{figure}[htbp]
\vspace{-10pt}
    \centering
    \includegraphics[width=0.35\textwidth]{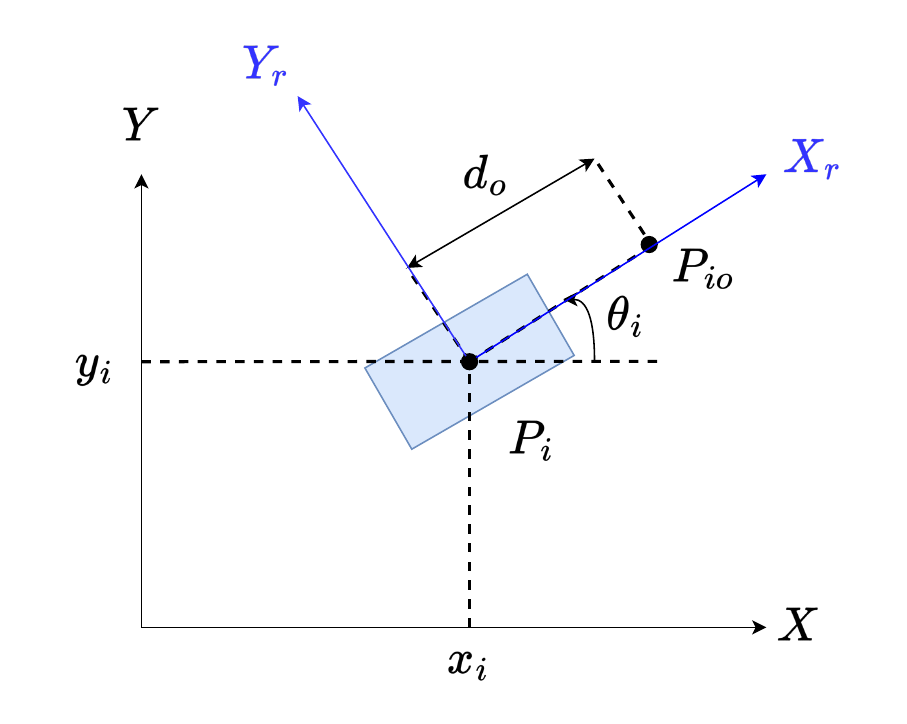}
    \caption{Unicycle model in 2D plane.}
    \label{fig:unicycle}
\end{figure}

\subsection{Multi-unicycle System}
In this paper, we consider an unknown source $(x^*,y^*)$ located in the 2D plane that transmits a 2D field distribution satisfying the following assumption.
\begin{assumption}[Source Signal Field]\label{ass:J}
    \rm{The source signal field function $J(x,y)$ is a twice-differentiable, radially unbounded, strongly concave function whose maximum coincides with the unknown source location $(x^*, y^*)$.}
\end{assumption}

Consider a group of $n$ unicycle agents that are initialized randomly in this 2D field distribution, for each agent $i\in\{1,2,...n\}$, its kinematic dynamics is described by 
\begin{equation}\label{eq:unicycle_model}
\begin{bmatrix}
\dot{x}_i \\
\dot{y}_i  \\
\dot{\theta}_i
\end{bmatrix}=\begin{bmatrix}
v_i\cos(\theta_i) \\
v_i\sin(\theta_i)  \\
\omega_i
\end{bmatrix},
\end{equation}
where $(x_i(t), y_i(t))$ denotes the position of agent $i$ in the Cartesian coordinate system, and $\theta_i(t)$ is the heading angle with respect to the $X$-axis of the global frame of reference as illustrated in Figure~\ref{fig:unicycle} (see $P_i$). 
\begin{assumption}[Agent Sensing Capabilities]\label{ass:sensing}
    Each agent is equipped with sensor systems that can measure the local source signal strength and the bearing angle relative to its neighboring agents.
\end{assumption}

We consider the following agent assignment setup. Within the group of robots, one agent is randomly assigned as the leader and maintains the role afterward. The main goal of the leader is to safely traverse the field, search and approach the source as quickly as possible, relying solely on local information. The rest of the robots are assigned as the followers, whose task is to form a cohesive flock with their neighbors' signal average based on the relative source signal.

\subsection{Flocking Topology}\label{sec:flock_topo}
All agents in the group are assumed to form a communication topology, including the leader and followers. This topology among the agents is modeled as an undirected graph $\mathcal{G} = (\mathcal{E}, \mathcal{V})$, where $\mathcal{V}=\left \{ 1,2,...,{ n} \right \}$ is the vertex set and $\mathcal{E} \subseteq  \left \{  (i,j): i\in\mathcal{V}, j\in\mathcal{N}_i\right \}$ is the edge set. The set of neighbors for agent $i$ is denoted as $\mathcal{N}_i$, which contains all neighbors that agent $i$ can sense and communicate with. We consider an undirected graph $\mathcal{G}$ in this work, i.e., 
$(i,j) \in \mathcal{E} \Leftrightarrow  (j,i) \in \mathcal{E}$ holds, and its adjacency matrix $A = \left[ a_{ij}\right]$ is symmetric with nonzero elements $a_{ij} = 1$ for $ (j,i) \in \mathcal{E}$, else $a_{ij}=0$. Given the communication graph, each agent $i$ is able to share its local sensed source signal gradient with its neighbor agents $j \in \mathcal{N}_i$ and aims to maintain a desired gradient error norm $d^*_{\nabla \bm{J}}\in \mathbb{R}_+$ with the neighbors' source gradient centroid.

Once the leader agent $L$ is assigned at the initial time instant, it maintains this role throughout. Accordingly, the set of followers is denoted by $\mathcal{V}_f:= \mathcal{V} \setminus L$. Note the leader can be considered a neighboring agent if it belongs to the neighbor set $\mathcal{N}_i$ of a follower agent $i\in\mathcal{V}_f$. The communication is limited to connected nodes, such that the leader's information is solely available to its neighboring followers, and the communication topology is not an all-to-all form.
\begin{assumption}[Inter-agent Information Exchange]\label{ass:communication}
   The information exchange between agents occurs if and only if $(i,j)\in \mathcal{E}$. Each agent can communicate with its neighboring agents  $j\in \mathcal{N}_i$ along the edge of the communication graph, exchanging information such as the gradient measurement $\nabla \bm{J}_j$, velocity $v_j$,  heading $\theta_{j}$ and bearing angle $\phi_{ji}$ in its local North-East reference frame.
\end{assumption}

\subsection{Control Barrier Function}\label{sec:CBF}
Let us briefly recall well-known results from the literature on control barrier functions for certifying the safety of general nonlinear systems. Note that for the kinematic model of the unicycle in \eqref{eq:unicycle_model}, the dynamics of agent can be rewritten in the general form of a nonlinear affine control system as follows 
\begin{equation}\label{eq:affine_sys}
   \dot{\bm{\xi}}=f(\bm{\xi})+g(\bm{\xi})\bm{u},
\end{equation}
where the state $\bm{\xi}\in \mathbb{R}^q$ for the unicycle agent will be discussed in further detail in Section \ref{set:connectivity}, and $f(\bm{\xi}):\mathbb{R}^q \rightarrow \mathbb{R}^q, g(\bm{\xi}):\mathbb{R}^q \rightarrow \mathbb{R}^{q\times m}$ are locally Lipschitz continuous in $\bm{\xi}$, and the control input  $\bm{u} \in \mathcal{U}_{\text{adm}}\subset \mathbb{R}^m$ is constrained in the set of admissible inputs. 
For a given continuously differentiable function $h(\bm{\xi}):\mathbb{R}^q \rightarrow \mathbb{R}$, we define the following three sets to study the safety of the systems:
\begin{subequations}
\begin{align} \label{eq:safeset0}
    \mathcal{S}_h & = \left \{ \bm{\xi} \in \mathbb{R}^q : h(\bm{\xi})\geq 0\right \} \\
\label{eq:safeset1}
    \partial \mathcal{S}_h &= \left \{\bm{\xi} \in \mathbb{R}^q : h(\bm{\xi}) = 0 \right \} \\
\label{eq:safeset2}
    \text{Int}(\mathcal{S}_h) & = \left \{\bm{\xi}\in \mathbb{R}^q : h(\bm{\xi}) > 0 \right \}.
\end{align}
\end{subequations}
Whenever it is clear from the context, we omit the dependence on $h$ in these notations. 

\begin{definition}\label{def:CBF}
    \rm{(\textit{Control Barrier Function}) Given a set $\mathcal{S}$ defined in \eqref{eq:safeset0}-\eqref{eq:safeset2}, $h(\bm{\xi})$ is a \textit{Control Barrier Function (CBF)} for system \eqref{eq:affine_sys} if there exists a class $\mathcal{K}_{\infty}$ function $\alpha$ such that 
    \begin{equation}\label{eq:CBF_def}
         \sup_{\bm{u} \in \mathbb{R}^{m}}[L_{f}h(\bm{\xi})+L_{g}h(\bm{\xi})\bm{u}+\alpha(h(\bm{\xi}))]\geq 0, \quad \forall \bm{\xi} \in \mathcal{S}.
    \end{equation}
    where $L_{f}$, $L_{g}$ denote the Lie derivatives along $f$ and $g$. }
\end{definition}

Given the control barrier function $h$ satisfying \eqref{eq:CBF_def}, for all $\bm{\xi}\in \mathcal{S}$, we can define the set
\begin{equation}
    K(\bm{\xi}) = \left \{ \bm{u} : 
    L_{f}h(\bm{\xi}) + L_{g}h(\bm{\xi})\bm{u} + \alpha(h(\bm{\xi}))\geq 0 \right \}.
\end{equation}
Then for any locally Lipschitz function $\bm{k}: \rline^q\to\rline^m$ such that $\bm{k}(\bm{\xi})\in K(\bm{\xi})$, the set $\mathcal{S}$ is {\it forward invariant} with respect to the closed-loop system \eqref{eq:affine_sys} with $\bm{u}=\bm{k}(\bm{\xi})$, i.e. for all $\bm{\xi}(t_0) \in \mathcal{S}$, the solution $\bm{\xi} (t) \in \mathcal{S}$ for all $t \geq t_0$. We refer interested readers to the related discussion in \cite{Ames_TAC}.

\subsection{Problem Formulation }\label{sec:prob}
Within the source field $J(x,y)$ satisfying Assumption \ref{ass:J}, the control objective of the assigned leader is to safely search the source's location. Simultaneously, the followers must realize a cohesive flock relative to their neighbors' centroid, while avoiding potential inter-agent collisions and connectivity breaks. The setup of the multi-agent system is summarized as follows.
\begin{enumerate}
    \item All agents can locally measure the source field gradient and communicate its local information only with their neighbors. None of them has access to global information (e.g., source position or global 2D field). 
    \item Once the leader agent is assigned, its main task is to seek the source using the local field gradient measurement. The leader agent will only communicate its local information to its connected neighbors and not to the rest of the followers. In this setup, the followers do not know the identity of the leader, as it is regarded only as a neighbor of a subset of followers. 
\end{enumerate}
Building on the above setup, we formulate a distributed control design problem for the multi-agent system as follows. 

\textbf{\emph{Safe Source-seeking (leader) and Flocking-cohesion (followers) Control Problem:}} Consider a group of $n$ unicycle robots \eqref{eq:unicycle_model} traversing across a source field $J(x,y)$ satisfying Assumption \ref{ass:J} and given a safe set  $\mathcal{S}\subset \mathbb{R}^q$ for the system \eqref{eq:affine_sys}, design a distributed control law $\bm{u}_i$ for each follower agent $i\in \mathcal{V}_f$, and a control policy for the leader $L$, such that the following motion tasks can be achieved.
\begin{enumerate}
    \item  \textbf{Source-seeking (Leader Task):} The leader robot converges to the source location $(x^*, y^*)$, i.e.,
    \begin{equation}\label{eq:convergence-0}
    \lim_{t\to\infty} \left\| \begin{bmatrix}
        x_{\text{L}}(t)-x^*\\y_{\text{L}}(t)-y^*
    \end{bmatrix}
    \right\| = 0,
    \end{equation}
    where $(x_{\text{L}} , y_{\text{L}})$ is the position of the leader.
    \item  \textbf{Flocking-cohesion (Follower Task):} For each follower agent $i\in\mathcal{V}_f$, we define its flocking center as the average of the field gradient of all its neighbors (i.e., $\textbf{c}_i = \frac{1}{{N}_i}\sum_{j\in \mathcal{N}_i} \nabla \bm{J}_{j}$). The flocking-cohesion task for each follower is to converge and maintain a desired space relative to its flocking center.     
    \item \textbf{Inter-agent Collision Avoidance and Connectivity Preservation (Group Task):} The multi-agent system is {\it safe} at all time (i.e., $\bm{\xi}(t)\in \mathcal{S}, \,\forall t\geq t_0$),
     where the set $\mathcal{S}$ is characterized by the minimum inter-agent safe margin and the maximum communication range.
     It refers that each agent maintains a safe space relative to its neighbors, preventing inter-agent collisions and ensuring the communication graph $\mathcal{G}(t)$ remains connected. Equivalently, no group splitting or fragmentation is allowed throughout the motion evolution.
\end{enumerate}

\section{Flocking-Cohesion Control Design}\label{sec:contr_flocking}
In this section, we introduce the source-signal-guided flocking pattern and then present two distributed flocking-cohesion controllers for the follower agents. The first one is designed based on a scalar flocking error, where an offset must be assigned along the axis of the agent to tackle the nonholonomic constraint. As an alternative, the second one is presented to avoid the usage of offset points (which introduces undesired measurement error). By leveraging an orientation-based flocking error vector, the second method ensures that the unicycle agents converge to the desired configuration while not violating the velocity constraints. 

\subsection{Source-seeking Control Law of the Leader Agent}
As introduced, the assigned leader agent $L$ is driven by the projected gradient-ascent control law, which is presented in our previous work \cite{Li_SS} and is given by  
\begin{equation}\label{eq:SS}
\bm{u}_\text{L} = \begin{bmatrix} 
 v_\text{L}\\ 
 \omega_\text{L}
\end{bmatrix}=\begin{bmatrix} 
 k_v \left\langle  {\bm{o}}(\theta_\text{L}),\nabla \bm{J}(x_\text{L},y_\text{L}) \right\rangle  \\ 
 -k_\omega \left\langle {\bm{o}}(\theta_\text{L}),\nabla \bm{J}^\perp(x_\text{L},y_\text{L}) \right\rangle
\end{bmatrix},
\end{equation}
where ${\bm{o}}(\theta_\text{L}) = [\cos(\theta_\text{L}), \,\sin(\theta_\text{L})]$ is the unit orientation vector of leader, $\nabla \bm{J}(x_\text{L},y_\text{L}) =  [\frac{\partial J}{\partial x}(x_\text{L},y_\text{L}), \,\frac{\partial J}{\partial y}(x_\text{L},y_\text{L})]$ is the measured gradient of the source field, $\nabla \bm{J}^\perp(x_\text{L},y_\text{L}) =  [- \frac{\partial J}{\partial y}(x_\text{L},y_\text{L}),\,
\frac{\partial J}{\partial x}(x_\text{L},y_\text{L}) ] $ is the corresponding orthogonal vector, 
$k_v,k_\omega >0$ denote control gains in the concave field.

\subsection{Source-signal-guided Flocking Pattern} \label{sec:flocking_pattern}
Unlike the conventional flocking patterns that primarily rely on the relative positions or inter-agent distances, this work explores the use of inter-agent relative field gradients as a novel mechanism for coordinated group flocking. Given the radially unbounded and strongly concave source field function $J(x,y)$ as in Assumption \ref{ass:J}, its corresponding Hessian is uniformly negative definite with quadratic lower- and upper bound (i.e., $\exists A, B >0, -AI \leq \nabla^2 \bm{J}(x,y) \leq -BI, \forall (x,y)\in\mathbb{R}^2$). Take the inter-agent pair $(i,j)\in \mathcal{E}$ for example, their gradient difference $\mu_{ij}$ can be linearly coupled to the corresponding Euclidean distance $d_{ij}$ as $Ad_{ij} \leq \mu_{ij} \leq Bd_{ij}$, where 
\begin{align}
 \mu_{ij}(\nabla \bm{J}_{i}, \nabla \bm{J}_{j})  &= \left \|  \nabla \bm{J}_{j}(x_j,y_j) - \nabla \bm{J}_{i}(x_i,y_i)\right \| \label{eq:mu_ij}
   \\   d_{ij}& = \left \| P_j(x_j,y_j) - P_i(x_i,y_i)\right \|, \label{eq:P_dij}
\end{align}
with $P_i(x_i,y_i)$ denoting the position of agent $i$. In this case, the signal-gradient-based flocking mechanism is a physical-level variant of the traditional distance-based flocking. Considering the agents navigate within a signal field where obstacles exist, the signal-based measurements and communication (whether referencing the source position or exchanging between neighboring agents) will remain unaffected by environmental obstructions. This stands in contrast to the distance-sensing or relative position-based flocking methods, whose performance can be severely compromised by such obstacles. In the rest of the paper, we will use this signal gradient flocking pattern to introduce the flocking-cohesion controller, inter-agent collision avoidance, and connectivity preservation.
\begin{remark}
The strong concavity assumption on the source field $J(x,y)$ (cf. Assumption \ref{ass:J}) ensures a global linear relation between the signal gradient difference and inter-agent distance. This assumption can be relaxed by restricting the analysis to a compact set, where $\nabla^2 \bm{J}$ is uniformly bounded and the strongly concavity condition holds locally.
\end{remark}

\subsection{Orientation-free Flocking-cohesion Controller}\label{sec:ori_un_flock}
\begin{figure}[htbp]
    \centering \includegraphics[width=0.4\textwidth]{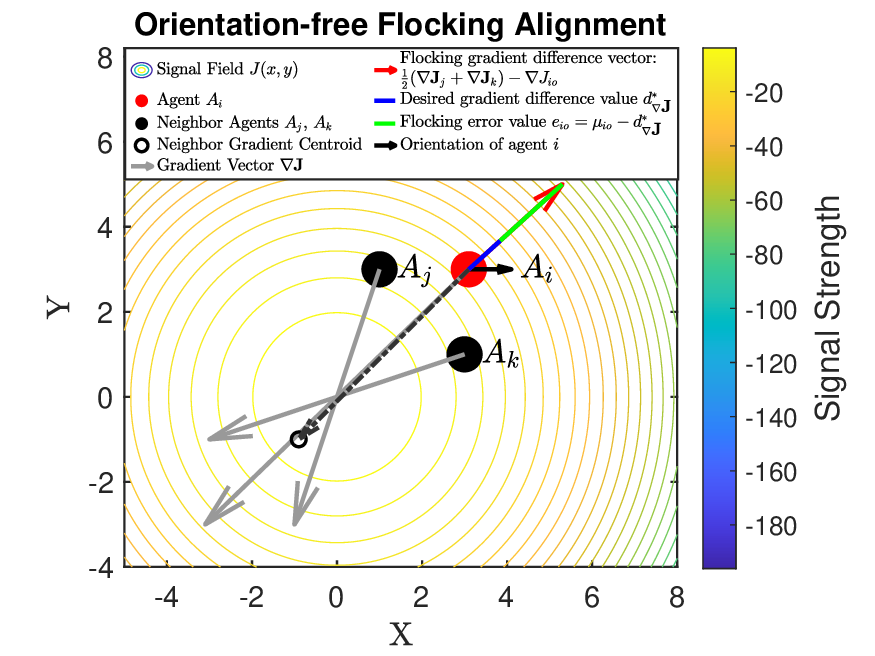}
    \vspace{-10pt}
     \caption{Illustration of orientation-free flocking alignment in a quadratic signal field. Agent $i$ (solid red dot) is controlled to flock with its two neighbors $j$ and $k$ (solid black dots). The dark dashed arrow points to the centroid of the neighbor's gradient vectors (dark hollow circle). Flocking error (scalar, denoted as a green line) is defined to coordinate the gradient difference (red arrow line) between the agent $i$ and this neighbor centroid in a desired distance $d^*_{\nabla \bm{J}}$ (solid blue line).}
    \label{fig:2D_free}
\end{figure}
As commonly adopted for controlling unicycle-type robot that has non-holonomic constraint $ \dot{x}\sin(\theta) - \dot{y}\cos(\theta) = 0$, there are two well-known feedback linearization methods which convert the unicycle model into a single-integrator\cite{Yamamoto} \cite{Lawton} and double integrator \cite{d1992dynamic}, respectively. One of the limitations of the latter approach is that the linear velocity of the robot needs to be nonzero to avoid singularity in the dynamic feedback linearization. It restricts the motion of the multi-unicycle system. Hence, we first demonstrate the flocking cohesion by converting the unicycle model into a single integrator in this subsection. The unicycle model is feedback linearized by considering an offset point $P_{io}$ shifted from the robot's center point along the longitudinal robot $X_r$-axis (denoted in the blue line in Figure~\ref{fig:unicycle}) as 
\begin{equation}\label{eq:double_integrator}
    P_{io} = (x_i + d_o\cos(\theta_{i}), y_i + d_o\sin(\theta_{i})),
\end{equation}
where $d_o\in \mathbb{R}_{>0}$ denotes a small distance from the robot center to its offset point, the offset point is prescribed in its North-East reference frame $\left\{ X,Y\right\}$. In the rest of the paper, the subscript $io$ refers to the offset point for the $i$-th agent. 

Given the \textit{source-signal-guided flocking pattern} introduced in Section \ref{sec:flocking_pattern}, we now formulate the flocking-cohesion task for the multi-agent system. For this section, we consider using the inter-agent gradient-based space function $\mu$ that is based on the local source gradient difference between the agent $i$ and its neighbors' centroid, where each local field gradient vector is obtained in the local reference frame sharing the same orientation of North and East. Specifically, the flocking field gradient distance {$\mu_{io}\in\mathbb{R}_+$} and the flocking error {$e_{io}\in\mathbb{R}$} for follower agent $i\in\mathcal{V}_f$ are defined by
\begin{equation}\label{eq:ori_uncon_error}
\begin{aligned}
    \mu_{io}(\nabla \bm{J}_{{io}}, \nabla \bm{J}_{\mathcal{N}_i}) & := \left \| \left(  \frac{1}{{N}_i}\sum_{j\in \mathcal{N}_i}  \nabla \bm{J}_{j}\right) - \nabla \bm{J}_{io} \right \| 
 \\ e_{io}  & := \mu_{io}(\nabla \bm{J}_{{io}}, \nabla \bm{J}_{\mathcal{N}_i}) - d^*_{\nabla \bm{J}}
\end{aligned}
\end{equation}
where ${N}_i=\text{dim}(\mathcal{N}_i)$, $\nabla \bm{J}_{io} $ denotes the source field gradient measured on the offset point $P_{io}$ of agent $i$. Therefore, the flocking cohesion task aims at maintaining a desired gradient difference $d^*_{\nabla \bm{J}}$ between the agents' offset point $P_{io}$ and its neighbors' gradient centroid. Figure~\ref{fig:2D_free} illustrates the flocking cohesion using this setup in the signal field. Note that the offset points are omitted in plots for visualization clarity.

As shown above, the distance offset $d_o$ in \eqref{eq:double_integrator} should be chosen sufficiently small such that the point $P_{io}$ is close to the center point of agent $i$ and the approximation $\nabla \bm{J}_{io} \approx \nabla \bm{J}_{i} $ holds. However, it is noted that the source's field gradient difference between the sensor's position and the offset point cannot be negligible in practice. An alternative flocking scheme, which does not introduce an approximation error, will be proposed later in Section \ref{sec:ori_con_flock}. Additionally, as the flocking error does not use the orientation information, there is no consensus on the final orientation of the agents when they have converged to the flocking configuration. 

For simplifying the analysis of the closed-loop system, let us describe the source-gradient difference between agent $i$ and its neighbors $\mathcal{N}_i$ in polar form as follows. Let the angle of the source-gradient difference be defined by $ \beta_i =  \text{tan}^{-1}\frac{ \sum_{j\in \mathcal{N}_i}\frac{1}{N_i} \left( \nabla {J}_{j,y} - \nabla {J}_{io,y} \right)}{ \sum_{j\in \mathcal{N}_i} \frac{1}{N_i}\left( \nabla {J}_{j,x}- \nabla {J}_{io,x}\right)}$, such that the source-gradient difference can be expressed as $\frac{1}{N_i} \sum_{j \in \mathcal{N}_i} (\nabla \bm{J}_j - \nabla \bm{J}_{io}) = \left\| \frac{1}{N_i} \sum_{j \in \mathcal{N}_i} (\nabla \bm{J}_j - \nabla \bm{J}_{io}) \right\| \, \bm{o}_{e_{io}} $ with 
\begin{equation}\label{eq:oei_vector}
    \bm{o}_{e_{io}} =  \begin{bmatrix}\cos(\beta_{i}) & \sin(\beta_{i}) \end{bmatrix}.
\end{equation}
Note that the 2-argument arctangent function $\text{atan2}(\cdot,\cdot)$ is used for computing the angle $\beta_i\in (-\pi,\pi]$. For each follower, we propose the distributed flocking controller that is defined in its local North-East reference frame, as well as the orientation and source gradient measurement. In this case, we recall that the Assumption \ref{ass:sensing} and \ref{ass:communication} hold.

Denote the orientation of agent and its orthogonal vector as 
\begin{equation}\label{eq:orientation}
    \bm{o}_i = \begin{bmatrix}
             \cos(\theta_i) & \sin(\theta_i)
            \end{bmatrix},\;\;
    {\bm{o}}^\perp_i = \begin{bmatrix} \sin(\theta_i) & -\cos(\theta_i) \end{bmatrix}, 
\end{equation}
the distributed orientation-free flocking-cohesion controller $\bm{u}_i= [v_i,\, \omega_i]^\top$ is proposed for each follower agent $i \in \mathcal{V}_f$:
\begin{subequations}\label{eq:ori_free_controller}
\begin{align}
         v_i &= K_f e_{io} \bm{o}_i\Big(\nabla^2 \bm{J}_{io}\Big)^{-1}  \bm{o}^\top_{e_{io}}
         +\frac{1}{N_i} \sum_{k\in \mathcal{N}_i}  v_k  \bm{o}_i\bm{P}_{ik}  \bm{o}^\top_k \label{eq:ori_free_error_v} \\
    \omega_{i} &=   -\frac{K_f e_{io}}{d}\  {\bm{o}}^\perp_i  \Big(\nabla^2 \bm{J}_{io}\Big)^{-1}  \bm{o}^\top_{e_{io}}
 -\frac{1}{{N}_i}\sum_{k\in \mathcal{N}_i }  \frac{v_k}{d}  {\bm{o}}^\perp_i\bm{P}_{ik} \bm{o}^\top_k
 \label{eq:ori_free_error_w}
\end{align}
\end{subequations}
where ${N}_i=\text{dim}(\mathcal{N}_i)$, $\bm{o}_{e_{io}}$ is defined in \eqref{eq:oei_vector},  the control parameters are given by $ K_f>0$ and $ \bm{P}_{ik} = \Big(\nabla^2 \bm{J}_{io}\Big)^{-1}\nabla^2 \bm{J}_{k}$. 

\begin{remark}\label{remark:contraction_mapping}
Consider the implicit function for the longitudinal velocity in the flocking controller \eqref{eq:ori_free_controller} by $v = \mathcal{M}(v,\rho)$, where $v = \mathop{\text{col}}\limits_{i\in\mathcal V}(v_i)$ is a column vector and $\rho$ contains all remaining variables. One can check that for a given $\rho$, the resulting $\mathcal{M}$ defines a local contraction mapping w.r.t. $v$, for instance, when the neighboring agents are close to each other such that $\nabla^2\bm{J}_k \approx \nabla^2\bm{J}_{io}$ for all $k\in\mathcal N_i$. 
Correspondingly, when $\rho$ has slow dynamics, by time-scale separation, the solution $v$ can be obtained {\rm distributedly} by using the update law $v(\eta) = \mathcal{M}(v(\eta-1),\rho)$, where $\eta$ is the update time step taken place in the fast time-scale. In this case, for each agent, the update at time $\eta$ is based on the information from its neighbors at time $\eta-1$. By the contraction mapping theorem, there exists a fixed point $v^*$ such that $\lim_{\eta\rightarrow\infty}v(\eta) = v^* = \mathcal{M}(v^*,\rho)$. 
\end{remark}

\begin{figure}[htbp]
    \centering
    \includegraphics[width=0.4\textwidth]{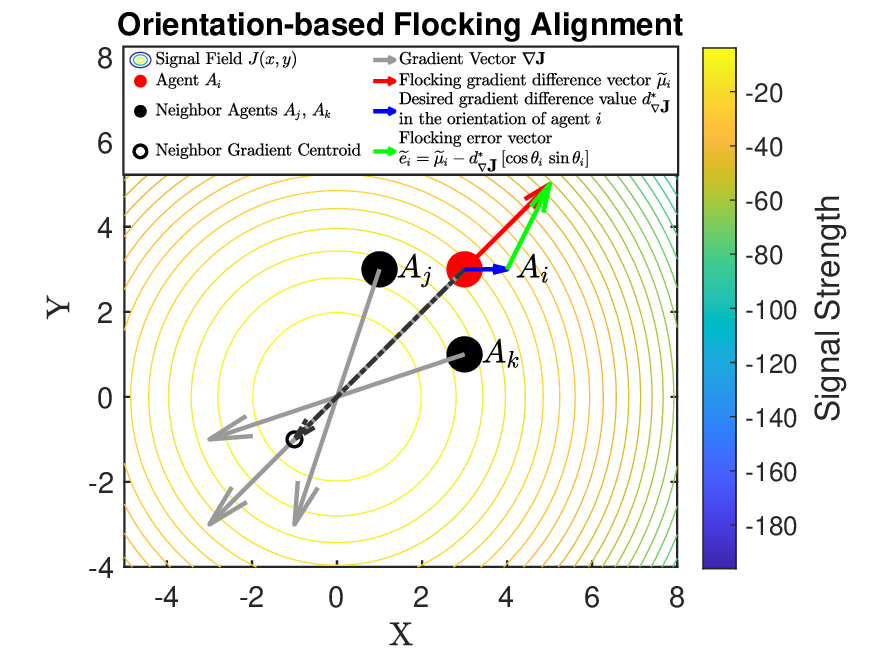}
    \vspace{-12pt}
    \caption{Illustration of an orientation-based flocking alignment in a quadratic signal field. The red arrow line denotes the difference vector between the agent $i$ and its neighbors' averaging signal gradient. The flocking error vector (green arrow line) is defined to align the agent $i$'s orientation (blue arrow line) with this averaging gradient difference vector (red arrow line) in a desired distance $d^*_{\nabla \bm{J}}$ (c.f. the scalar flocking error in the orientation-free flocking setup in Figure~\ref{fig:2D_free}). }
    \label{fig:2D_based}
\end{figure}
\begin{theorem}[Orientation-free Flocking Cohesion]\label{thm:ori_uncon_flock}
Let $\mathcal{G}$ be a static connected undirected graph representing a multi-agent system of $n$ unicycle agents \eqref{eq:unicycle_model}, each equipped with local sensor systems measuring the field gradient $\nabla \bm{J}$ that satisfies Assumption \ref{ass:J}. Suppose that the leader is the neighbor of at least one follower (i.e. $L \in \mathcal{N}_{i\in\mathcal{V}_f}$) in the graph $\mathcal{G}$. 

Then the proposed distributed orientation-free flocking cohesion control law $\bm{u}_i = [v_i,\, \omega_i ]^\top$, with controlled longitudinal \eqref{eq:ori_free_error_v} and angular \eqref{eq:ori_free_error_w} velocity for all follower agents $i \in \mathcal{V}_f$, and with the source-seeking controller \eqref{eq:SS} for the leader $L$, can achieve both the flocking-cohesion and source-seeking tasks (i.e., \eqref{eq:convergence-0} holds and the flocking error defined in \eqref{eq:ori_uncon_error} converges $e_{io}(t)\to 0$ as $t\to\infty$ for all $i$).  
\end{theorem}
\begin{proof}
    See Appendix \ref{Appendix:thm_ori_uncon_flock}.
\end{proof}

\subsection{Orientation-based Flocking-cohesion Controller}\label{sec:ori_con_flock}
As introduced in \textit{Orientation-free Flocking-cohesion}, the flocking error $e_{io}$ for each agent is defined as a scalar function that describes the discrepancy between the gradient distance measure {$\mu_{io} \in \mathbb{R}_{+}$} and the desired constant $d^*_{\nabla \bm{J}}\in\mathbb{R}_{>0}$ as in \eqref{eq:ori_uncon_error}. Accordingly, the approach taken in Theorems \ref{thm:ori_uncon_flock} relies on the feedback linearization of the unicycle for nonholonomic constraint. In practice, the assignment of offset point $P_{io}$ results in approximation error between $\nabla \bm{J}_{io}$ and $\nabla \bm{J}_{i}$. 

Alternatively, to address the flocking problems without using feedback linearization and to avoid the resulting approximation error between the offset point and agent's center-of-mass, we introduce a vector-based form of relative information for flocking cohesion. For each follower, the flocking alignment is expressed in vector form, with the source relative gradient measurement denoted by $\widetilde{\bm{\mu}}_i\in \mathbb{R}^{1\times 2}$, and the flocking cohesion error given by the vector $\widetilde{\bm{e}}_i \in \mathbb{R}^{1\times 2}$, as follows
\begin{align}\label{eq:ori_con_error}
\widetilde{\bm{\mu}}_i(\nabla \bm{J}_{i}, \nabla \bm{J}_{\mathcal{N}_i})  
& := \left( \frac{1}{{N}_i}\sum_{j\in \mathcal{N}_i}  \nabla \bm{J}_{j}\right) - \nabla \bm{J}_{i}  
\\ \widetilde{\bm{e}}_{i} & :=  \widetilde{\bm{\mu}}_i(\nabla \bm{J}_{i}, \nabla \bm{J}_{\mathcal{N}_i}) - d^*_{\nabla \bm{J}}  \begin{bmatrix}\cos(\theta_i) & \sin(\theta_i) \end{bmatrix} \notag
\end{align}
where the error vector $\widetilde{\bm{e}}_{i}$ represents the deviation between the observed relative field gradient vector $\widetilde{\bm{\mu}}_i$ and the desired one $d^*_{\nabla \bm{J}} [\cos(\theta_i),\, \sin(\theta_i)]$, with $d^*_{\nabla \bm{J}} \in \mathbb{R}_{>0}$
denoting the desired gradient distance.
An illustration of the vector form of relative information used in this orientation-based flocking cohesion approach is shown in Figure~\ref{fig:2D_based}.

While introducing the new relative gradient vector $\widetilde{\bm{\mu}}_i$ and $\widetilde{\bm{e}}_i$, we maintain the same system setup as in Section \ref{sec:ori_un_flock} and the Assumption \ref{ass:communication} still holds. Accordingly, a new distributed flocking-cohesion controller $\bm{u}_i= [v_i,\, \omega_i]^\top$ for each follower agent $i\in\mathcal{V}_f$ is given as follows 
\begin{subequations}\label{eq:ori_based_controller}
\begin{align}
v_i &= k_{fv} \widetilde{\bm{e}}_{i} \nabla^2 \bm{J}_{i}  \bm{o}_i^\top
       + \frac{1}{N_i}\sum_{k\in \mathcal{N}_i} v_k
         \frac{ \bm{o}_i\nabla^2\bm{J}_{k} \bm{o}^\top_k}{ \bm{o}_i\nabla^2\bm{J}_{i}\bm{o}^\top_i} \label{eq:ori_con_error_v} \\
\omega_i &= -\frac{k_{f\omega}}{d^*_{\nabla \bm{J}}} \widetilde{\bm{e}}_{i}  (\bm{o}^\perp_i)^\top
       + \frac{1}{N_i}\sum_{k\in \mathcal{N}_i}\frac{v_k}{d^*_{\nabla \bm{J}}}
         \frac{ \bm{o}^\perp_i \bm{Q}_i\nabla^2\bm{J}_{k} \bm{o}^\top_k}{ \bm{o}_i\nabla^2\bm{J}_{i} \bm{o}^\top_i} \label{eq:ori_con_error_w}
\end{align}
\end{subequations}
where $\bm{o}_i, \bm{o}^\perp_i$ are defined in \eqref{eq:orientation}. It is assumed that $\nabla \bm{J}_{k}$ and $ \nabla \bm{J}_{i}$ are shared between agents $i,k$ if $(i,k)\in \mathcal{E}$, as given in Assumption \ref{ass:communication}. The flocking control gains are defined to be $k_{fv},k_{f\omega}>0$ and $ \bm{Q}_i=  \sbm{0  & 1 \\-1 &0 }\nabla^2\bm{J}_{i} \sbm{0  & 1 \\-1 &0 }$. Similar to Remark~\ref{remark:contraction_mapping}, 
the computation of velocity input in the flocking controller \eqref{eq:ori_con_error_v} and \eqref{eq:ori_con_error_w} can be done in a distributed way.

\begin{theorem}[Orientation-based Flocking Cohesion]\label{thm:ori_con_flock}  Let us consider the multi-agent systems dynamics and graph assumptions as in Theorem~\ref{thm:ori_uncon_flock} where the $n$-unicycle systems \eqref{eq:unicycle_model} are communicating with a static connected undirected graph $\mathcal{G}$. Assume that the second-order and third-order partial derivative of $J_i(x_i,y_i)$ is bounded for all $ \left(x_i, y_i\right)$. Consider the distributed orientation-based flocking control laws $\bm{u}_i= [v_i,\, \omega_i]^\top$ given in \eqref{eq:ori_con_error_v}-\eqref{eq:ori_con_error_w} for all follower agents $i\in \mathcal{V}_f$. Then the closed-loop system achieves both the flocking-cohesion and source-seeking tasks (i.e., \eqref{eq:convergence-0} holds and the flocking error defined in \eqref{eq:ori_con_error} converges $\widetilde{\bm{e}}_i \to \bm{0}_{1\times 2}$ as $t\to \infty$ for all $i$). 
\end{theorem}
\begin{proof}
    See Appendix \ref{Appendix:thm_ori_con_flock}.
\end{proof}
\vspace{0.1cm}

\section{Inter-agent Safety and Connectivity} \label{set:connectivity}
In the analysis of Theorem~\ref{thm:ori_uncon_flock} and \ref{thm:ori_con_flock}, the communication graph $\mathcal{G}$ is assumed to be static and connected for all time to achieve flocking cohesion and avoid splitting or fragmentation \cite{Ji, Olfati-2, Jadbabaie, Zavlanos}. When the communication graph is dynamic, the connectivity assumption becomes critical during the flocking evolution. Though the widely adopted distance-based potential function method can guarantee connectedness, the constraint on the input can introduce a limiting factor that may disrupt connectivity. 
In this section, a distributed CBF-QP framework is proposed for the multi-agent system with a dynamic communication graph. For the followers, the objective is to achieve safe flocking-cohesion while ensuring the communication graph remains connected among inter-agent pairs. For the leader, the optimization is to realize safe source-seeking while maintaining the control inputs within an admissible input set.

\subsection{Dynamic Communication Graph and Connectivity}
As described in the problem formulation and the flocking pattern design,  all agents locally measure the source field gradient $\nabla \bm{J}_i(x_i, y_i)$ and achieve flocking behavior by aligning with the source signal gradient. Suppose that none of the agents have range sensors and consider the spatial interpretation in Section \ref{sec:flocking_pattern}, we adopt the inter-agent gradient-based space function $ \mu_{ij}$ defined in \eqref{eq:mu_ij} as a proxy of distance.
Accordingly, we assume that each agent $i$ is only able to interact with the rest when $ \mu_{ij}$ is within the limited communication range $r$. To ensure flocking cohesion, this further implies that the communication range must satisfy $r >d^*_{\nabla \bm{J}}$.
Let us define the communication error $ \delta_{\text{com-ij}}$ by
\begin{equation}\label{eq:error_com}
        \delta_{\text{com-ij}} =  r - { \mu_{ij}}(\nabla \bm{J}_{i}, \nabla \bm{J}_{j}),
\end{equation}
this setup leads to dynamic neighbor sets $\mathcal{N}_i(t)$, whose composition depends on the spatial communication error $\delta_{\text{com-ij}}$ as $ \mathcal{N}_i(t)  = \left \{ j \neq i\in \mathcal{V} : \delta_{\text{com-ij}}(t)  > 0 \right \}$, and the dynamic edge set $\mathcal{E}(t) =   \left \{  (i,j): i\in\mathcal{V}, j\in\mathcal{N}_i(t)\right \}$. Thus the dynamic communication topology of the multi-agent systems is described by a dynamic undirected graph $\mathcal{G}(t) = (\mathcal{V},\mathcal{E}(t))$ and the associated switching times are represented by the time sequence $\left\{ t_k\right\}_{k\in\mathbb{N}}$. Throughout this section, the connectivity preservation is analyzed within the framework of the dynamic graph $\mathcal{G}:=\mathcal{G}(t)$. For clarity, the time argument $t$ is omitted unless explicitly required in the rest content.
\begin{remark}[Inter-agent Connectivity]\label{remark:connectivity}
Consider the dynamic communication graph $\mathcal{G}(t)$ for the multi-agent system, the agents $i$ and $j$ are said to be \textit{connected} at time $t$ if the undirected edge $(i,j)\in\mathcal{E}(t)$ exists.
\end{remark}

\subsection{Inter-agent Safety and Connectivity}
\label{sec:CBF_conn}
\subsubsection{Safety Analysis}
To simplify the analysis, we assume throughout the paper that all agents are point-mass agents such that the inter-agent collision occurs only when the Euclidean distance between two agents falls below a small threshold. In practice, this assumption can be relaxed to agents that occupy an area or volume, where a fixed minimum safe separation must be maintained to avoid inter-agent collision. Though our approach uses the field gradient information that may not be directly linked to the Euclidean distance, the safe separation can be coupled to the gradient-based space measure  $\mu_{ij}$ defined in \eqref{eq:mu_ij} if the Hessian of the field $J$ is uniformly negative definite. In this case, a collision event of any pair of agents (i.e., $P_i(x_i,y_i)= P_j(x_j,y_j)$ in \eqref{eq:P_dij}) during the motion evolution implies that $\mu_{ij}(\nabla \bm{J}_{i}(x_i,y_i), \nabla \bm{J}_{j}(x_j,y_j)) = \left \|  \nabla \bm{J}_{j}(x_j,y_j) - \nabla \bm{J}_{i}(x_i,y_i)\right \| = 0$ for the point-mass agent. 
\begin{remark}[Safety]\label{reamrk:safety}
   Given the Hessian assumption of the source field $J$, let us consider the inter-agent gradient-based space function $\mu_{ij}$ defined in \eqref{eq:mu_ij}, the multi-agent system is said to be \textit{safe} at time $t$ if the separation error $\delta_{\text{sep-ij}}$ is 
\begin{equation}\label{eq:sep}
  \delta_{\text{sep-ij}} = \mu_{ij} - d_r >0, \quad \forall(i,j)\in\mathcal{E}(t)
\end{equation}
where $d_r \in \mathbb{R}_+$ denotes the minimum safe margin in the gradient space. The system is said to be \textit{collision-free} over the time interval $\left [0,T \right]$ if the safety condition \eqref{eq:sep} holds for all agent pairs and for all $t\in \left [0,T \right]$.
\end{remark}

\subsubsection{Inter-agent zeroing control barrier function (ZCBF)}
In the general unicycle-model-based zeroing control barrier function (ZCBF) strategy, a virtual offset point is typically introduced to address the mixed relative degree problem. In this section, we build upon our recent work \cite{Li_CBF} that ensures the relative degree of the distance- and orientation-based ZCBF for the unicycle robot is one and uniform for both input variables. Accordingly, no virtual offset point is required as in prior literature, and the inter-agent collision avoidance can directly be defined with respect to the agents' center point-of-mass. 

Consider the extended state space for each unicycle follower agent $i\in\mathcal{V}_f$ in \eqref{eq:unicycle_model} as follows
\begin{equation}\label{eq:ext_dynamics}
\dot{\bm{\xi}_i } 
= \begin{bmatrix} 
  \dot{x}_i\\ 
  \dot{y}_i\\ 
  \dot{v}_i\\
  \ddot{x}_i \\
  \ddot{y}_i
\end{bmatrix}
=\underbrace{\begin{bmatrix} 
  v_i\cos(\theta_i)\\ 
  v_i\sin(\theta_i)\\ 
  0\\ 
  0\\ 
  0
\end{bmatrix}}_{f_i(\bm{\xi}_i)}+\underbrace{\begin{bmatrix} 
  0 & 0 \\ 
  0 & 0 \\ 
  1 & 0 \\ 
  \cos(\theta_i) & -v_i\sin(\theta_i)\\ 
  \sin(\theta_i) & v_i\cos(\theta_i)
\end{bmatrix}}_{g_i(\bm{\xi}_i)}\underbrace{\begin{bmatrix} 
  a_i\\ 
  \omega_i
\end{bmatrix}}_{\bm{u}_i}
\end{equation}
where the control input $\bm{u}_i$ is with acceleration $a_i$ and angular velocity $\omega_i$, and the extended state is defined as $ \bm{\xi}_i = [
  {\xi}_{i,1},\, 
  {\xi}_{i,2},\,
  {\xi}_{i,3},\,
  {\xi}_{i,4},\,
 {\xi}_{i,5} ]^\top =  [x_i,\,  y_i ,\,
  v_i,\,
  \dot{x}_i,\,
  \dot{y}_i]^\top$. For the multi-agent system at time $t$, we propose the inter-agent ZCBF $h_{ij}(t)$ for each follower $i$ and its neighbor $j\in\mathcal{N}_i(t)$ within the communication range $r$ as
  \begin{subequations}\label{eq:CBF_hij}
    \begin{align}
          h_{ij} & = D_{ij}(e^{-P_{ij}} + e^{-P_{ji}})  \label{eq:hij} \\ 
        \text{with} \quad  D_{ij} & =  \underbrace{(r - \mu_{ij})}_{\delta_{\text{com-ij}}}\underbrace{(\mu_{ij}-d_r)}_{\delta_{\text{sep-ij}}},  \label{eq:Dij}
    \end{align}
\end{subequations}
where $P_{ij}, P_{ji}$ are scalar functions representing the orientation projections between agents $i$ and $j$, which will be formally defined below. Note that $\mu_{ij}$ is the gradient difference norm defined in \eqref{eq:mu_ij}, $\delta_{\text{com-ij}}, \delta_{\text{sep-ij}}$ represent the communication and separation error as in \eqref{eq:error_com}, \eqref{eq:sep}, respectively. 
\begin{remark}
As introduced in Section \ref{sec:flocking_pattern}, the signal-gradient-based flocking is regarded as a physical-level variant of distance-based flocking. Given the further interpretation of \textit{inter-agent connectivity} and \textit{safety} in Remark~\ref{remark:connectivity} and \ref{reamrk:safety}, the limited communication range $r$ and the minimum safe margin $d_r$ jointly define spatial constraints that regulate interactions among agents during flocking evolution.
\end{remark}

Accordingly, the non-negative space function $D_{ij}$ in \eqref{eq:Dij} characterizes whether the neighboring agents $i$ and $j$ are within their communication range (i.e. if ${ \mu_{ij}} < r$ then $(i,j)$ is an edge in $\mathcal{E}(t)$), and whether they are keeping a safe relative range (i.e. ${ \mu_{ij}}>d_r$). It is equal to zero when their gradient difference norm reaches the maximum communication range (i.e. ${ \mu_{ij}} \to r$) or when these two agents enter the minimum safe range  (i.e. ${ \mu_{ij}} \to d_r$). The pre-set ranges are defined as $r>d_r>0$ such that $h_{ij} >0$ if and only if $d_r<{ \mu_{ij}}<r$.

On the other hand, with the extended state $\bm{\xi}_i$ defined in \eqref{eq:ext_dynamics}, the unit orientation vector of agent $i$ can be rewritten as $\bm{o}_i = [
    \frac{\xi_{i,4}}{\xi_{i,3}},\, \frac{\xi_{i,5}}{\xi_{i,3}}]=[\frac{\dot{x_i}}{v_i},\, \frac{\dot{y_i}}{v_i}]$. Considering the connected neighboring pair $(i,j)\in \mathcal{E}(t)$, we note that the corresponding unit bearing vector for agent $i$ and $j$ is obtained by using their individual bearing angle $\phi_{ij}, \phi_{ji}$ as
$\bm{b}_{ij} = [
        \cos(\phi_{ij}),\, \sin(\phi_{ij})]$ and $ \bm{b}_{ji} =[
        \cos(\phi_{ji}),\,  \sin(\phi_{ji})]$, respectively.
Using these notations, we construct the orientation projection functions for the inter-agent ZCBF \eqref{eq:CBF_hij} as
$ P_{ij}  =  \left \langle {\bm{o}}_{i}, {\bm{b}}_{ij} \right \rangle +v_i \epsilon$ and $ P_{ji} =  \left \langle {\bm{o}}_{j}, {\bm{b}}_{ji} \right \rangle +v_j \epsilon$,
where $\left \langle \cdot,\cdot \right \rangle$ denotes the usual inner-product operation. Note $\epsilon \in \mathbb{R}_{>0}$ is chosen as a sufficiently small offset and designed to ensure $\left \| L_{g_i}h_{ij} \right \| \neq 0$ and $ \left \| L_{g_j}h_{ji} \right \| \neq 0$ for a valid ZCBF.
\begin{proposition}[Uniform Relative Degree]\label{pro:relative_degree}
    Let the unicycle agent $i$ be governed by the extended dynamics \eqref{eq:ext_dynamics}. For any neighboring pairs $(i,j)\in\mathcal{E}$, the inter-agent control barrier function $h_{ij}$ in \eqref{eq:hij} has relative degree one with respect to the control inputs $a_i$ and $\omega_i$, uniformly over the domain where $d_r< \mu_{ij}<r$.    
\end{proposition}
\begin{proof}
See Appendix \ref{Appendix:Uniform relative degree}. 
\end{proof}
\vspace{0.1cm}
Note this property will later be used in the proof of optimal solution (c.f. Theorem~\ref{thm:feasibility} below).

\subsubsection{Distributed Control Barrier Function-based Quadratic Programming (CBF-QPs)}
We impose a time-invariant admissible input set $\mathcal U_{\text{adm}}$ for all agents in the implementation, given by
\begin{equation}\label{set:adm_conn}
\begin{aligned}
\mathcal{U}_{\text{adm}}
  = \Bigl\{ \bm{u}_{i\in \mathcal{V}} =
    \begin{bmatrix}
      u_{i,1} \\[2pt] u_{i,2}
    \end{bmatrix}
    \in \mathbb{R}^2 \Bigm|
    & a_{\min} \leq u_{i,1} \leq a_{\max}, \\
    & \omega_{\min} \leq u_{i,2} \leq \omega_{\max}
  \Bigr\}.
\end{aligned}
\end{equation}
Then, with the nominal control inputs (i.e., source-seeking $\bm{u}_{\text{ss}}$ is given in \eqref{eq:SS} for the leader $L$, flocking-cohesion $ \bm{u}_{\text{flock-i}}$ is given in \eqref{eq:ori_free_controller} or \eqref{eq:ori_based_controller} for the follower $i\in\mathcal{V}_f$), we introduce the distributed CBF-QPs for inter-agent safety (collision avoidance) and connectivity preservation, as follows.
\begin{itemize}
    \item {\bf \textit{For Leader agent} $L$:}
\end{itemize}
\begin{subequations}\label{QP_framework:leader}
\begin{equation}\label{QP:leader}
\begin{aligned}
   &  \begin{bmatrix}
\bm{u}^*_{\text{L}} \\
\bm{\Upsilon}^*_{\text{L}}  
\end{bmatrix} = \mathop{\mathrm{argmin}}
\frac{1}{2}\left(
\left\| \bm{u}_{\text{L}} - \bm{u}_{\text{ss}} \right\|^2
+ 
\left\| \bm{\Upsilon}_{\text{L}} - \mathop{\mathrm{col}}\limits_{k\in\mathcal{N}_{\text{L}}}(\gamma_{\text{ref-Lk}}) \right\|^2
\right)
\end{aligned}
\end{equation}
\begin{align}\label{QP:leader_QP_CBF}
\text{s.t.} \quad &
L_{f_{\text{L}}}h_{\text{Lk}} + 
L_{g_{\text{L}}}h_{\text{Lk}} \bm{u}_{\text{L}}  + \gamma_{\text{Lk}} \alpha(h_{\text{Lk}}) \geq 0, \quad 
\forall k \in \mathcal{N}_{\text{L}}(t)\nonumber
 \\& \qquad \hspace{1.8cm}\text{(Inter-agent CBF Constraint)}
\end{align}
\begin{equation}\label{QP:leader_QP_adm}
\qquad  \bm{u}_{\text{L}} \in \mathcal{U}_{\text{adm}} \hspace{0.90cm}  \text{(Admissible Input Constraint)} \end{equation} 
\begin{equation}\label{QP:leader_QP_gamma}
\qquad  \bm{\Upsilon}_{\text{L}} \in \mathbb{R}_+^{N_{\text{L}}}  \hspace{2.3cm} \text{(Weight Constraint)}
\end{equation} 
\end{subequations}
\begin{itemize}
    \item {\bf \textit{For Follower agent $i\in\mathcal{V}_f$}:}
\end{itemize}
\begin{subequations}\label{QP_framework:follower}
\begin{equation}\label{QP:follower}
\begin{aligned}
\begin{bmatrix}
\bm{u}^*_i \\
\bm{\Upsilon}^*_i  
\end{bmatrix}
&= \mathop{\mathrm{argmin}}
\frac{1}{2}\left(
\left\| \bm{u}_i - \bm{u}_{\text{flock-i}} \right\|^2
+ \left\| \bm{\Upsilon}_i - \mathop{\mathrm{col}}\limits_{j\in\mathcal{N}_i}(\gamma_{\text{ref-ij}}) \right\|^2
\right)
\end{aligned}
\end{equation}
\begin{align}\label{QP:follower_QP_CBF}
\text{s.t.} &\quad
 L_{f_i}h_{ij} + L_{g_i}h_{ij} \bm{u}_i + \gamma_{ij} \alpha(h_{ij}) \geq 0, \quad \forall j \in \mathcal{N}_i(t) \nonumber
 \\& \qquad \hspace{2.2cm} \text{(Inter-agent CBF Constraint)} 
\end{align}
\begin{equation}\label{QP:follower_QP_adm}
\qquad \bm{u}_i \in \mathcal{U}_{\text{adm}}  \hspace{1.0cm}   \text{(Admissible Input Constraint)}
\end{equation}
\begin{equation}\label{QP:follower_cons_gamma}
 \qquad \bm{\Upsilon}_i \in \mathbb{R}_+^{N_i} \hspace{2.4cm}  \text{(Weight Constraint)}
\end{equation}
\end{subequations}
Let us take the optimization framework \eqref{QP_framework:follower} of followers for illustration. In the quadratic programming problem \eqref{QP:follower}, with ${N}_i=\text{dim}(\mathcal{N}_i(t))$, the column vector $\mathop{\text{col}}\limits_{j\in\mathcal{N}_i}( \gamma_{\text{ref-ij}}) \in \mathbb{R}^{N_{i}}$ contains the reference weight parameter $\gamma_{\text{ref-ij}}\in\mathbb{R}_+$ for agent $i$ with respect to its neighbor $j\in\mathcal{N}_i(t)$. Correspondingly, $\bm{\Upsilon}^*_i  = \mathop{\text{col}}\limits_{j\in\mathcal{N}_i}(\gamma^*_{ij}) \in \mathbb{R}_+^{N_{i}}$ is an optimized stacked vector that composes of the updated weight parameters $\gamma^*_{ij}$. In constraint condition \eqref{QP:follower_QP_CBF}, $\alpha(\cdot)$ denotes an extended class $\mathcal{K}_{\infty}$ function. Accordingly, the derived optimal solution $\bm{u}^*_i$ must satisfy one admissible control input set $\mathcal{U}_{\text{adm}}$ that restricts the physical action of robot as in \eqref{set:adm_conn}, and another CBF constraint set $\mathcal{U}_{\text{cbf-i}}(t)$ that restricts the inter-agent safety (collision-avoidance) and connectivity-preserving condition \eqref{QP:follower_QP_CBF} as
\begin{align}\label{set:CBF-cons}
\mathcal{U}_{\text{cbf-i}}(t)
  := \Bigl\{ \bm{u}_i \in \mathbb{R}^2 \;\Bigm|\;
     H_{ij} \geq 0, \,\, \forall j \in \mathcal{N}_i(t)  \Bigr\},
\end{align}
where $H_{ij} = L_{f_i} h_{ij} + L_{g_i} h_{ij} \bm{u}_i + \gamma_{ij} \alpha(h_{ij})$.
\begin{remark}[CBF-QP Feasibility Condition]\label{def:feasibility}
   Consider the inter-agent CBF $h_{ij}(\bm{\xi}_i(t), \bm{\xi}_j(t))$ defined in \eqref{eq:CBF_hij}, built on the extended state $\bm{\xi}_i(t), \bm{\xi}_j(t)$ for agent pair $(i,j)\in\mathcal{E}(t)$ at time $t$, the CBF-based quadratic programming (QP) problem \eqref{QP_framework:follower} is said to be \textit{feasible} if $\bm{u}^*_i \in \mathcal{U}_{\text{adm}}\bigcap\mathcal{U}_{\text{cbf-i}}(t)\neq \emptyset$. In other words, the solution $\bm{u}^*_i$ exists at time $t$.
\end{remark}

Ideally, the QP feasibility can be achieved by selecting a suitable extended class $\mathcal{K}_{\infty}$ function $\alpha(h_{ij})$. To do so, a state-dependent weight $\gamma_{ij}$ is proposed for the edge $(i,j)$ and it is updated concerning the reference weight parameter $\gamma_{\text{ref-ij}}$, as shown in \eqref{QP:follower}. 
With the above setup for the multi-agent systems, let us define a \textit{safe spatial communication set} $\mathcal{S}_{ij}$ for the connected agent pair $(i,j)\in\mathcal{E}(t)$ at time $t$ as
\begin{equation}\label{set:sij}
    \mathcal{S}_{ij} = \left \{ (\bm{\xi}_i, \bm{\xi}_j)\in \mathbb{R}^{5} \times \mathbb{R}^{5} \ | \ h_{ij}(\bm{\xi}_i,\bm{\xi}_j)> 0 \right \},
\end{equation}
which implies that the agent $i$ is at a safe and communicable distance from its neighbor $j\in\mathcal{N}_i(t)$ (i.e., $d_r<\mu_{ij}<r$), such that the agent pair $(i,j)\in\mathcal{E}(t)$ exists. Therefore, given the definition of \textit{inter-agent connectivity} and \textit{safety} in Remark~\ref{remark:connectivity} and \ref{reamrk:safety}, for the time $t$, the agents $i$ and $j$ are said to be \textit{connected} and \textit{collision-free} if $(\bm{\xi}_i, \bm{\xi}_j)\in \mathcal{S}_{ij}$. 
\begin{theorem}[QP Feasibility]\label{thm:feasibility}
    Let us consider the distributed CBF-based quadratic programming problems \eqref{QP_framework:leader}-\eqref{QP_framework:follower} for the multi-agent system. For a given time $t>0$, suppose that $(\bm{\xi}_i(t), \bm{\xi}_j(t))\in\mathcal{S}_{ij}$ holds for all $(i,j)\in\mathcal E(t)$, where $\mathcal E(t)$ denotes the set of connected edges. Then, the QP is feasible at time $t$. 
\end{theorem}
\begin{proof}
    See Appendix \ref{Appendix:feasibility}.
\end{proof}
\vspace{0.1cm}

\begin{theorem}[Inter-agent Safety and Connectivity]\label{thm:connectivity}
Consider the multi-agent system consist of $n$ unicycle agents, each defined by the extended state \eqref{eq:ext_dynamics} in $\Xi\subset \rline^2\times (0,\infty) \times \rline^2$, with globally Lipschitz $f$ and locally Lipschitz $g$. The followers $i\in\mathcal{V}_f$ are driven by the Lipschitz continuous solutions $\bm{u}^*_i$, which are obtained from the corresponding CBF-QP \eqref{QP_framework:follower} subject to the inter-agent safety and connectivity constraints \eqref{QP:follower_QP_CBF} and the admissible control input constraint \eqref{QP:follower_QP_adm}. Similarly, let the leader agent $L$ be driven by the optimal solutions $\bm{u}^*_{\text{L}}$ obtained from its corresponding CBF-QP \eqref{QP_framework:leader}. Suppose the initial graph $\mathcal{G}(t_0)$ is connected such that the assigned leader is initially the neighbor of at least one follower agent. Then the following properties hold:
\begin{description}
\item[{\bf P1}.] The inter-agent safe spatial communication set $\mathcal{S}_{ij}$,  defined in \eqref{set:sij}, based on the minimum safe range $d_r$ and maximum communication range $r$ ($r>d^*_{\nabla \bm{J}}>d_r>0$), is forward invariant. \item[{\bf P2}.] The dynamic communication graph $\mathcal{G}(t)$ remains connected for all time $t\geq t_0$ (i.e., the graph connectivity-preservation holds).
\item[{\bf P3}.] All trajectories of the system are guaranteed to be inter-agent collision-free for all time $t\geq t_0$.
\end{description}
\end{theorem}
\begin{proof}
    See Appendix \ref{Appendix:thm_connectivity}.
\end{proof}
\vspace{0.1cm}

In Section \ref{sec:contr_flocking}, the convergence of both flocking-cohesion control laws (where the flocking error $e_{io} \rightarrow 0$ in orientation-free method \eqref{eq:ori_free_controller} and $\widetilde{\bm{e}}_i \rightarrow \bm{0}_{1,2}$ in the orientation-based method \eqref{eq:ori_based_controller}) have been given in Theorem~\ref{thm:ori_uncon_flock} and \ref{thm:ori_con_flock}, respectively. For further discussing the existence of other undesired stationary states while involving the inter-agent safety and connectivity constraints, we need to consider the optimal control input $\bm{u}^*$ derived from the corresponding CBF-QPs when it is active. A corresponding lemma is proposed below.
\begin{lemma}[Steady-State Convergence]\label{lemma:equil}
 Given the initially connected multi-agent system in a strictly concave field $J(x,y)$, where the leader and followers are driven by the optimal $\bm{u}^*_\text{L}$ in \eqref{QP_framework:leader} and $\bm{u}^*_i$ in \eqref{QP_framework:follower}, respectively. Then the followers converges to the steady-state where $e_{io} = 0$ (or $\widetilde{\bm{e}}_i = \bm{0}_{1,2}$) as per Theorem~\ref{thm:ori_uncon_flock} (or Theorem~\ref{thm:ori_con_flock}), under the condition that the desired flocking configuration can be formed in geometry without violating the inter-agent safety and connectivity requirements. Otherwise, the followers will be stationary at the steady-state where $h_{ij}>0$ (safety and connectivity guaranteed) and $e_{io} \neq 0$ (or $\widetilde{\bm{e}}_i \neq \bm{0}_{1,2}$). 
\end{lemma}
\begin{proof}
    See Appendix \ref{Appendix:lemma_equil}.
\end{proof}

\section{Simulation Results} \label{sec:simulation}
In this section, we demonstrate a number of simulation results of multi-agent source-seeking and flocking-cohesion in a field 
that is given by a concave signal distribution $J(x,y) = - \sbm{x&y }\bm{H} \sbm{x\\y }$, where $\bm{H}=\bm{H}^\top>0$. Without loss of generality, the source of the signal distribution is located at $(0,0)$ with $\bm{H}=\bm{I}$ in the following simulations.

\begin{figure}[htbp]
	\centering
    \subfigure[]{
		\begin{minipage}[t]{0.24\textwidth}
		    \label{fig:flock_free_traj}
			\centering			
            \includegraphics[width=\textwidth]{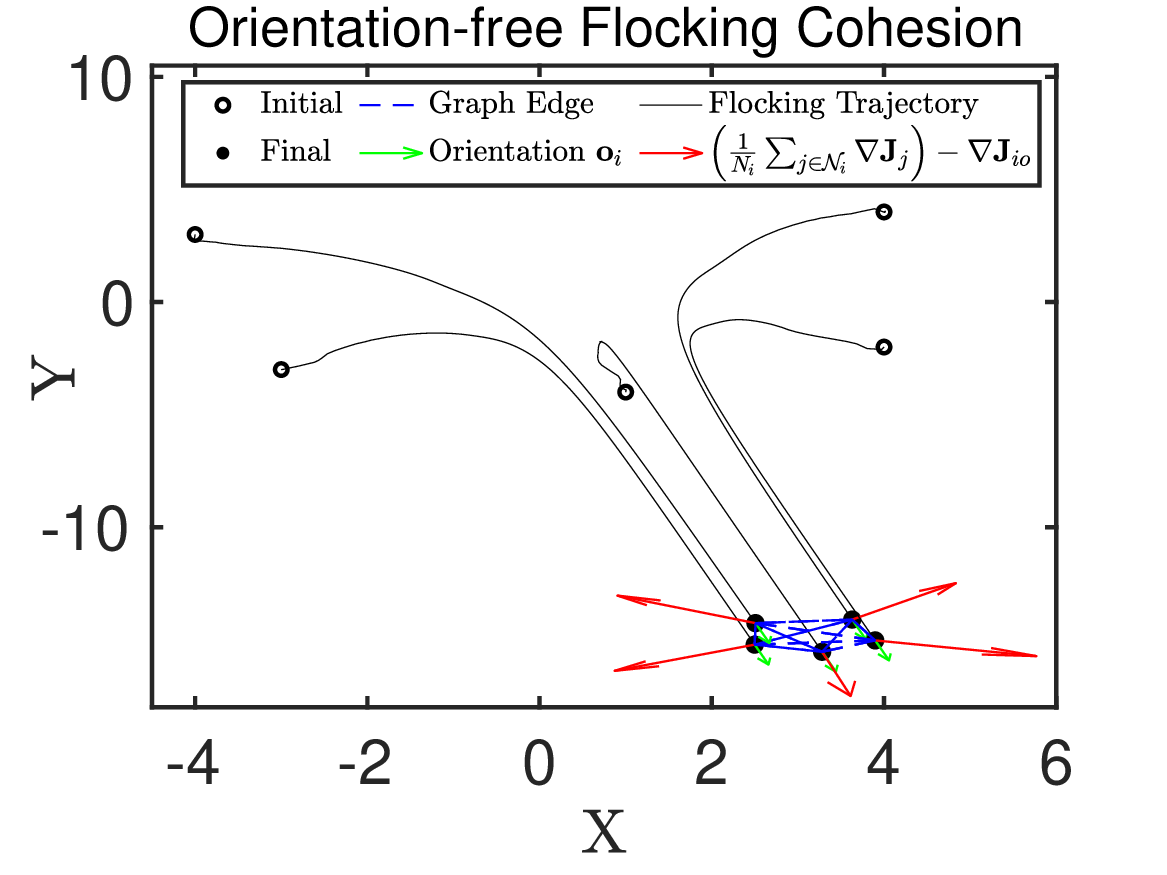}
		\end{minipage}%
	}%
	\subfigure[]{
		\begin{minipage}[t]{0.24\textwidth}
		 \label{fig:flock_free_e}
			\centering			
            \includegraphics[width=\textwidth]{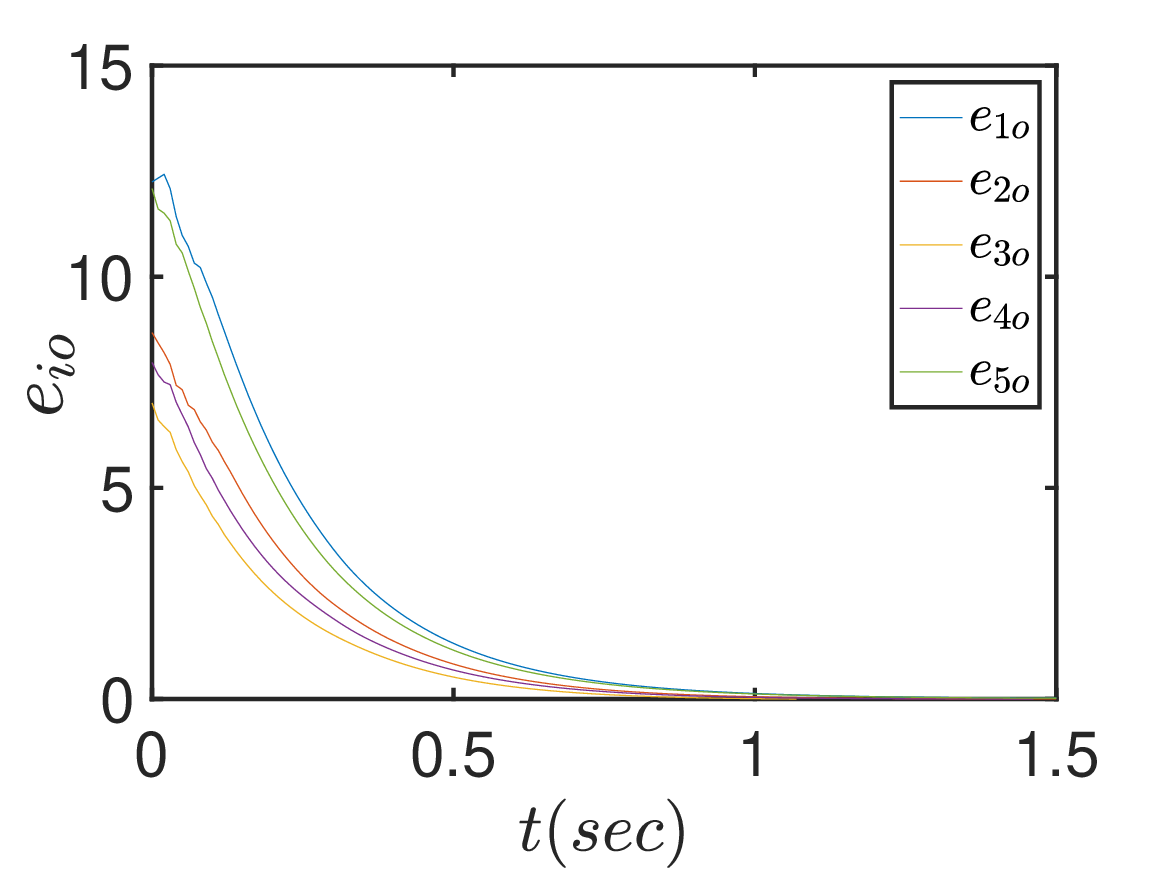}
		\end{minipage}%
	}%
\vspace{-8pt}
    
 \subfigure[]{
		\begin{minipage}[t]{0.24\textwidth}
		 \label{fig:flock_free_emu}
			\centering			
            \includegraphics[width=\textwidth]{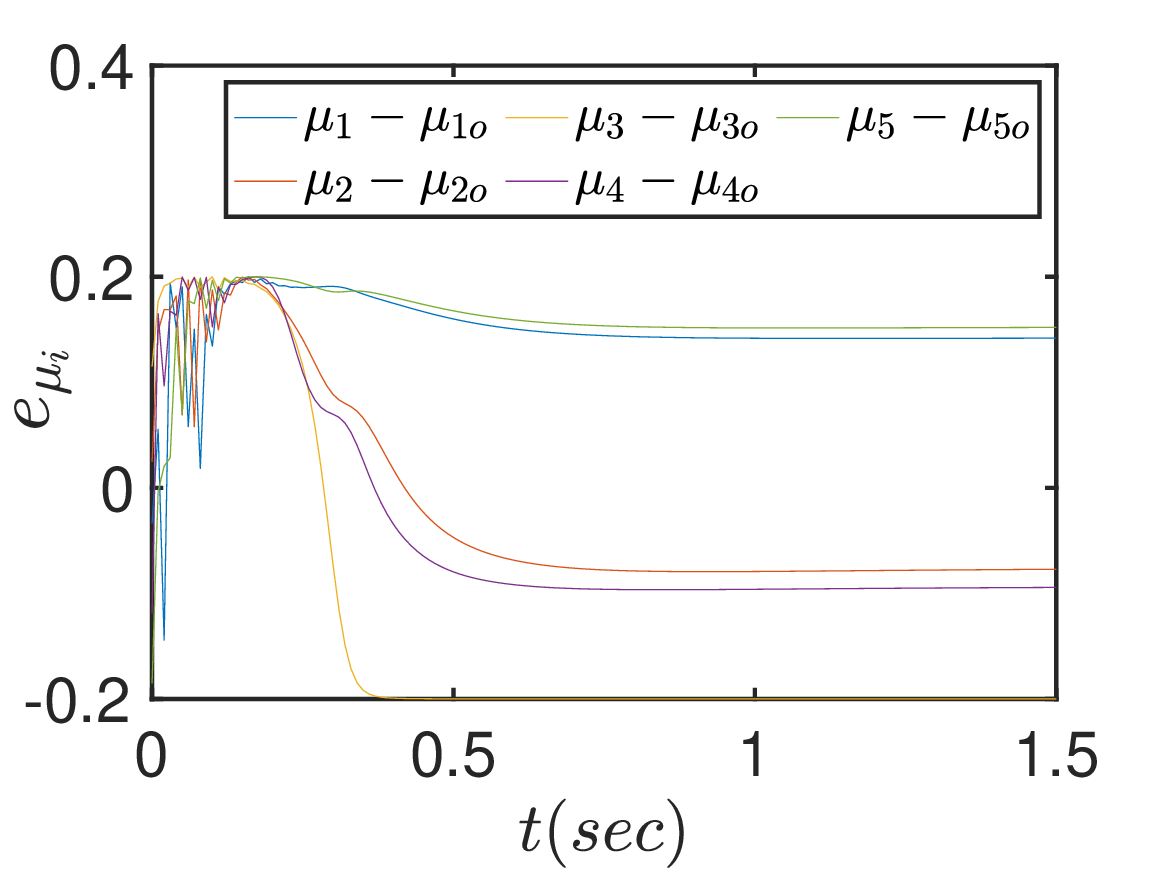}
		\end{minipage}%
	}%
  \subfigure[]{
		\begin{minipage}[t]{0.24\textwidth}
		 \label{fig:flock_free_emu_sum}
			\centering			
            \includegraphics[width=\textwidth]{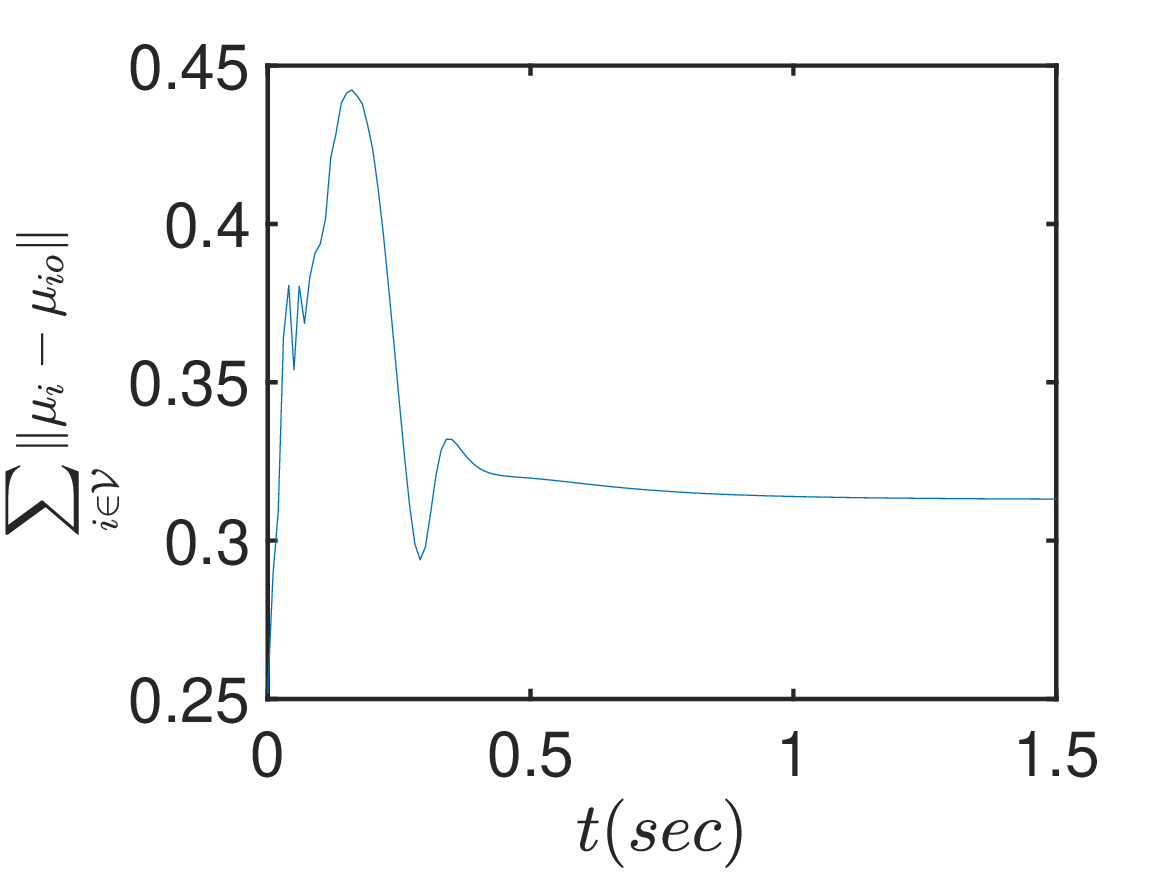}
		\end{minipage}%
	}%

	\centering
    \vspace{-6pt}
     \caption{Orientation-free flocking cohesion with the distributed controller in \eqref{eq:ori_free_error_v}-\eqref{eq:ori_free_error_w}, where the control gain is set to be $K_f = 5$. (a) Flocking cohesion trajectory; (b) Orientation-free flocking error between offset point $P_{io}$ of agent $i$ and its neighbors: $e_{io} = \mu_{io} - d^*_{\nabla \bm{J}}$ with $ \mu_{io}  = \left \| \left(  \frac{1}{{N}_i}\sum_{j\in \mathcal{N}_i}  \nabla \bm{J}_{j}\right) - \nabla \bm{J}_{io} \right \|$; (c) Flocking measurement error between offset $P_{io}$ and the agent's center $P_i$: $e_{\mu_i} = \mu_i - \mu_{io}$ with $\mu_i =\left \| \left(  \frac{1}{{N}_i}\sum_{j\in \mathcal{N}_i}  \nabla \bm{J}_{j}\right) - \nabla \bm{J}_{i} \right \| $; (d) Sum of the offset error norm: $\sum_{i\in\mathcal{V}} \left\| \mu_i - \mu_{io}\right\|$. }
	\label{fig:orientation_free_flocking}
\end{figure}

\begin{figure}[htbp]
\vspace{-8pt}
	\centering
    \subfigure[]{
		\begin{minipage}[t]{0.24\textwidth}
		    \label{fig:flock_based_traj}
			\centering			
            \includegraphics[width=\textwidth]{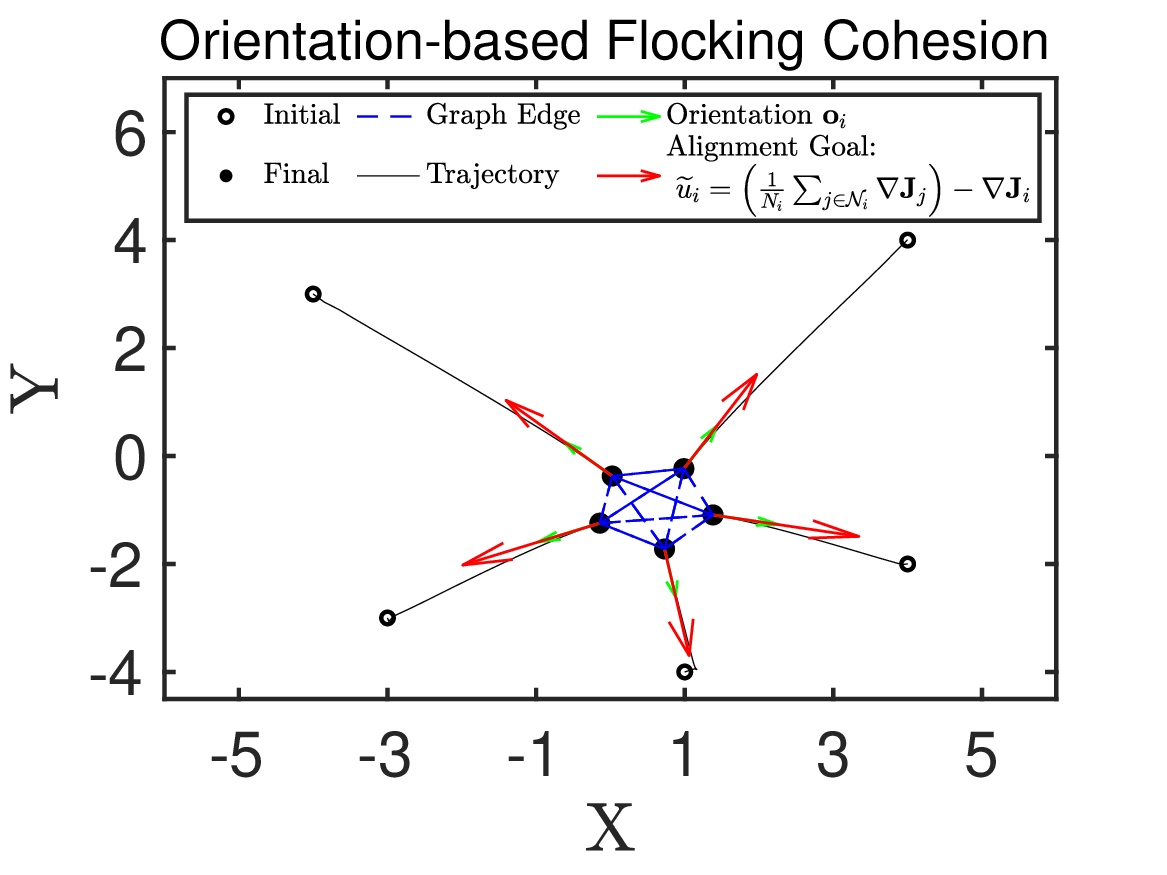}
		\end{minipage}%
	}%
 \subfigure[]{
		\begin{minipage}[t]{0.24\textwidth}
		 \label{fig:flock_based_e}
			\centering			
            \includegraphics[width=\textwidth]{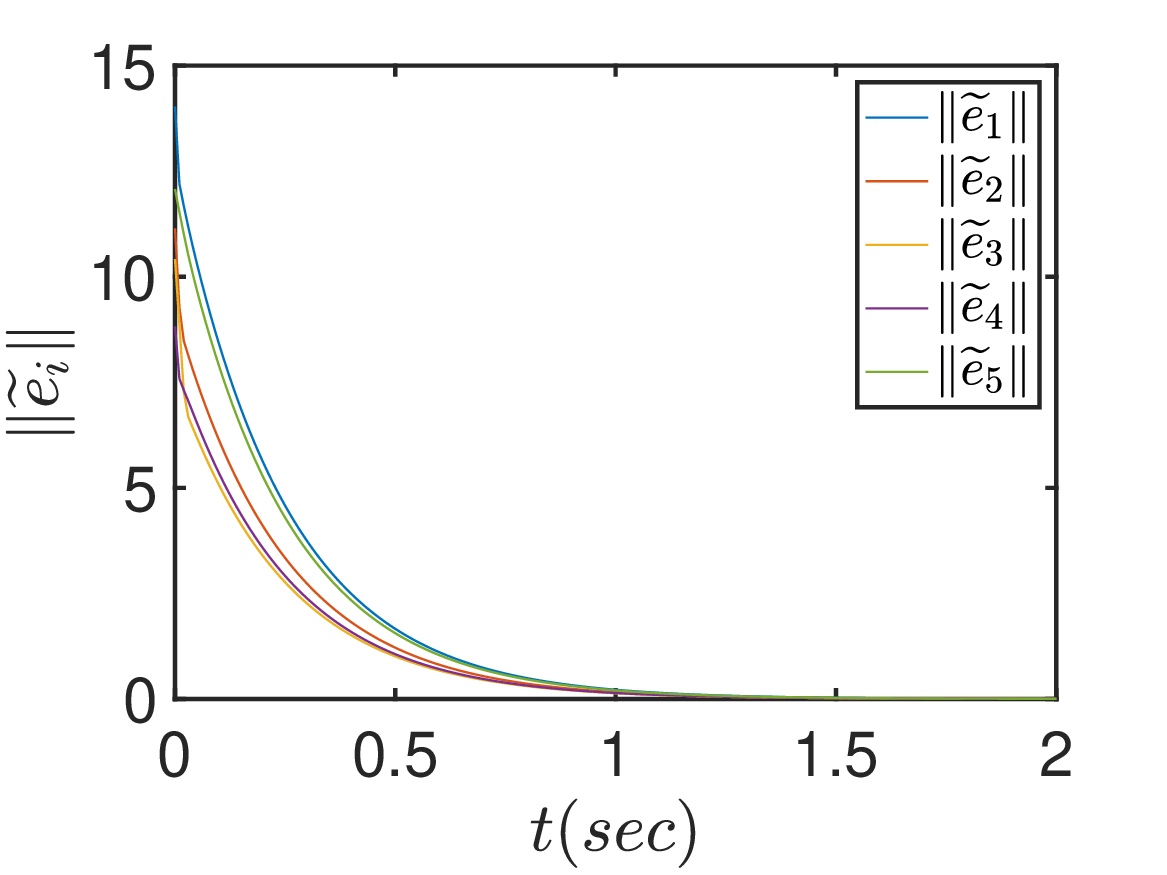}
		\end{minipage}%
	}%

	\centering
    \vspace{-6pt}
     \caption{Orientation-based flocking cohesion with the distributed controller in \eqref{eq:ori_con_error_v}-\eqref{eq:ori_con_error_w}, where the control gain is set to be $k_{fv} = 1, k_{f\omega} = 20$. (a) Flocking trajectory; 
     (b) Orientation-based flocking error norm of agent $i$: $\left\|\widetilde{\bm{e}}_{i} \right\| = \left\|\widetilde{\bm{u}}_i - d^*_{\nabla \bm{J}} [\cos(\theta_i),\,\sin(\theta_i) ]\right\|$.}
     \label{fig:orientation_based_flocking}
\end{figure}

\subsection{Flocking Cohesion}
For validating the two distributed flocking controllers $\bm{u}_i$ on unicycle robots, the undirected topology $\mathcal{G}$ is assumed to be complete and static for all $t\geq t_0$ in this subsection. Consider that the initial states of $5$ agents are set randomly within the source field,
the flocking-cohesion task is defined using the desired $d^*_{\nabla \bm{J}} = 2$ in \eqref{eq:ori_uncon_error} and \eqref{eq:ori_con_error}. 

Figure~\ref{fig:orientation_free_flocking} and Figure~\ref{fig:orientation_based_flocking} demonstrate the unicycle robots' flocking in a cohesion, where the black hollow circle $'\circ'$ and solid $'\bullet'$ represent the agents' initial and final positions, respectively. The agents' trajectories are visualized in solid black lines, the dashed blue line shows the complete communication graph. To demonstrate the difference between these two flocking cohesion schemes, each agent's orientation is shown in green arrow lines, and the averaging neighbor gradient difference vector is in red arrow lines. Note the concave signal field $J(x,y)$ is not plotted here for a clear visualization. 

It can be seen that both flocking controllers drive the multi-agent system to converge to a desired cohesion, as shown with the flocking error convergence in Figure~\ref{fig:flock_free_e} and Figure \ref{fig:flock_based_e}. Given the different formulations of flocking as in  \eqref{eq:ori_uncon_error} and \eqref{eq:ori_con_error}, the agents' orientations (green arrow) are aligned with their neighbors' signal gradient average (red arrow) in Figure~\ref {fig:flock_based_traj}, whereas there is no alignment requirement in Figure~\ref{fig:flock_free_traj}. Particularly, as the feedback linearization is applied in the orientation-free controller with offset point $P_{io}$ (with $d_o=0.1$), the flocking coordinates the signal measurements between $P_{io}$ and the agent $i$'s neighbors. Figure~\ref{fig:flock_free_emu} plots the flocking measurement error between $P_{io}$ and the agent's center $P_i$, and Figure~\ref{fig:flock_free_emu_sum} demonstrates the sum of the error norm. It can be seen that the measurement error exists in the whole motion evolution, whereas it is not a problem in our orientation-based flocking method, as shown in Figure~\ref{fig:orientation_based_flocking}.

\begin{figure}[htbp]
\centering
\subfigure[]{
	\begin{minipage}[t]{0.24\textwidth}
		 \label{fig:ori_based_collision_init}
			\centering			\includegraphics[width=0.9\textwidth]{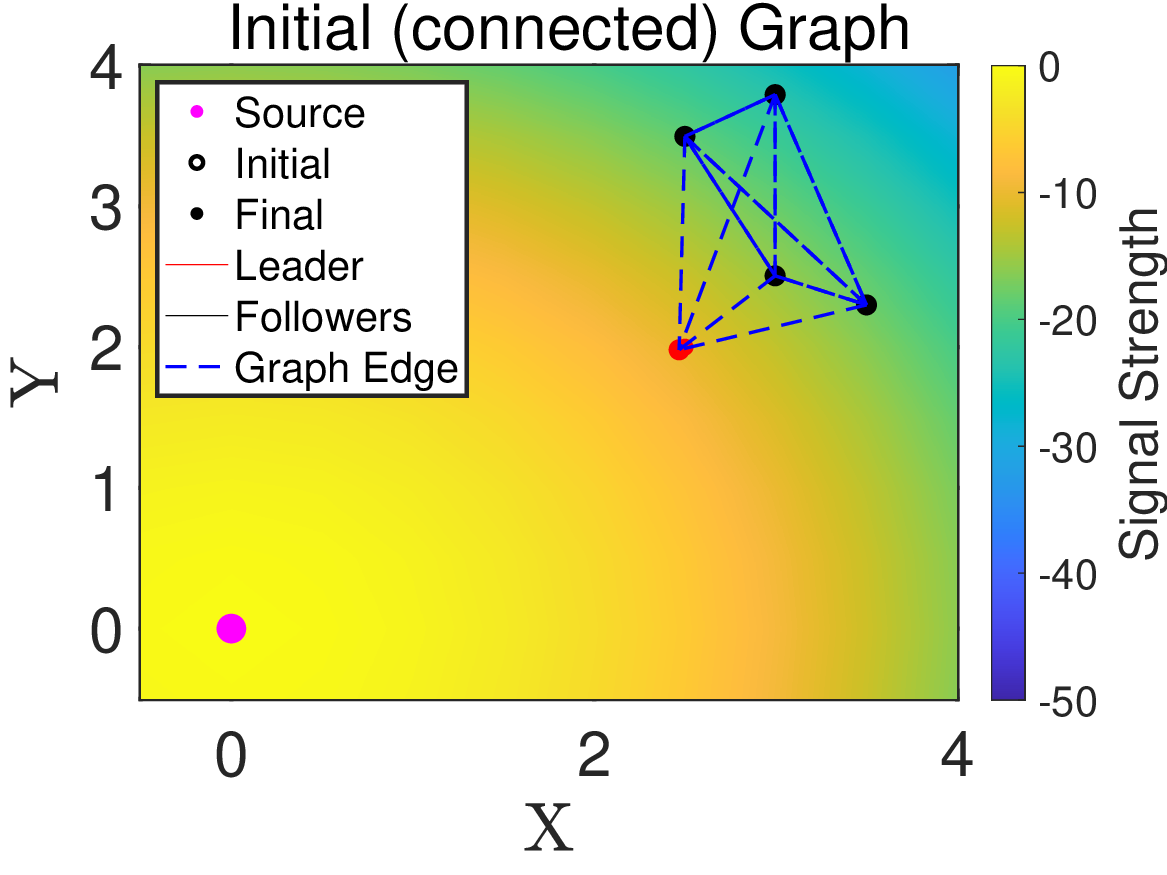}
		\end{minipage}%
	}%
    \subfigure[]{
	\begin{minipage}[t]{0.24\textwidth}
		 \label{fig:ori_based_collision_traj}
			\centering			\includegraphics[width=0.9\textwidth]{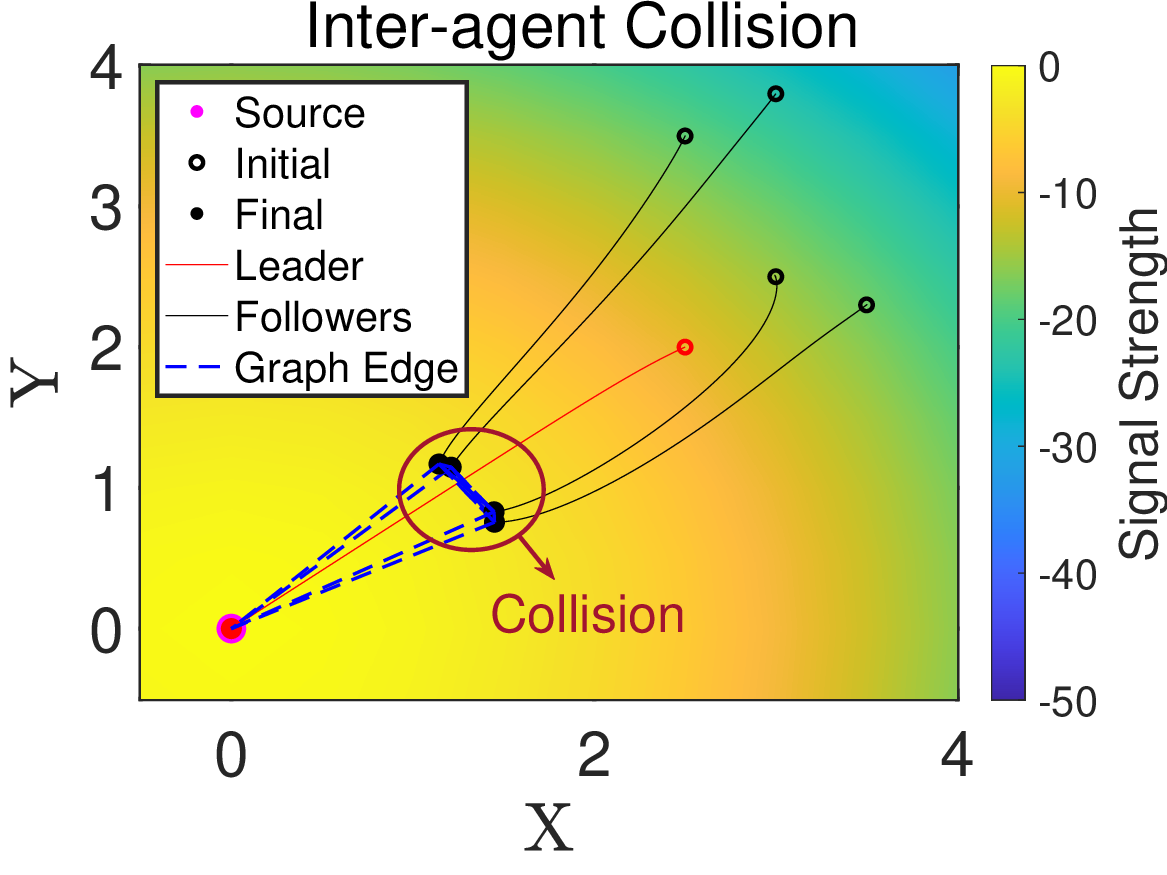}
		\end{minipage}%
	}%
    \vspace{-8pt}
    
    \subfigure[]{
		\begin{minipage}[t]{0.24\textwidth}
		    \label{fig:ori_based_collision_muij}
			\centering			\includegraphics[width=0.9\textwidth]{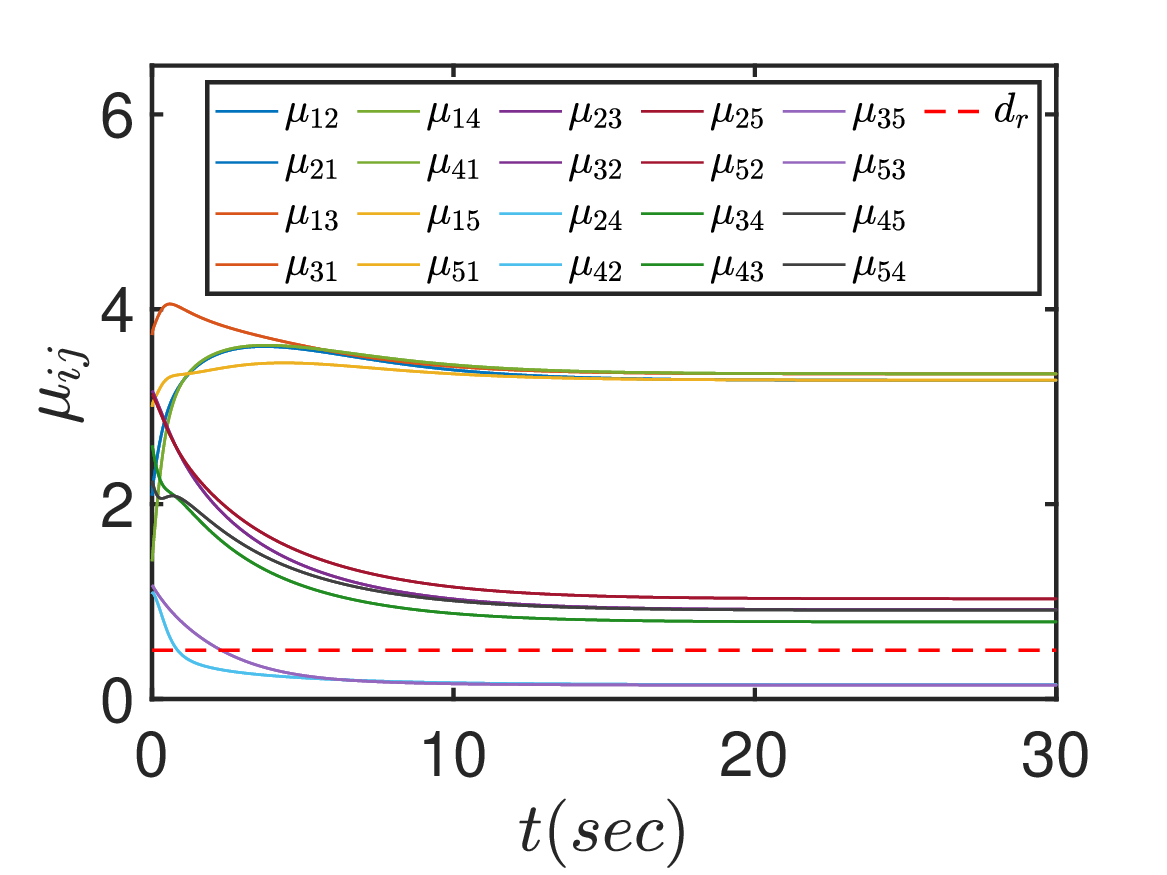}
		\end{minipage}%
	}%
    \subfigure[]{
		\begin{minipage}[t]{0.24\textwidth}
		    \label{fig:ori_based_collision_e}
			\centering			\includegraphics[width=0.9\textwidth]{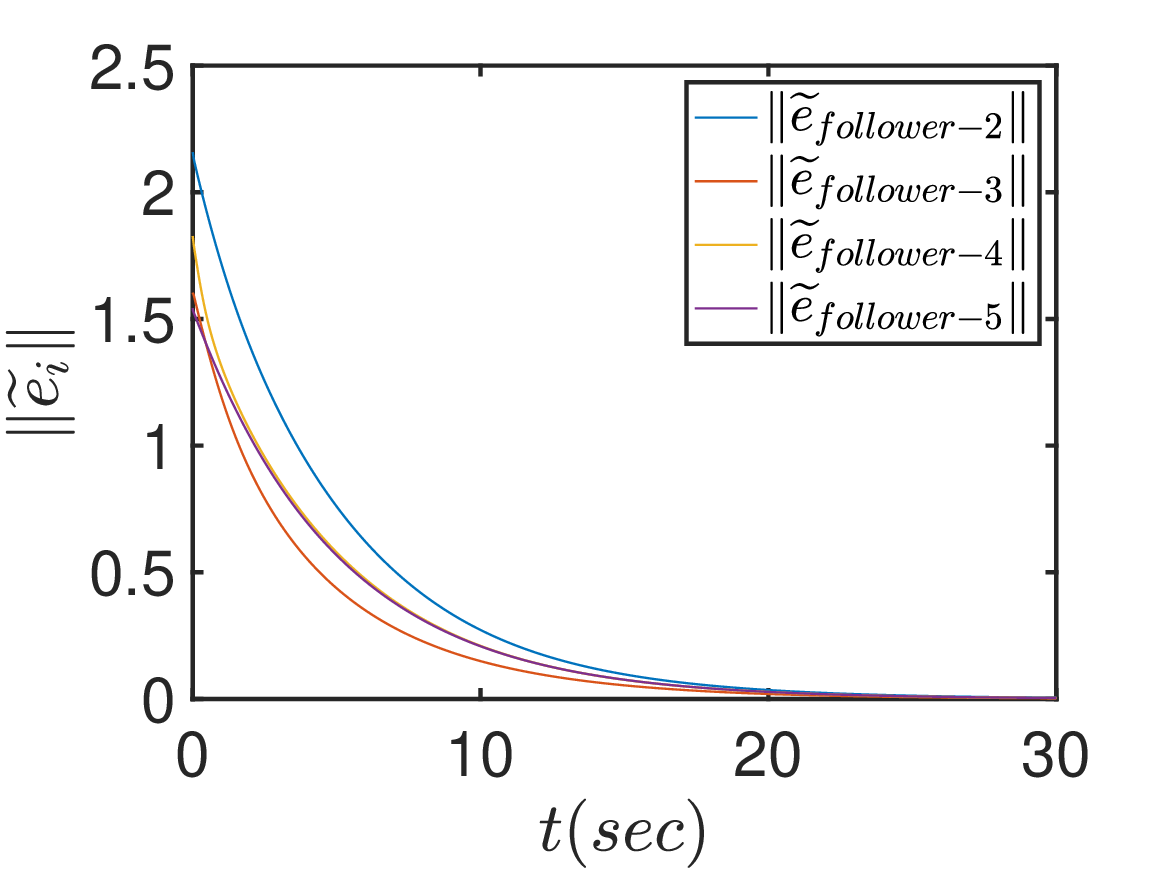}
		\end{minipage}%
	}%
  
	\centering
    \vspace{-6pt}
        \caption{Inter-agent collisions without safety guarantee: simulation results of source-seeking (leader) and orientation-based flocking cohesion (followers, $\bm{u}_{\text{flock-i}}$ is as in  \eqref{eq:ori_con_error_v}-\eqref{eq:ori_con_error_w})  
        with dynamic graph $\mathcal{G}(t)$. 
        The source-seeking gain and flocking gains are given by $k_v = 0.3, k_\omega = 1$ and $k_{fv} = 0.1, k_{f\omega} = 0.2$, respectively. (a) Initial states of agents with initially connected communication graph $\mathcal{G}(t_0)$; (b) Evolution trajectories with desired $d^*_{\nabla \bm{J}} = 1$; (c) Inter-agent signal gradient difference $u_{ij}$ for $(i,j)\in\mathcal{E}(t)$; (d) Orientation-based flocking error $\left\| \widetilde{\bm{e}}_i\right\|$.
        }
        \label{fig:ori_based_collision}
\end{figure}
\begin{figure}[htbp]
	\centering
  \subfigure[]{
		\begin{minipage}[t]{0.24\textwidth}
		    \label{fig:ori_based_safe_traj}
			\centering			\includegraphics[width=0.9\textwidth]{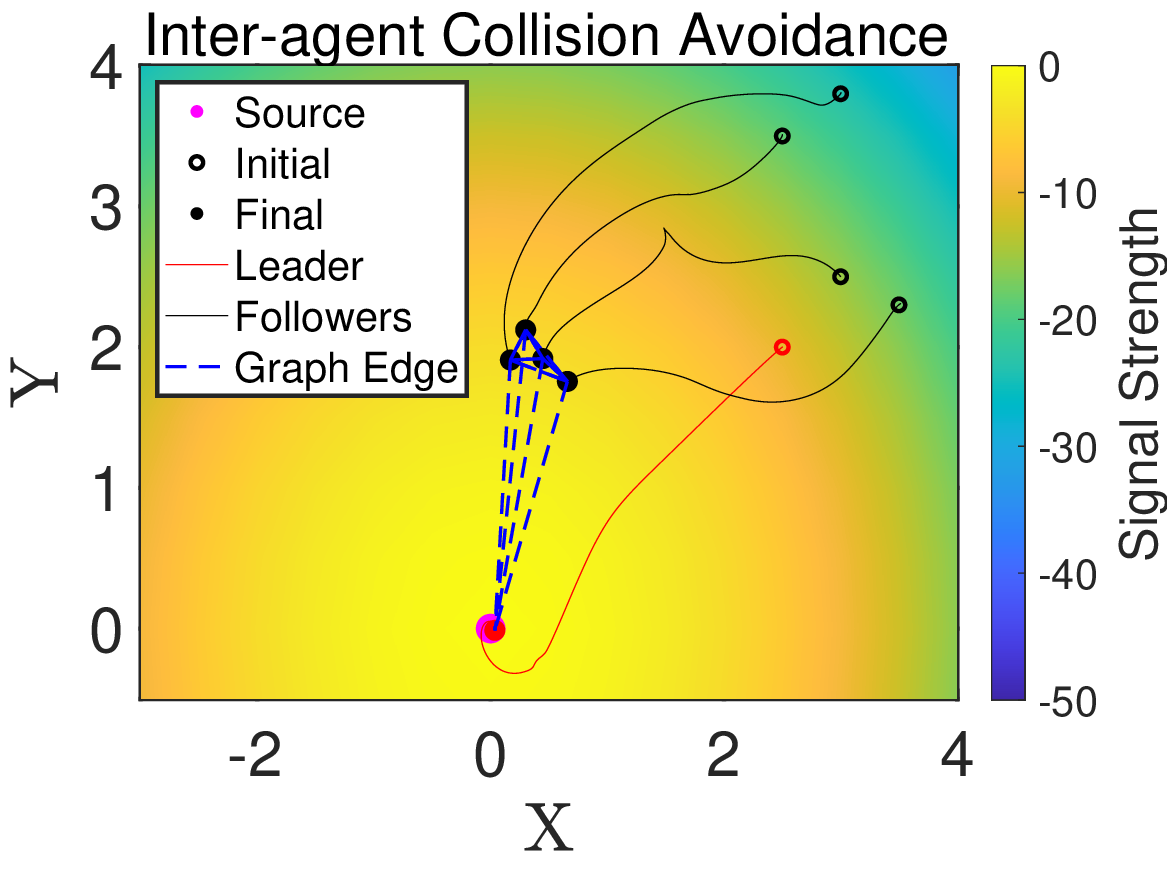}
		\end{minipage}%
	}%
  \subfigure[]{
		\begin{minipage}[t]{0.24\textwidth}
		    \label{fig:ori_based_safe_acc}
			\centering			\includegraphics[width=0.9\textwidth]{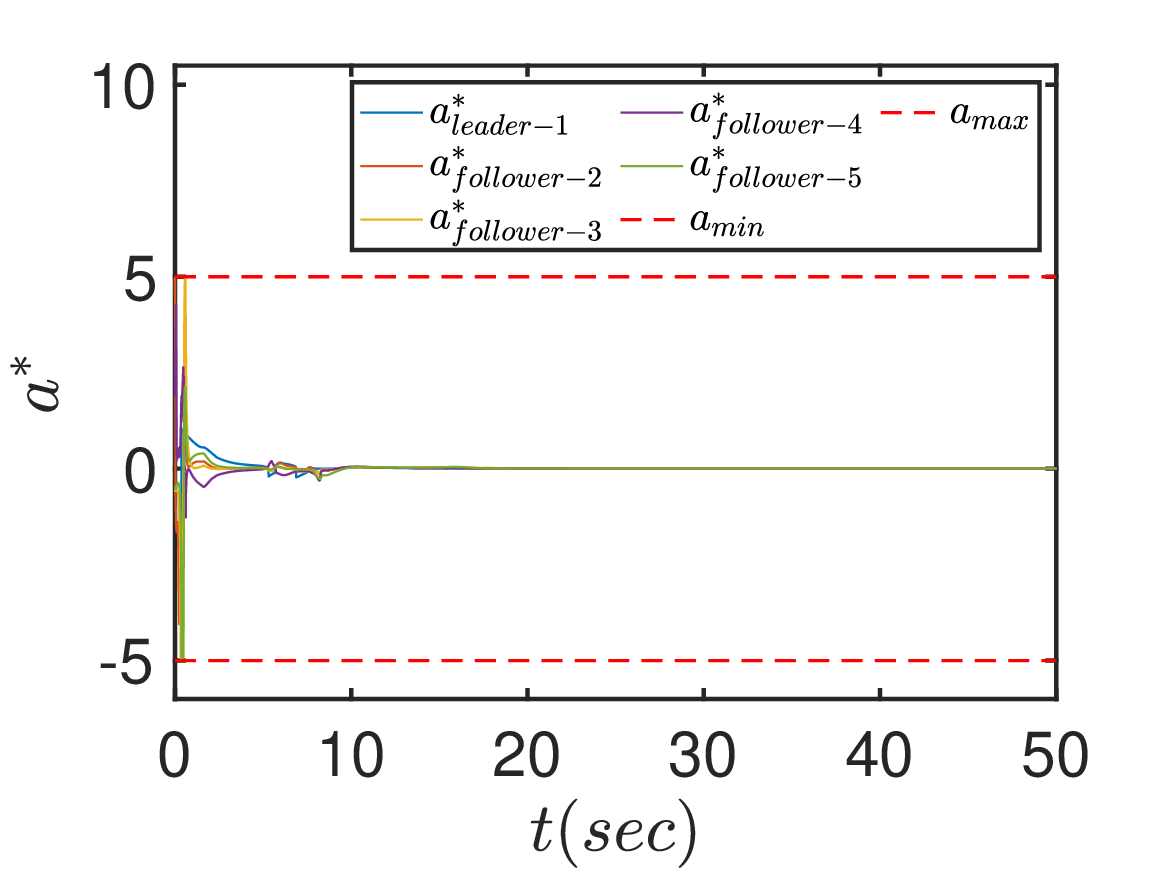}
		\end{minipage}%
	}%
    \vspace{-8.5pt}
    
  \subfigure[]{
		\begin{minipage}[t]{0.24\textwidth}
		    \label{fig:ori_based_safe_w}
			\centering			\includegraphics[width=0.9\textwidth]{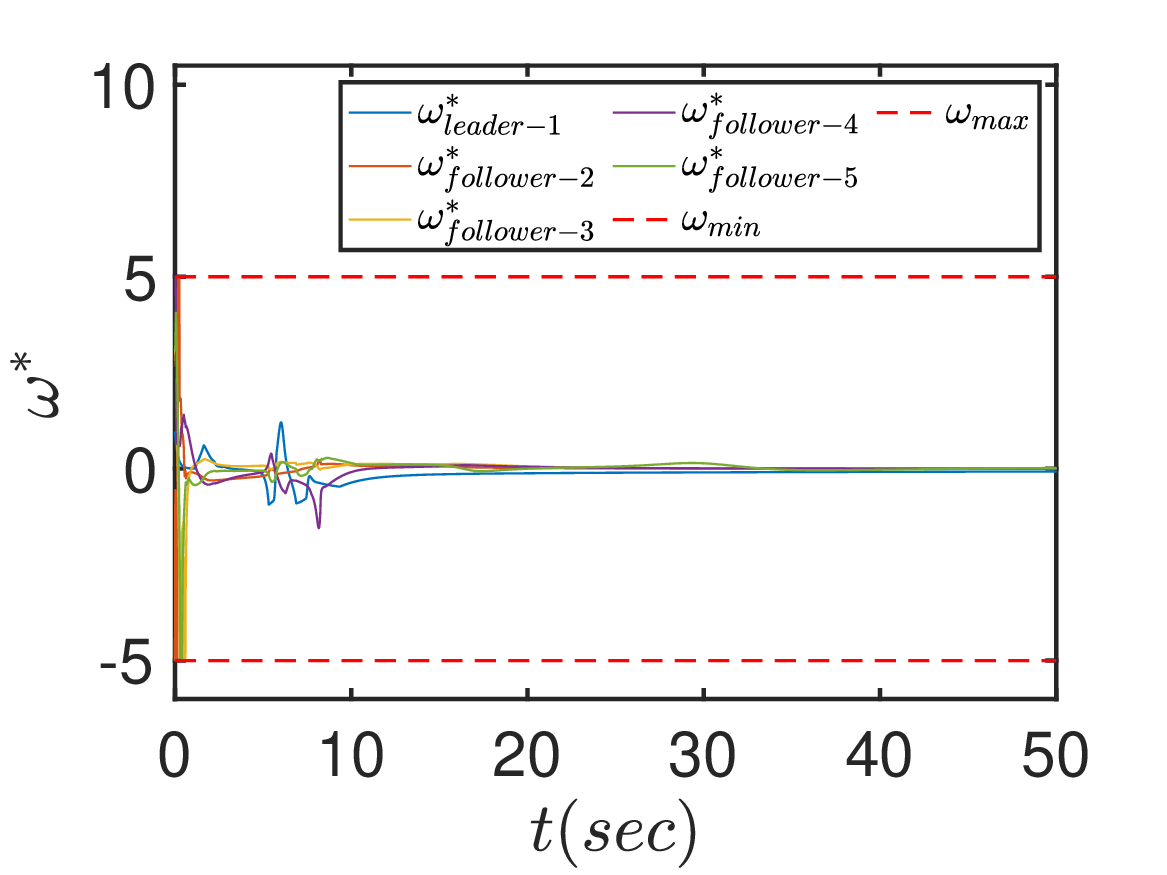}
		\end{minipage}%
	}%
    \subfigure[]{
		\begin{minipage}[t]{0.24\textwidth}
		    \label{fig:ori_based_safe_e}
			\centering			\includegraphics[width=0.9\textwidth]{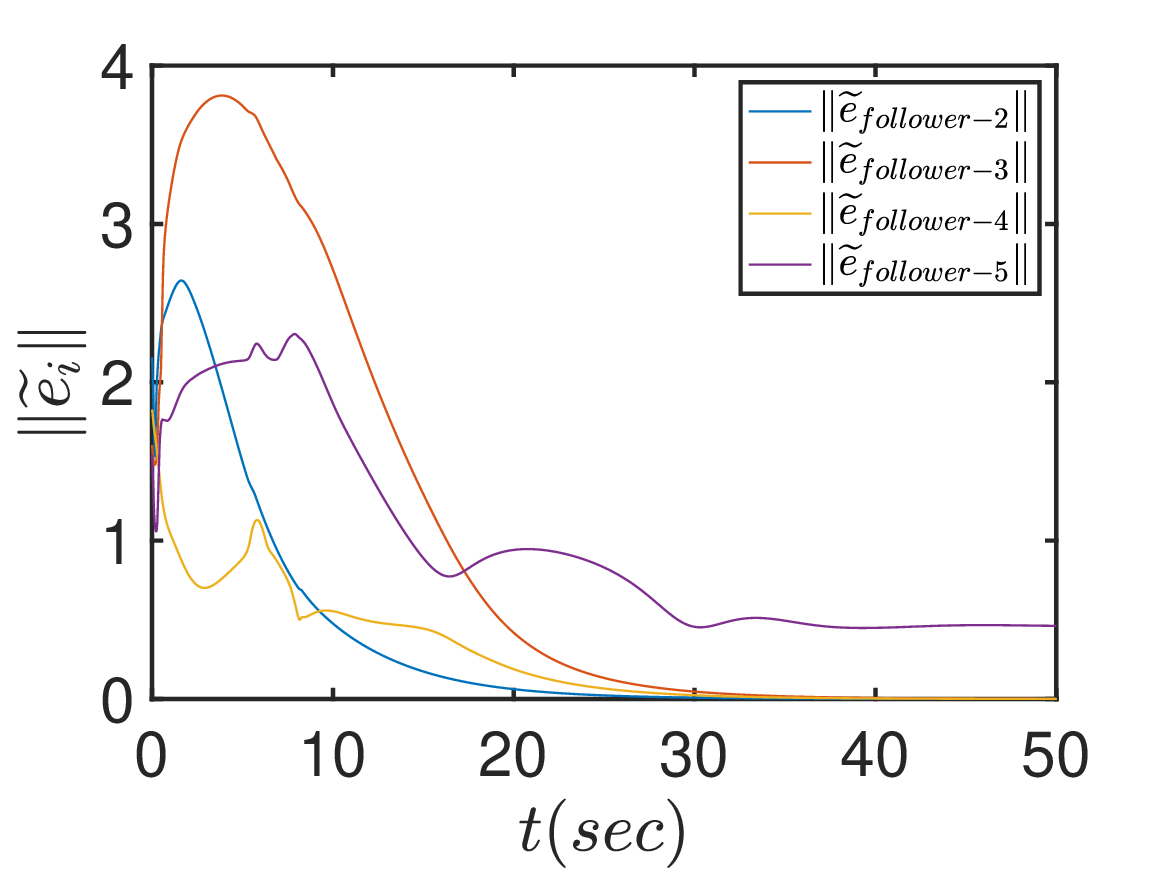}		\end{minipage}%
	}%
      \vspace{-8.5pt}
     
     \subfigure[]{
		\begin{minipage}[t]{0.24\textwidth}
		    \label{fig:ori_based_safe_mui}
			\centering			\includegraphics[width=0.9\textwidth]{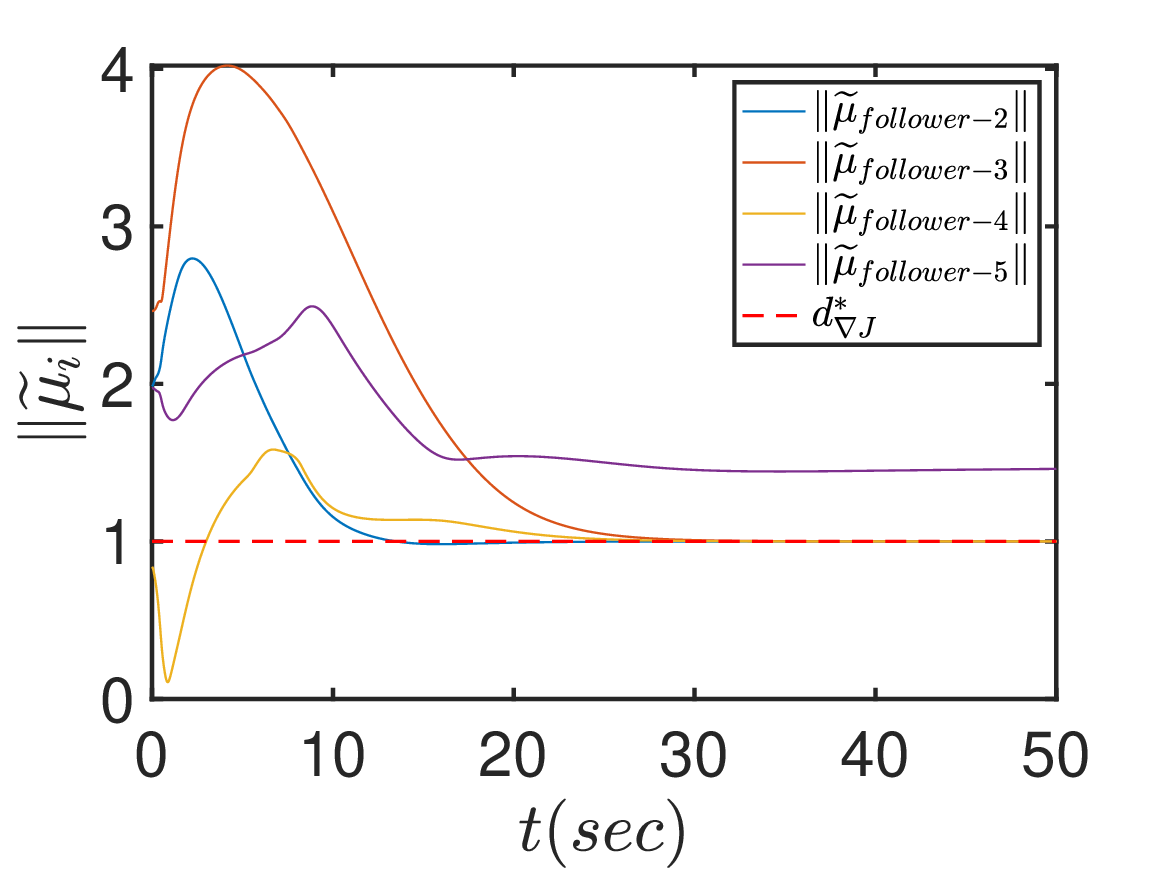}		\end{minipage}%
	}%
     \subfigure[]{
		\begin{minipage}[t]{0.24\textwidth}
		    \label{fig:ori_based_safe_muij}
			\centering			\includegraphics[width=0.9\textwidth]{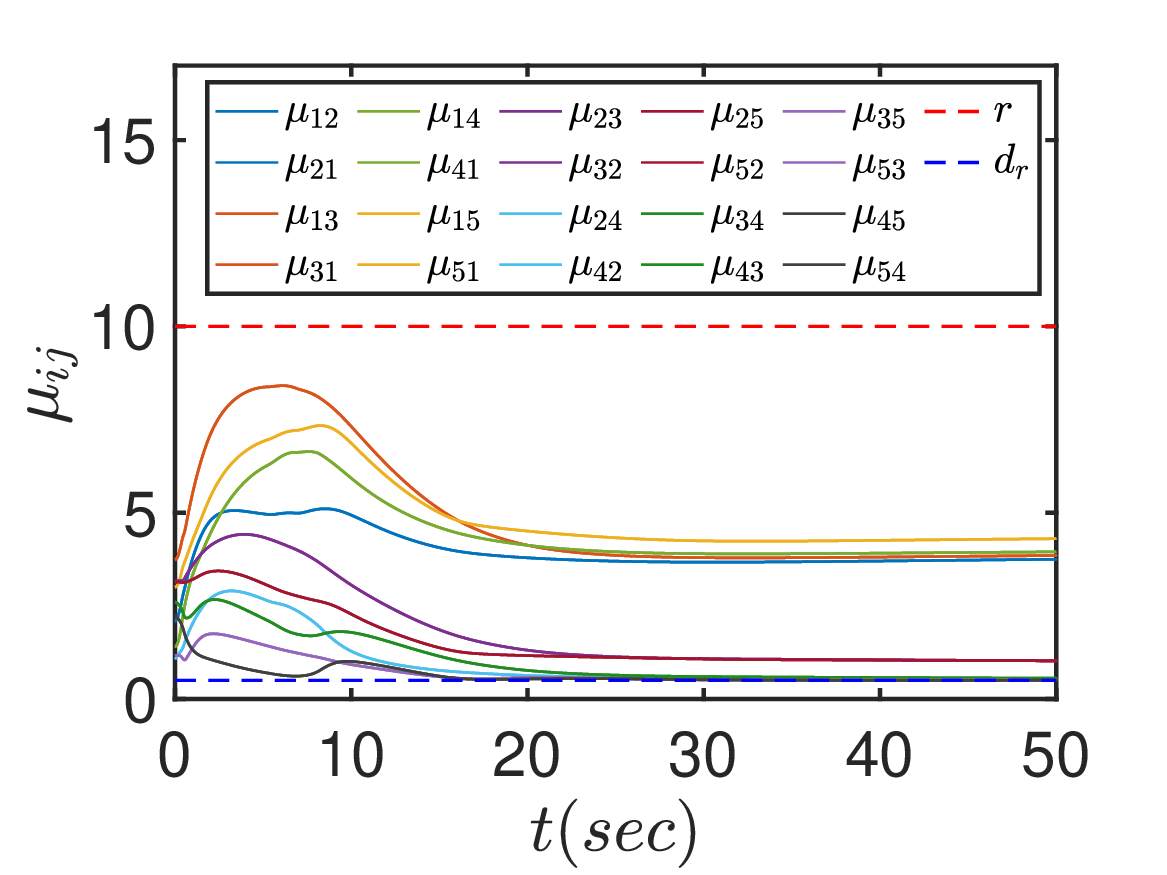}		\end{minipage}%
	}%
      \vspace{-8.5pt}
      
     \subfigure[]{
		\begin{minipage}[t]{0.24\textwidth}
		    \label{fig:ori_based_safe_gamma}
			\centering			\includegraphics[width=0.9\textwidth]{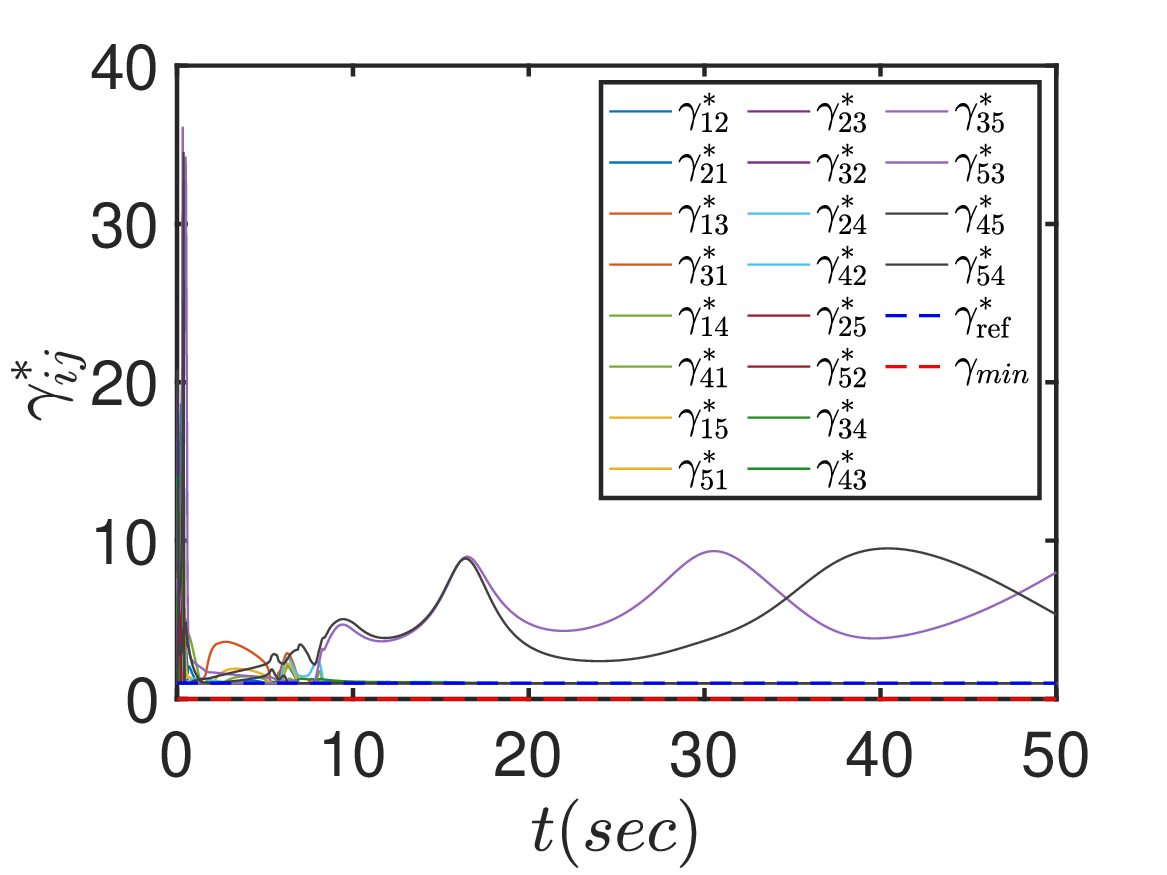}		\end{minipage}%
	}%
    \subfigure[]{
		\begin{minipage}[t]{0.24\textwidth}
		    \label{fig:ori_based_safe_hij}
			\centering			\includegraphics[width=0.9\textwidth]{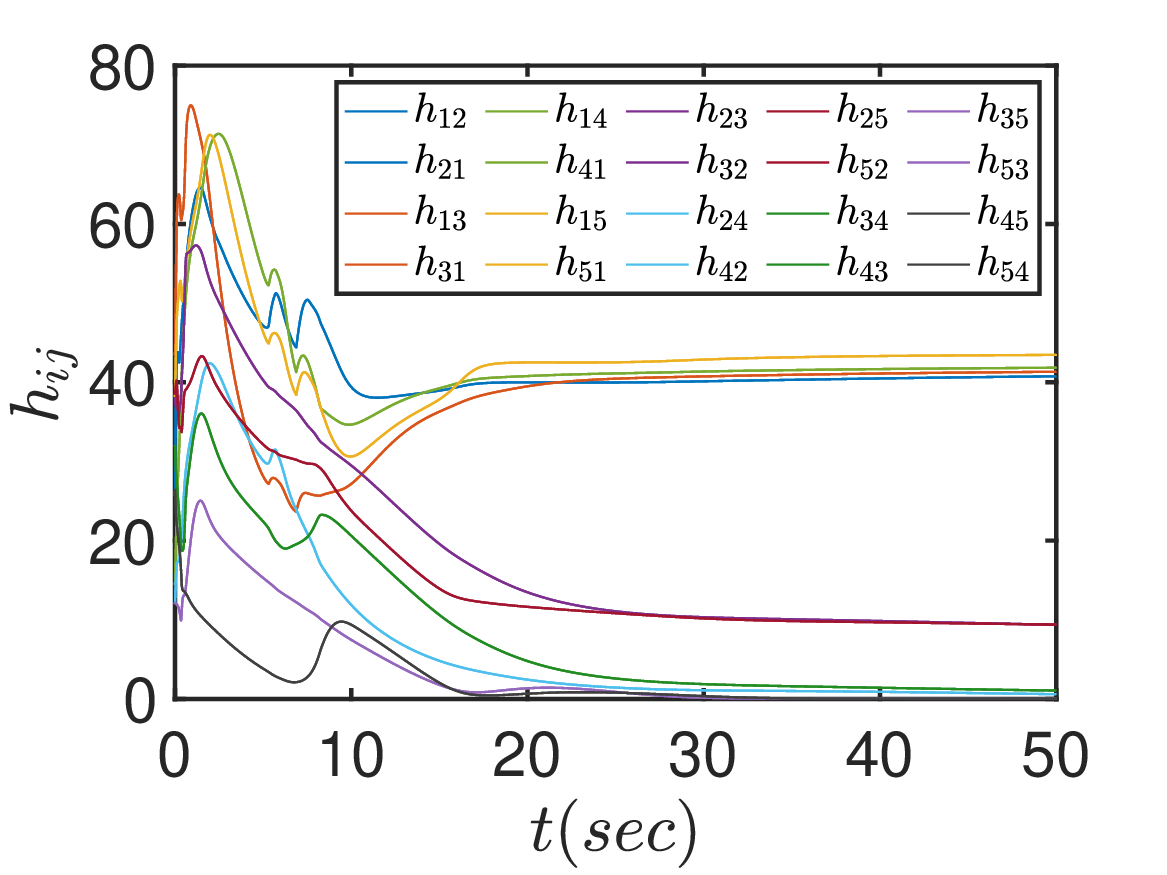}		\end{minipage}%
	}%
 
	\centering
    \vspace{-6pt}
         \caption{Inter-agent collision-free evolution: simulation results of source-seeking (leader) and orientation-based flocking-cohesion (followers,  $\bm{u}_{\text{flock-i}}$ is as \eqref{eq:ori_con_error_v}-\eqref{eq:ori_con_error_w}) based on CBF-QPs \eqref{QP_framework:leader}-\eqref{QP_framework:follower}. (a) Collision-free trajectories;
         (b)-(c) Optimal solutions of acceleration $a^*_i$ and angular velocity $\omega^*_i$; (d) Orientation-based flocking error $\left\| \widetilde{\bm{e}}_i\right\|$; (e) Signal gradient coordination $\left\| \widetilde{\bm{\mu}}_i\right\|$; (f) Inter-agent signal gradient difference $\mu_{ij}$; (g) Updated  weight $\gamma^*_{ij}$;
         (h) Inter-agent CBF $h_{ij}$. }
        \label{fig:ori_based_safe}
\end{figure}
\subsection{Connectivity-preserved Collision-free Source-seeking and Flocking-cohesion}
To validate the inter-agent collision avoidance (safety) and the graph connectivity preservation, we consider the source-seeking (leader) and flocking-cohesion (followers) tasks in this section. The undirected communication graph  is dynamic, and it is assumed that the agents can only communicate with each other if they are within the maximum communication range $r$. In the following simulations, the leader agent (red dot) is driven to locate the unknown signal source (solid magenta dot), while the follower agents (black dots) must achieve a safe and cohesive flocking with their neighbors.
\subsubsection{Inter-agent collision avoidance (safety)}
Figure~\ref{fig:ori_based_collision} and \ref{fig:ori_based_safe} provide validation for the inter-agent collision avoidance. The initial graph of the system is connected with the edges (shown in blue dashed lines) in Figure~\ref{fig:ori_based_collision_init}, and the dynamic graph switches according to the maximum communication range of $r=10$ uniformly for all agents. Figure~\ref{fig:ori_based_collision_traj}-\ref{fig:ori_based_collision_muij} plots the orientation-based flocking results without safety guarantees. It shows that the flocking is achieved when parts of agents reach a sufficiently close distance where $\mu_{ij} \rightarrow 0$ as in Figure~\ref{fig:ori_based_collision_muij}, followed by the convergence of flocking error in Figure~\ref{fig:ori_based_collision_e}.   

\begin{figure}[htbp]
	\centering
	\subfigure[$t = 0s$]{
		\begin{minipage}[t]{0.24\textwidth}
		 \label{fig:ori_fee_connectivity_initial}
			\centering			
            \includegraphics[width=0.9\textwidth]{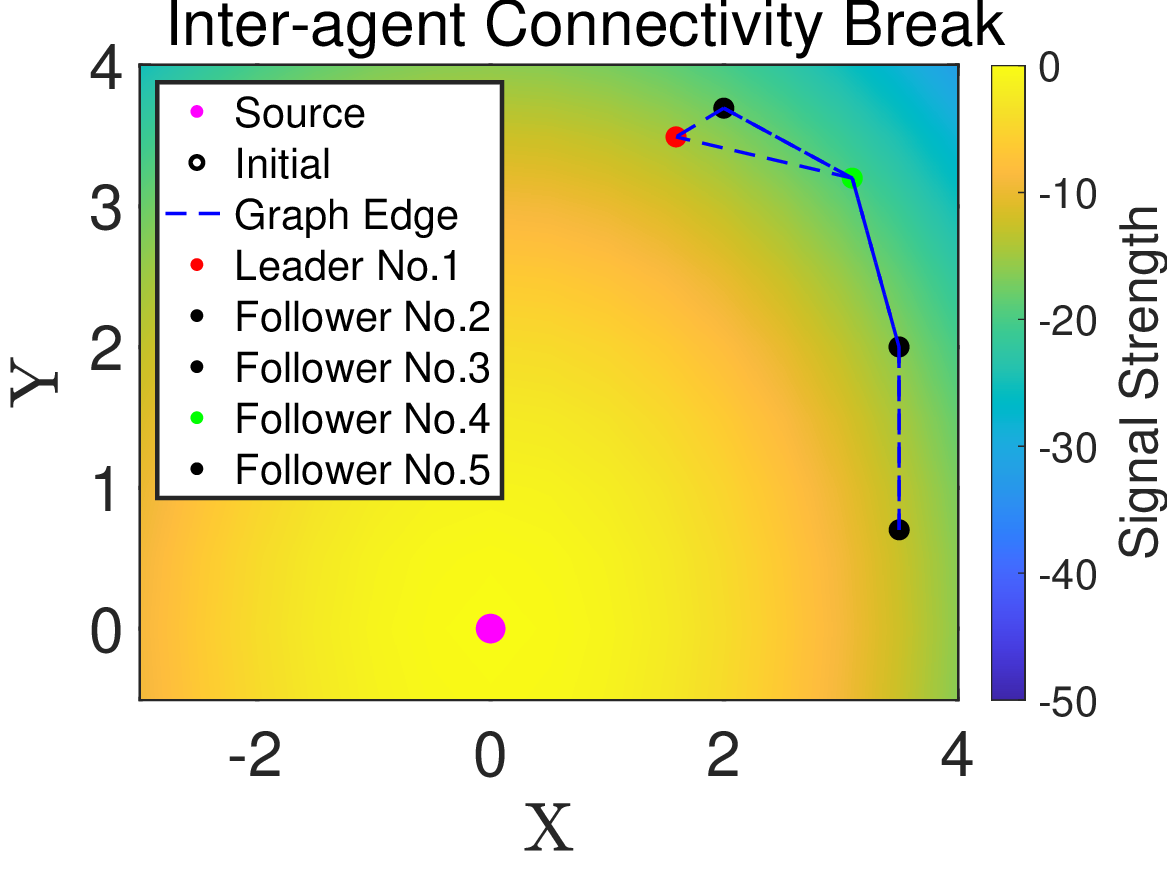}
		\end{minipage}%
	}%
 \subfigure[$t = 1s$]{
		\begin{minipage}[t]{0.24\textwidth}
		 \label{fig:ori_free_connectivity_break_1traj}
			\centering			
            \includegraphics[width=0.9\textwidth]{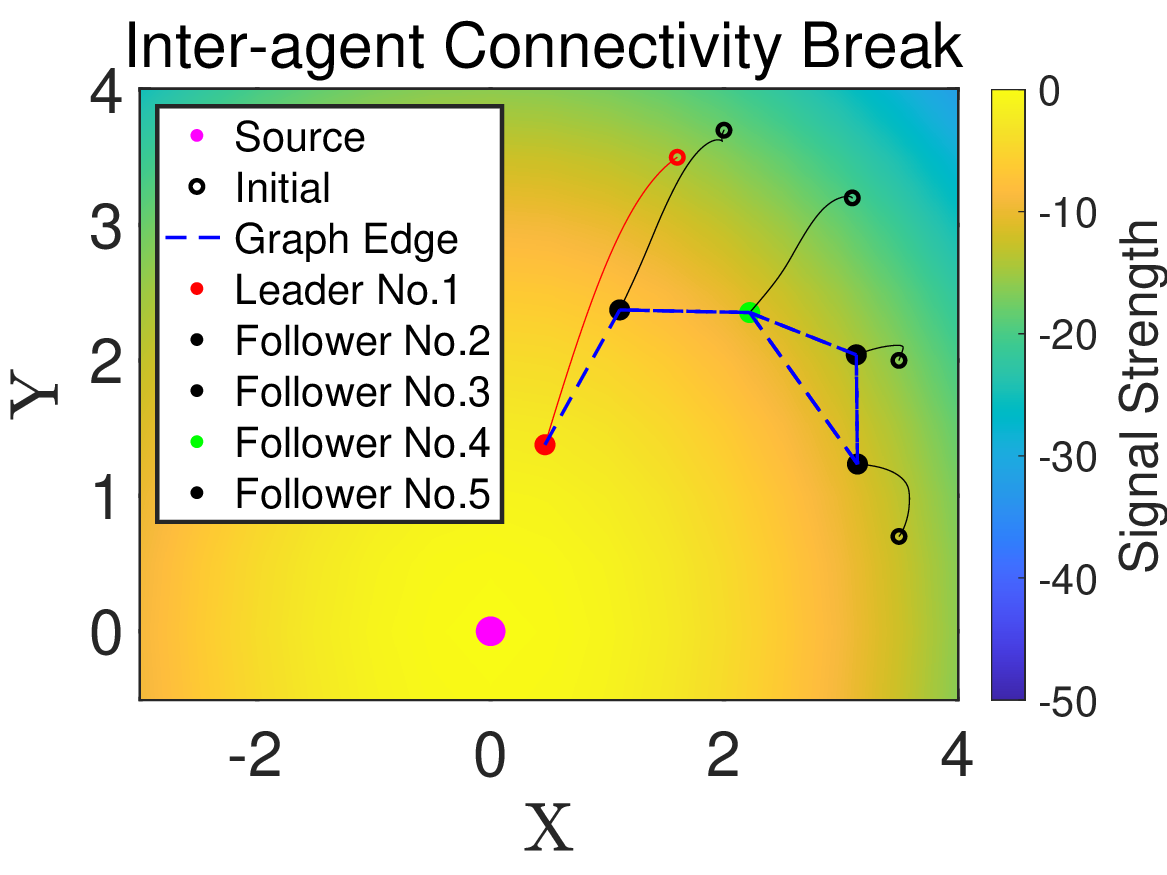}
		\end{minipage}%
	}%
    
\subfigure[$t = 10s$]{
		\begin{minipage}[t]{0.24\textwidth}
		 \label{fig:ori_free_connectivity_break_10traj}
			\centering			
            \includegraphics[width=0.9\textwidth]{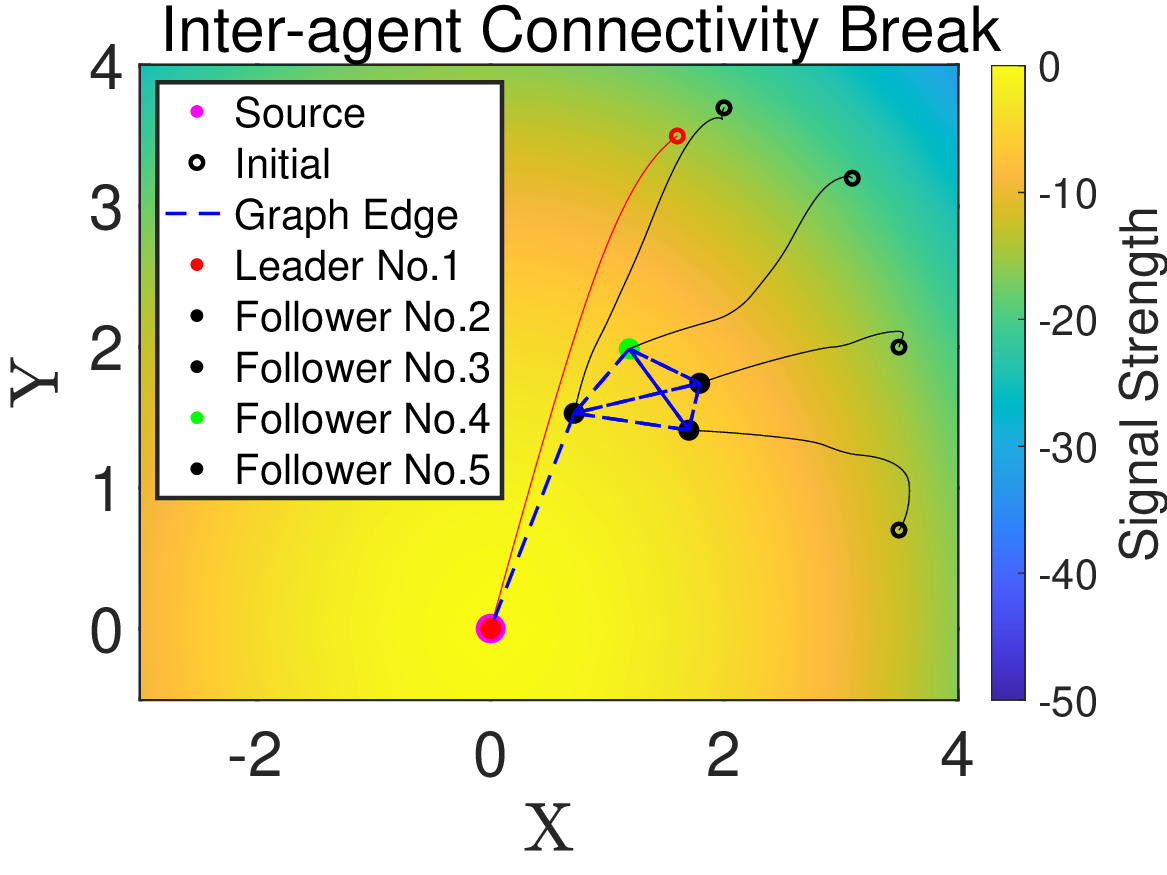}
		\end{minipage}%
       }%
\subfigure[$t = 10s$]{
		\begin{minipage}[t]{0.24\textwidth}
		 \label{fig:ori_free_connectivity_break_10eio}
			\centering			
            \includegraphics[width=0.9\textwidth]{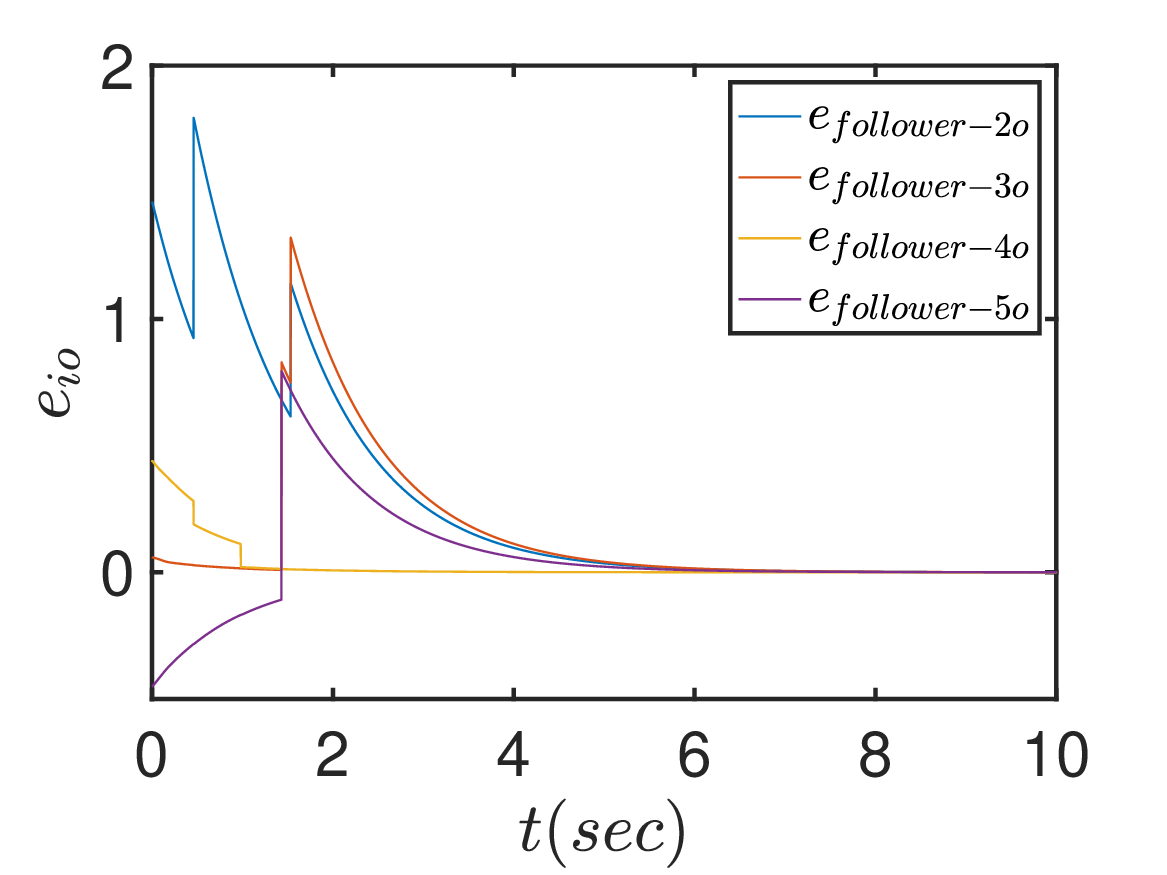}
		\end{minipage}%
	}%
	\centering
     \vspace{-10pt}
         \caption{Inter-agent connectivity break without preservation guarantee: simulation results of source-seeking (leader) and orientation-free flocking-cohesion (follower, $\bm{u}_{\text{flock-i}}$ is as in \eqref{eq:ori_free_error_v}-\eqref{eq:ori_free_error_w}) with dynamic graph $\mathcal{G}(t)$. The maximum communication range is set as $r=4$.
         The graph connectedness and motion trajectory evolution are plotted in (a-c) at $t=0s, 1s, 10s$. The orientation-free flocking error is shown in (d). }
        \label{fig:ori_free_connectivity_break}
\end{figure}
Therefore, to separate the agents and avoid collisions during evolution, additional inter-agent collision-avoidance (safety) constraints are deployed in the distributed CBF-QPs for agents. The resulting collision-free motion trajectories are shown in Figure~\ref{fig:ori_based_safe_traj}, all the connected inter-agent pairs are able to maintain a safe range $u_{ij}$ between the minimum safe margin $d_r=0.5$ and the maximum sensing hood $r=10$ for collision-free and connectivity-preserved flocking motion, shown in Figure~\ref{fig:ori_based_safe_muij}. Given that the initial connected graph $\mathcal{G}(0)$ is complete, Figure~\ref{fig:ori_based_safe_gamma} presents the evolution of the inter-agent weight parameter $\gamma^*_{ij} = \gamma_{\text{ref-ij}} + \lambda_j \alpha(h_{ij})$ in QP \eqref{QP_framework:leader}-\eqref{QP_framework:follower}, where the reference parameter is preset as $\gamma_{\text{ref-ij}} = 1$. The optimal $\gamma^{*}_{ij}$ remains non-negative, ensuring non-empty intersection between the admissible set $\mathcal{U}_{\text{adm}}$ and CBF constraint set $\mathcal{U}_{\text{cbf-i}}(t)$ at all time. As a result, it renders the non-negative CBFs for each connected pair (in Figure~\ref{fig:ori_based_safe_hij}) and the optimal solution $a^*_i, \omega^*_i$ constrained within the admissible set $\mathcal{U}_{\text{adm}}$ (shown in Figure~\ref{fig:ori_based_safe_acc}-\ref{fig:ori_based_safe_w}), validating the feasibility in Theorem~\ref{thm:feasibility}. 
However, unlike the trajectory in Figure~\ref{fig:ori_based_collision_traj} where flocking is achieved, Figure~\ref{fig:ori_based_safe_e}-\ref{fig:ori_based_safe_mui} show that followers converge to an undesired steady-state, characterized by some of $\left\| \widetilde{\bm{e}}_i\right\|\neq 0 $ and $\omega^*_i = 0$. This reflects the geometric conflict between the desired flocking configuration and the safety \& connectivity requirement, as stated in Lemma~\ref{lemma:equil}. 

\begin{figure}[ht]
	\centering
	 \subfigure[]{
		\begin{minipage}[t]{0.24\textwidth}
		 \label{fig:ori_free_connectivity_preservation_traj}
			\centering			
            \includegraphics[width=0.9\textwidth]{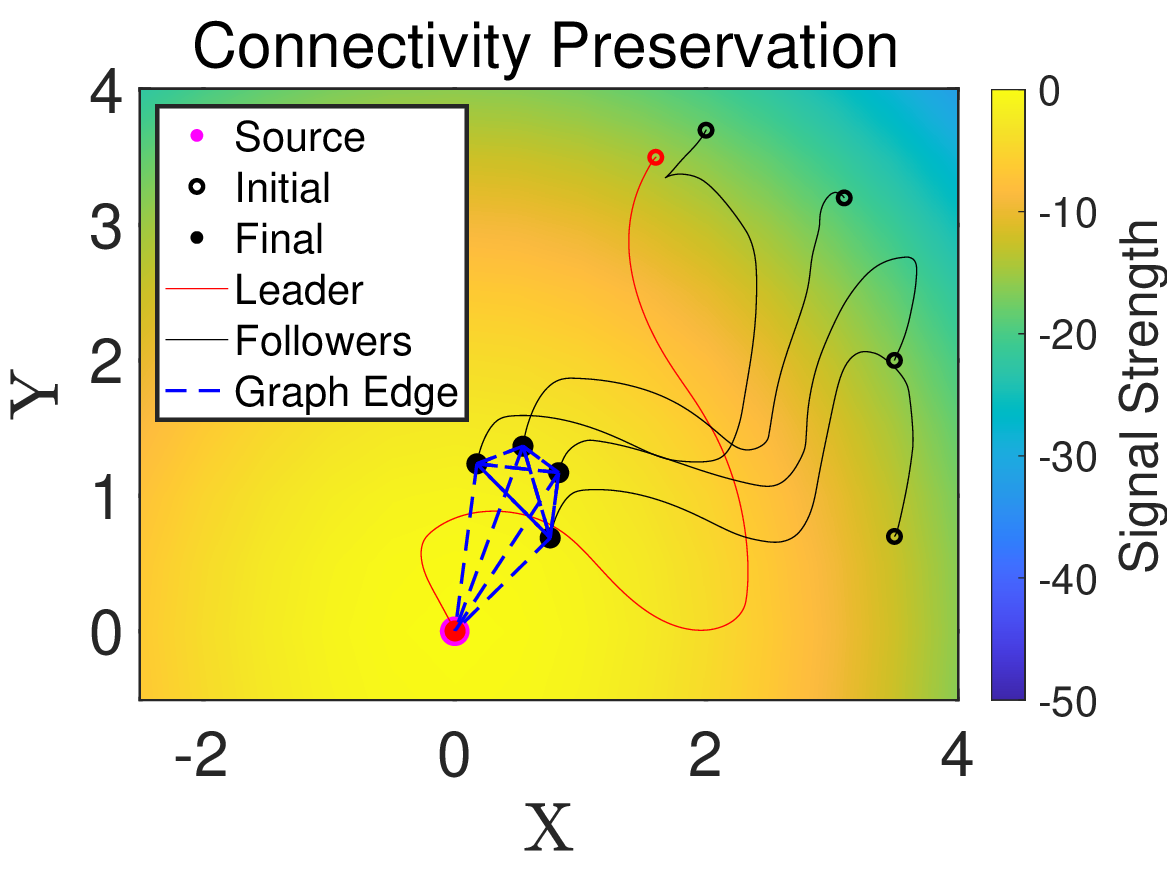}
		\end{minipage}%
	}%
  \subfigure[]{
		\begin{minipage}[t]{0.24\textwidth}
		 \label{fig:ori_free_connectivity_preservation_a}
			\centering			
            \includegraphics[width=0.9\textwidth]{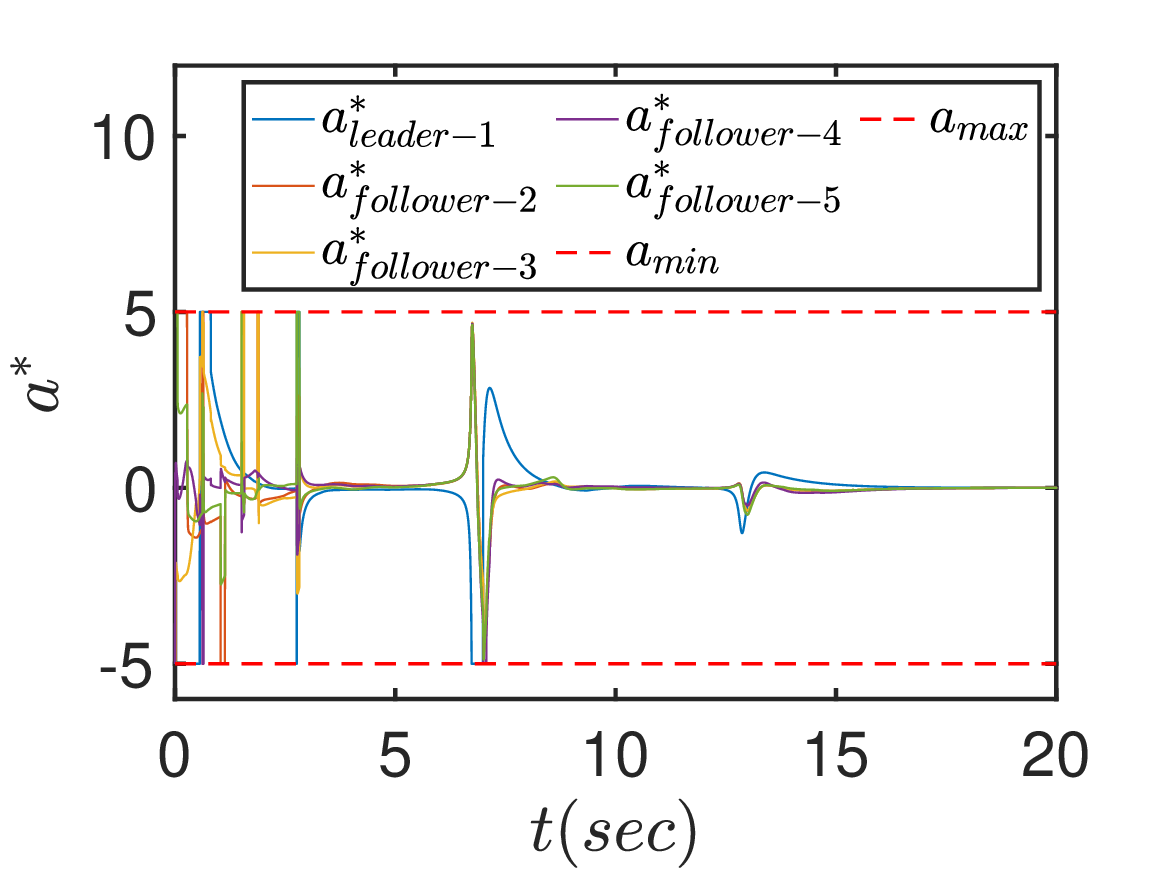}
		\end{minipage}%
	}%
    \vspace{-8pt}
    
	\subfigure[]{
		\begin{minipage}[t]{0.24\textwidth}
		 \label{fig:ori_free_connectivity_preservation_w}
			\centering			
            \includegraphics[width=0.9\textwidth]{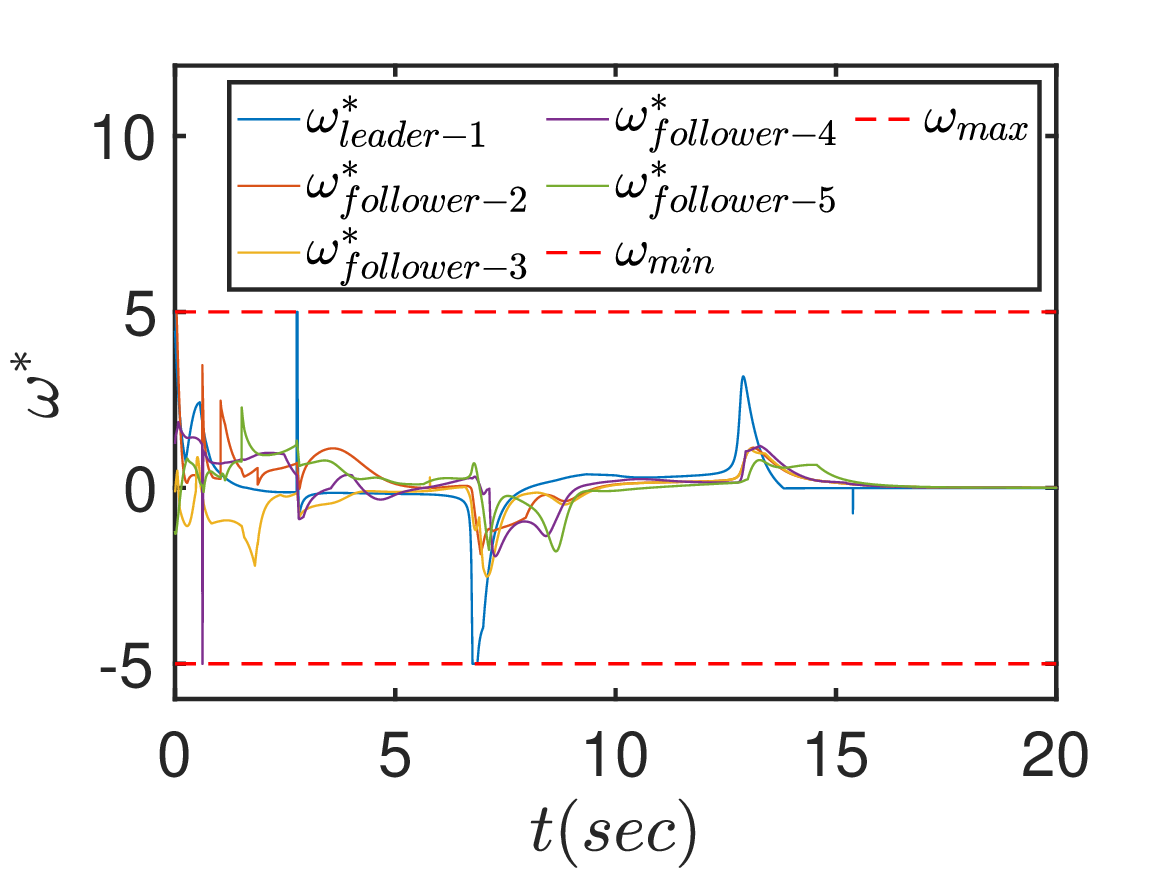}
		\end{minipage}%
	}%
    \subfigure[]{
		\begin{minipage}[t]{0.24\textwidth}
		 \label{fig:ori_free_connectivity_preservation_cbf}
			\centering			
            \includegraphics[width=0.9\textwidth]{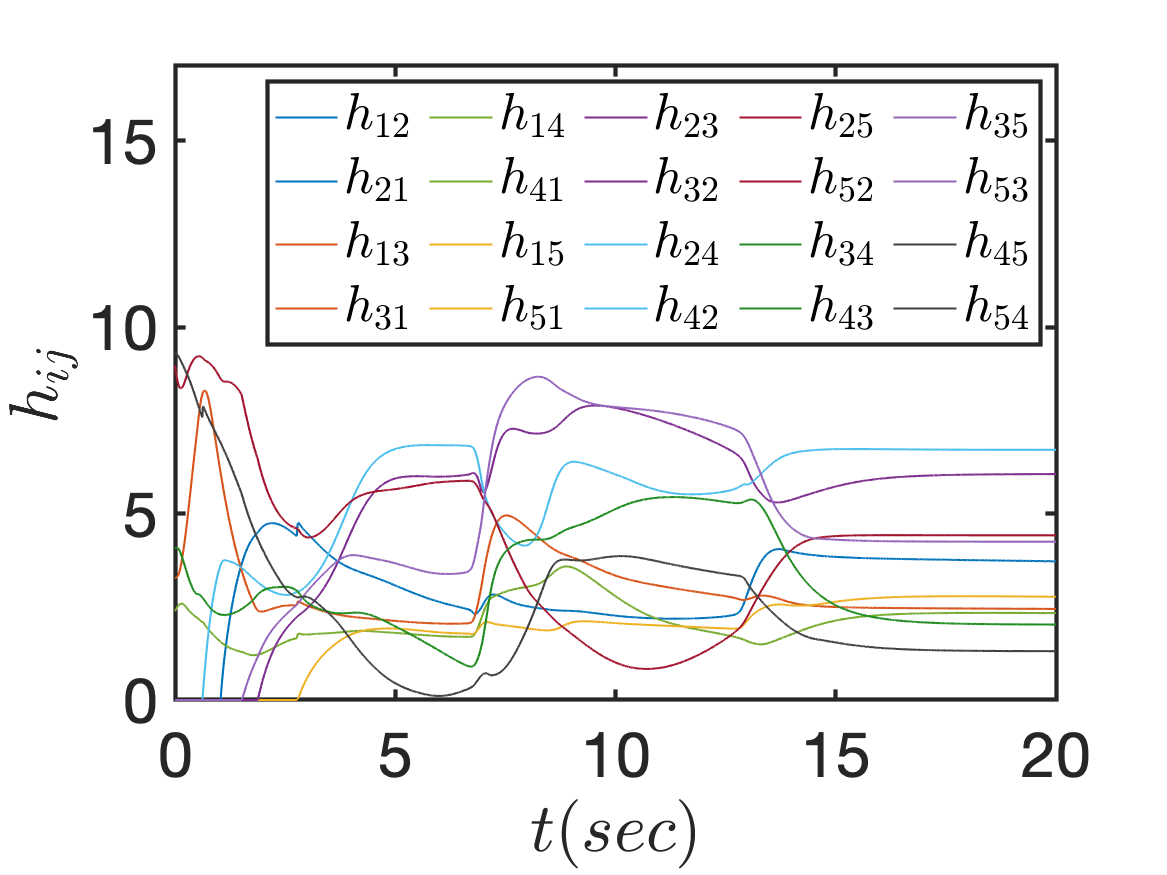}
		\end{minipage}%
	}%
     \vspace{-8pt}
     
	\subfigure[]{
		\begin{minipage}[t]{0.24\textwidth}
		 \label{fig:ori_free_connectivity_preservation_eio}
			\centering			
            \includegraphics[width=0.9\textwidth]{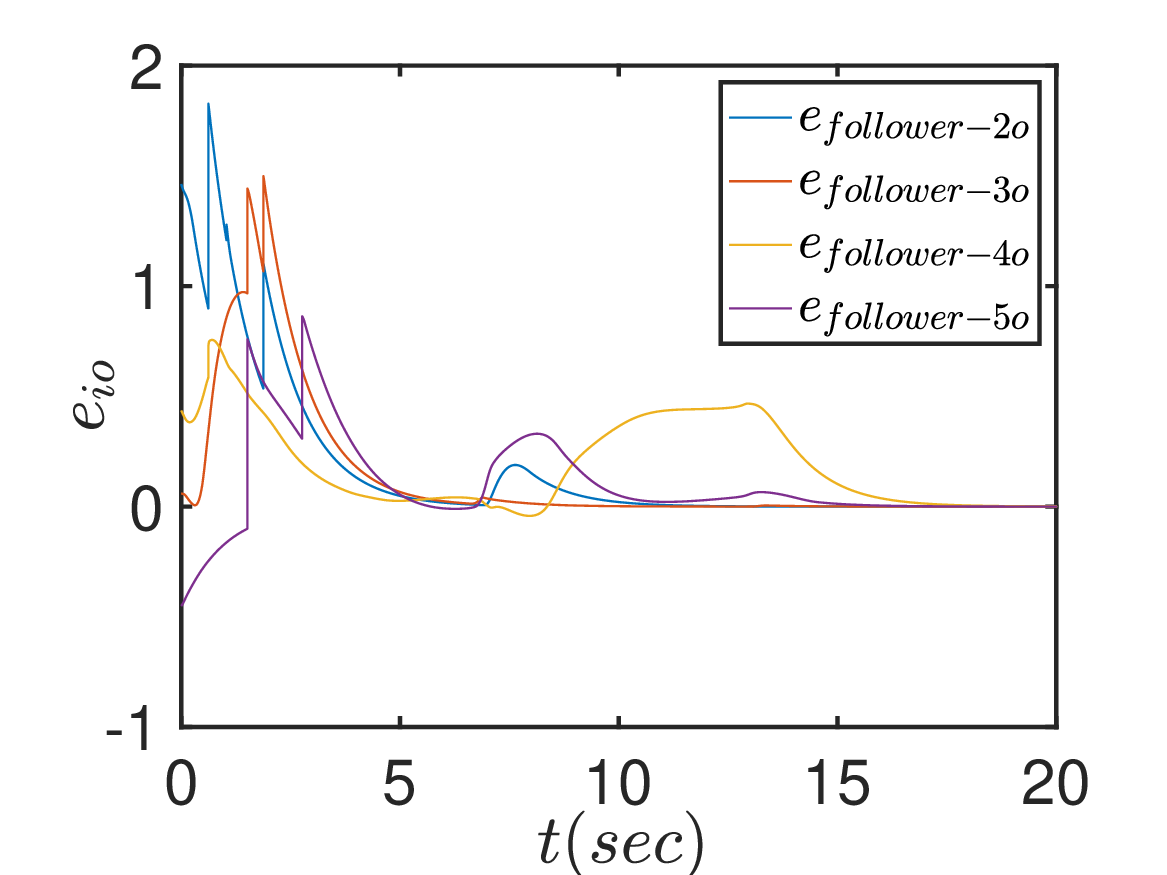}
		\end{minipage}%
	}%
    \subfigure[]{
		\begin{minipage}[t]{0.24\textwidth}
		 \label{fig:ori_free_connectivity_preservation_muij}
			\centering			
            \includegraphics[width=0.9\textwidth]{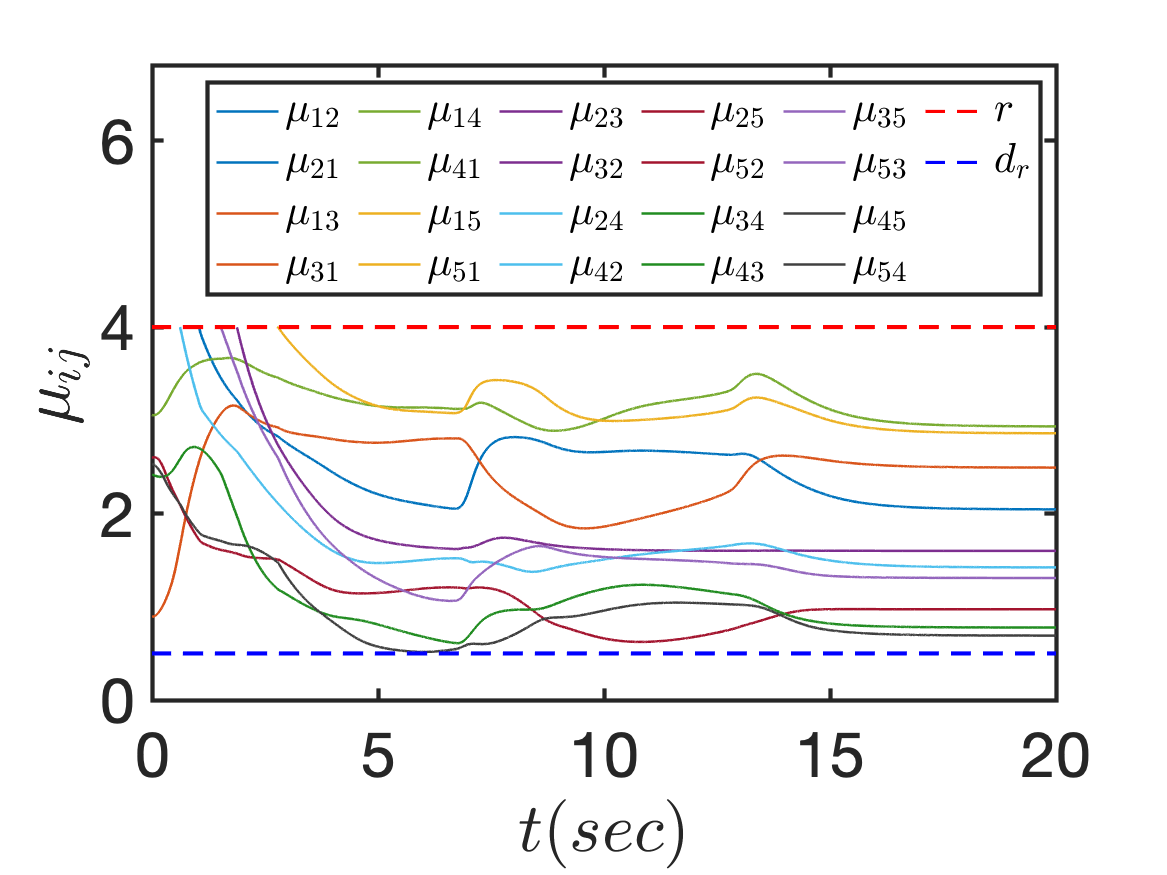}
		\end{minipage}%
	}%

	\centering
     \vspace{-6pt}
         \caption{Inter-agent connectivity-preserved evolution: simulation results of source-seeking (leader) and orientation-free flocking-cohesion (follower, $\bm{u}_{\text{flock-i}}$ is as  \eqref{eq:ori_free_error_v}-\eqref{eq:ori_free_error_w}) based on CBF-QPs \eqref{QP_framework:leader}-\eqref{QP_framework:follower}. (a) Evolution trajectories with dynamic graph $\mathcal{G}(t)$; (b-c) Optimal solution of acceleration $a^*_i$ and angular velocity $\omega^*_i$; (d) Inter-agent CBF $h_{ij}$ for connected agent pair $(i,j)\in\mathcal{E}(t)$; (e) Orientation-free flocking error $e_{io}$; (f) Signal gradient difference $\mu_{ij}$.}
        \label{fig:ori_free_connectivity_preservation}
\end{figure}
\subsubsection{Inter-agent connectivity preservation}
Figure~\ref{fig:ori_free_connectivity_break} and Figure~\ref{fig:ori_free_connectivity_preservation} provide validation of the connectivity preservation property, by considering the leader-follower scenario initialized with a connected but incomplete (dynamic) undirected topology $\mathcal{G}(0)$. As shown in Figure~\ref{fig:ori_fee_connectivity_initial}, the inter-agent connectedness (in blue dashed lines) is initialized based on the maximum communication range limits $r=4$. Figure~\ref{fig:ori_free_connectivity_break} demonstrates the results without inter-agent connectivity preservation guarantee, where the initially connected edge between the leader and the follower No.~$4$ breaks around $t=1s$, resulting a final flocking configuration as in Figure~\ref{fig:ori_free_connectivity_break_10traj}, where the follower group is connected to the leader by only one edge.

As a comparison, Figure~\ref{fig:ori_free_connectivity_preservation} presents the results with CBF-QPs deployed for all agents. The source-seeking and flocking-cohesion tasks are safely achieved without inter-agent connectivity break. Additionally, as a further validation of the QP feasibility in Theorem~\ref{thm:feasibility}, we plot the evolution of the solution $a^*_i, \omega^*_i$ in Figure~\ref{fig:ori_free_connectivity_preservation_a}-\ref{fig:ori_free_connectivity_preservation_w}, which both remain within the admissible limits. Note that the inter-agent space $\mu_{ij}$ is plotted only when the edge $(i,j)$ exists in the dynamic communication graph (that is, when the agents enter the maximum sensing hood $r=4$). For example, the evolution of 
$\mu_{12}$ (blue line), $\mu_{23}$ (purple line), $\mu_{24}$ (light blue line), $\mu_{35}$ (light purple line), $\mu_{51}$ (yellow line) appear around  $t=1\sim3s$ in Figure~\ref{fig:ori_free_connectivity_preservation_muij}, 
corresponding to the formation of new edges $(1,2), (2,3), (2,4), (3,5), (5,1)$. Notably, the leader and follower No.~$4$ remain connected ($\mu_{14}$ in light green line). These results confirm that $d_r < \mu_{ij} < r$ holds for all connected pairs $(i,j)\in\mathcal{E}(t)$, validating the effectiveness of our feasibility-guaranteed CBF-QPs framework.

\section{Conclusions}\label{sec:conclusion}
In this article, we analyze the distributed flocking control problem for a multi-agent system with non-holonomic constraints, presenting a safe connectivity-preserved group source-seeking and flocking-cohesion algorithm in an unknown environment covered by a certain signal distribution. Two distributed cohesive flocking controllers are proposed by sharing their local source gradient measurements between connected agents, which reduces the workload of relative distance measurement and calculation. Specifically, the proposed flocking controllers enable the agents with any initial states in the connected undirected network to converge to the flocking cohesion. In order to guarantee a safe group motion that involves the conflict of flocking cohesion and separation rules, the potential collisions can be avoided by distributing safety control barrier certificates to each agent, which only considers the objects in the limited communication range.

\appendices
\section{Flocking Cohesion \& Source Seeking}
\subsection{Proof of Theorem~\ref{thm:ori_uncon_flock}: Orientation-free Flocking}\label{Appendix:thm_ori_uncon_flock}
\begin{proof}\label{sec:ori_uncon_proof}
We first prove the flocking property of the system. With the linearized model defined at offset point $P_{io}$ in \eqref{eq:double_integrator}, the time derivative of the flocking error $e_{io}$ in \eqref{eq:ori_uncon_error} 
is calculated as
$\dot{e}_{io} =  {\bm{o}}_{e_{io}} \left [\left(\frac{1}{{N}_i} \sum_{j\in \mathcal{N}_i} \nabla^2 \bm{J}_{j}  \begin{bmatrix}\dot x_{j} \\ \dot y_{j} \end{bmatrix} \right ) - \nabla^2 \bm{J}_{io}  \begin{bmatrix}\dot x_{io} \\ \dot y_{io} \end{bmatrix} \right]$, where $\bm{o}_{e_{io}}$ is given in \eqref{eq:oei_vector}. Then, by substituting the proposed orientation-free flocking controller \eqref{eq:ori_free_error_v} and \eqref{eq:ori_free_error_w} into above, and using the orientation vectors $\bm{o}_i$ and $\bm{o}^\perp_i$ as in \eqref{eq:orientation} (which satisfy the relation  $\bm{I} = {\bm{o}}^\top_i {\bm{o}}_i+ {({\bm{o}_i}^{\perp})^\top} {\bm{o}}^\perp_i$, with $\bm{I}$ is a $2\times 2$ identity matrix), we obtain that
 $\dot e_{io} = {\bm{o}}_{e_{io}} \left [ \left(\frac{1}{{N}_i} \sum_{j\in \mathcal{N}_i} \nabla^2 \bm{J}_{j}   \begin{bmatrix}\dot x_{j} \\ \dot y_{j} \end{bmatrix} \right) 
- \left(\frac{1}{{N}_i} \sum_{k\in \mathcal{N}_i} \nabla^2 \bm{J}_{k}  \begin{bmatrix}\dot x_{k} \\ \dot y_{k} \end{bmatrix} \right) \right]
- \bm{o}_{e_{io}} \nabla^2 \bm{J}_{io} (\nabla^2 \bm{J}_{io})^{-1}{\bm{o}}^\top_{e_{io}} e_{io} K_f = - {\bm{o}}_{e_{io}} \bm{I}{\bm{o}}^\top_{e_{io}} e_{io} K_f  = -e_{io}K_f$.
Since the flocking error $e_{io}$ in above is a linear stable autonomous system, $e_{io}$ converges to zero exponentially, i.e., $e_{io}(t)=\exp(-K_ft)e_{io}(0)$. Therefore, the group of $n$ agents achieves the flocking cohesion task. Note that one can use $V_f=\frac{1}{2}\sum_{i \in \mathcal{V}_f} e_{io}^2$ as the Lyapunov function for showing the flocking cohesion. In this case, its time-derivative satisfies 
$\dot V_f (e)  = \sum_{i \in \mathcal{V}_f} e_{io} \dot  e_{io} = -K_f\sum_{i \in \mathcal{V}_f} e_{io}^2$, where one can also deduce the exponential stability of $e_{io}=0$. 

With regard to the source-seeking property of the whole dynamic flocking system, we only need to analyze the leader agent's motion as it is always connected to the flocking network. Following the approach in our previous work\cite{Li_SS}, let us consider the leader with the extended state variables 
$\bm{z}_{\text{L}} =  [{{z}}_{\text{L},1},\,
{{z}}_{\text{L},2},\,
{{z}}_{\text{L},3},\,
{{z}}_{\text{L},4} ]^\top= [x_{\text{L}},\, y_{\text{L}},\, \cos(\theta_{\text{L}}),\, \sin(\theta_{\text{L}})]^\top$ 
and use the following Lyapunov function
\begin{equation}\label{eq:Vs}
      V_s(\bm{z}_{\text{L}}) = J^* - J_{\text{L}}({z}_{\text{L},1}, {z}_{\text{L},2}) + \frac{1}{2}{z}^2_{\text{L},3} + \frac{1}{2}{z}^2_{\text{L},4},
\end{equation}
where $J^*$ is the maxima of $J$ at the source, and its time-derivative follows $\dot V_s = - k_v  \left \langle  \nabla \bm{J}_{\text{L}} ,  \begin{bmatrix}\cos(\theta_{\text{L}}) & \sin(\theta_{\text{L}}) \end{bmatrix}  \right \rangle ^2 
     \leq 0$.
Accordingly, following the same argumentation as in \cite[Proposition III.1]{Li_SS}, the gradient-ascent controller $\bm{u}_\text{L}= [v_\text{L},\,\omega_\text{L}]^\top$ in \eqref{eq:SS}  guarantees the boundedness of the closed-loop leader's state trajectory and the convergence of the position to the source location $(x^*,y^*) $ for any initial conditions. 

Since the leader agent $L$ belongs to at least one follower agents' neighbor set $\mathcal{N}_i$ in the connected graph $\mathcal{G}$, we can conclude that the group achieves both the source-seeking task \eqref{eq:convergence-0} and the flocking-cohesion task.
\end{proof}\vspace{0.1cm}

\subsection{Proof of Theorem~\ref{thm:ori_con_flock}: Orientation-based Flocking}\label{Appendix:thm_ori_con_flock}
\begin{proof}
Consider the Lyapunov function for the closed-loop systems as $ V = V_f + V_s$, 
where $V_s$ defined in \eqref{eq:Vs} is the source-seeking Lyapunov function of the leader, and $V_f$ is the corresponding Lyapunov function for the flocking of the followers, which will be defined shortly below. 

As the source-seeking property of the leader has been shown in Appendix \ref{sec:ori_uncon_proof}, we now analyze the flocking property with respect to the new error variable $\widetilde{\bm{e}}_{i}$ in \eqref{eq:ori_con_error}. The time derivative of the error vector is computed as $\dot{\widetilde{\bm{e}}}^\top_i   
=\frac{1}{{N}_i}\sum_{j\in \mathcal{N}_i}  \nabla^2 \bm{J}_{j} \bm{o}^\top_j v_j
  - \nabla^2 \bm{J}_{i}  \bm{o}^\top_i v_i + d^*_{\nabla \bm{J}} \left(\bm{o}^\perp_i\right)^\top \dot \theta_i$. Substituting the 
  flocking controller \eqref{eq:ori_con_error_v} and \eqref{eq:ori_con_error_w}, we have  
\begin{align}\label{eq:e_dot}
 \dot{\widetilde{\bm{e}}}^\top_i  
 &  =  \frac{1}{{N}_i}\sum_{j\in \mathcal{N}_i}  \nabla^2 \bm{J}_{j}        \bm{o}^\top_j v_j- \frac{1}{{N}_i}\sum_{k\in \mathcal{N}_i}  \nabla^2 \bm{J}_{k}  \bm{o}^\top_k v_k  \nonumber \\
 & \quad-k_{fv} \nabla^2 \bm{J}_i  \bm{o}^\top_i \widetilde{\bm{e}}_i \nabla^2 \bm{J}_i  \bm{o}^\top_i
    - k_\omega \left(\bm{o}^\perp_i\right)^\top \widetilde{\bm{e}}_i  \left(\bm{o}^\perp_i\right)^\top \\
& =-k_{fv} \nabla^2 \bm{J}_i  \bm{o}^\top_i \widetilde{\bm{e}}_i \nabla^2 \bm{J}_i  \bm{o}^\top_i
       - k_{f\omega} \left(\bm{o}^\perp_i\right)^\top\widetilde{\bm{e}}_i \left(\bm{o}^\perp_i\right)^\top  \nonumber
\end{align}
Define $V_f =  \frac{1}{2}\sum_{i\in \mathcal{V}_f}\widetilde{\bm{e}}_{i}\widetilde{\bm{e}}^\top_{i}$ for flocking cohesion of the followers, following its time derivative as $\dot{V}_f = \sum_{i\in \mathcal{V}_f}\widetilde{\bm{e}}_{i}\dot{\widetilde{\bm{e}}}^\top_{i} = -k_{fv}\sum_{i\in \mathcal{V}_f} \left(\widetilde{\bm{e}}_i \nabla^2 \bm{J}_{i}  \bm{o}^\top_i\right)^2  -k_{f\omega} \sum_{i\in \mathcal{V}_f}  \left(\widetilde{\bm{e}}_i   \left(\bm{o}^\perp_i\right)^\top\right)^2  
 \leq 0$.
From this inequality, it follows immediately that $\dot V_f \in L_1(\rline_+)$.  
As $V_f$ is radially unbounded with respect to $\left \| \widetilde{\bm{e}}_{i} \right \|$, it follows from $\dot{V}_f \leq 0$ that the error vector $\widetilde{\bm{e}}_i$ is bounded for all $i$. 
Given the source-seeking proof in Appendix \ref{sec:ori_uncon_proof}, the leader $L$ converges to the source location $(x^*, y^*)$ and its state trajectory is bounded. Note that by hypothesis, the leader belongs to at least one follower agent's neighbor set $\mathcal{N}_i$ at any given time instance {$t \geq t_0$} in the connected graph $\mathcal{G}$, considering the bounded flocking error $\widetilde{\bm{e}}_i$, it implies that the followers have bounded state trajectories. Hence, with the unicycle dynamics in \eqref{eq:unicycle_model}, the control input of longitudinal velocity $v_i$ is bounded. Then, considering the control input of angular velocity $\omega_i$ given in \eqref{eq:ori_con_error_w} and the hypothesis of the theorem that the second-order and third-order partial derivative of $J_i$ is bounded, it is clear that $\omega_i$ is proven as bounded. By using this property,  let us consider $\ddot{\widetilde{\bm{e}}}_i$ which can be derived from $\dot{\widetilde{\bm{e}}}_i$ in \eqref{eq:e_dot}, its boundedness can be proven by the bounded $\dot{\widetilde{\bm{e}}}_i, \omega_i, \nabla^2 \bm{J}_i$ and $\nabla^3 \bm{J}_i$. With $\dot V_f$ in the above, it is now straightforward to show that $\ddot V_f$ is bounded, i.e., $\dot V_f$ is uniformly continuous.
Accordingly, using Barbalat's lemma, $\dot V_f \in L_1(\rline_+)$ and uniform continuity of $\dot V_f$ imply that  $\lim_{t\to \infty}  \dot V_f \to 0$. 

Let us now analyze the asymptotic behavior of the closed-loop systems when $\dot V_f=0$. From above calculation of $\dot{V}_f$, $\dot V_f=0$ if and only if $\widetilde{\bm{e}}_{i}=\bm{0}_{1\times 2}$. By contradiction, suppose that $\widetilde{\bm{e}}_{i}\neq \bm{0}_{1\times 2}$, $\widetilde{\bm{e}}_i\nabla^2 \bm{J}_i\bm{o}^\top_i=0$ and $\widetilde{\bm{e}}_i \left(\bm{o}^\perp_i\right)^\top= 0$. Since $J_i$ is a strictly concave function, $\nabla^2 \bm{J}_i$ is a symmetric negative-definite matrix. Therefore $\widetilde{\bm{e}}_i\nabla^2 \bm{J}_i\ \bm{o}^\top_i=0$ and $\widetilde{\bm{e}}_i \left(\bm{o}^\perp_i\right)^\top= 0$ hold if and only if $\widetilde{\bm{e}}_i=\bm{0}_{1\times 2}$. This contradicts that $\widetilde{\bm{e}}_{i}\neq \bm{0}_{1\times 2}$. Hence, in the asymptote when $\dot V_f=0$, we have $\widetilde{\bm{e}}_i=\bm{0}_{1\times 2}$, corresponding to the scenario where each error vector converges to zero, i.e., the group achieves flocking cohesion.
\end{proof}

\section{Proof of Proposition~\ref{pro:relative_degree}: Uniform Relative Degree}\label{Appendix:Uniform relative degree}
\begin{proof}
Consider the multi-unicycle system within a quadratic concave source field $J(x,y)=-x^2-y^2$, where the state of each agent is modeled upon the extended state space, defined as $\bm{\xi}_i$ in \eqref{eq:ext_dynamics}. Given the inter-agent CBF $h_{ij}$ in \eqref{eq:hij}, its Lie derivative along the vector field $g_i(\bm{\xi}_i)$ associated with unicycle agent $i$ is expressed by 
$ L_{g_i}h_{ij}(\bm{\xi}_i)  = \frac{\partial h_{ij}(\bm{\xi}_i)}{\partial \bm{\xi}_i}g_i(\bm{\xi}_i)  = [
     \frac{\partial h_{ij}(\bm{\xi}_i)}{\partial \xi_{i,3}} + \left<\bm{\nu}_i(\bm{\xi}_i), \bm{o}_i(\bm{\xi}_i)\right>,\,
      \left<\bm{\nu}_i(\bm{\xi}_i), \bm{\chi}_i(\bm{\xi}_i)\right>]$, with the partial derivative vector $\bm{\nu}_i(\bm{\xi}_i)= [
     \frac{\partial h_{ij}(\bm{\xi}_i)}{\partial \xi_{i,4}},\,
     \frac{\partial h_{ij}(\bm{\xi}_i)}{\partial \xi_{i,5}}]$, the state vector $\bm{\chi}_i(\bm{\xi}_i) = [ -\xi_{i,5},\, \xi_{i,4}]$, and the orientation vector $\bm{o}_i(\bm{\xi}_i)$, while $\left< \cdot, \cdot \right>$ denoting the inner product. By substituting the system dynamics and omitting the state argument  $\bm{\xi}_i$ for brevity, it is calculated that $ L_{g_i}h_{ij} = -\begin{bmatrix}
       D_{ij}\epsilon e^{-P_{ij} }
        &  D_{ij}e^{-P_{ij} } \left \langle {\bm{o}}_{i}, {\bm{b}}^\perp_{ij} \right \rangle
    \end{bmatrix}$,
where $D_{ij}$ is as in \eqref{eq:Dij}, ${\bm{b}}^\perp_{ij} $ is the orthogonal vector of ${\bm{b}}_{ij}$. It follows that $\left \| L_{g_i}h_{ij} \right \|\neq 0 $ is ensured such that $h_{ij}$ is a valid ZCBF in the range of $d_r<\mu_{ij}<r$ (i.e., $h_{ij}>0$). In particular, the relative degrees of $h_{ij}$ regarding the control input element $a_i$ and $\omega_i$ are both $1$, uniformly.
\end{proof}

\section{Proof of Theorem~\ref{thm:feasibility}: QP Feasibility}\label{Appendix:feasibility}
\begin{proof}
First, recall that the \textit{safe spatial communication set}  $\mathcal{S}_{ij}$ defined in \eqref{set:sij} follows $h_{ij}(\bm{\xi}_i, \bm{\xi}_j)>0$. Given the ZCBF defined in \eqref{eq:CBF_hij} and its uniform relative degree property (Proposition~\ref{pro:relative_degree}), it follows that $d_r<\mu_{ij}<r$, ensuring $\left \| L_{g_i}h_{ij} \right \|\neq 0 $. This guarantees that the ZCBF is valid over the set $\mathcal{S}_{ij}$, such that the CBF constraint conditions \eqref{QP:leader_QP_CBF} and \eqref{QP:follower_QP_CBF} are well-defined and enforceable. 

Subsequently, each QP solution $\bm{u}^*$ must satisfy two classes of constraints: (i) $\mathcal{U}_{\text{adm}}$ defined in \eqref{set:adm_conn} is for control input saturation; (ii) CBF constraint set $\mathcal{U}_{\text{cbf-i}}(t)$ given in \eqref{set:CBF-cons} is for ensuring inter-agent safety and connectivity preservation, respectively. According to the \textit{CBF-QP Feasibility Condition} in Remark~\ref{def:feasibility}, the optimal solution $\bm{u}^*$ exists only if the intersection of these two constraint sets is non-empty. In other words, \textit{infeasibility} can arise when these imposed constraints are in conflict, which falls into two main types:
\begin{enumerate}
    \item \textit{Conflict Type I}: Conflicts among CBF constraints imposed by neighboring agents;
    \item \textit{Conflict Type II}: Conflicts between $\mathcal{U}_{\text{adm}}$ and $\mathcal{U}_{\text{cbf-i}}(t)$.
\end{enumerate}
In the following proof, we will discuss these conflicts under various cases at time $t$, where at least one type of constraint is active for infeasibility analysis.
Since the leader and followers share the same QP framework (differing only in the nominal reference control inputs), we focus on the distributed QP in \eqref{QP_framework:follower} as a representative example for feasibility analysis. Before proceeding, let us state the \textit{active} constraint and clarify the QP feasibility condition in the following remarks.
\begin{remark}[Active Constraints in QP]\label{reamrk:active}
    In QP, a constraint is said to be \textit{active} at the optimal solution $\bm{u}$ if it holds with equality and contributes to the Karush-Kuhn-Tucker (KKT) conditions. Specifically, for an inequality constraint of the form $\mathcal{C}(\bm{u})\leq 0$, it is \textit{active} at $\bm{u}$ if $\mathcal{C}(\bm{u}^*)=0$.
\end{remark}
\begin{remark}[QP Feasibility Condition]
    The QP in \eqref{QP_framework:follower} for follower agent $i\in\mathcal{V}_f$ is guaranteed \textit{feasible} at time $t$, if 
    \begin{equation}\label{cond:feas}
    \mathcal{F}:=  \sup_{\bm{u}_i\in\mathcal{U}_{\text{adm}}} \left[ L_{f_i}h_{ij}+L_{g_i}h_{ij}\bm{u}_i+\gamma_{ij}\alpha(h_{ij})\right]  \geq 0
\end{equation}
holds for all neighbor agents $j \in \mathcal{N}_{\text{cbf-i}}(t)$ imposing CBF constraints on agent $i$.
\end{remark}
Note this provides an alternative interpretation of the feasibility condition discussed in Remark~\ref{def:feasibility}.

Now, given the condition of $(\bm{\xi}_i(t), \bm{\xi}_j(t))\in\mathcal{S}_{ij}$ for all $(i,j)\in\mathcal E(t)$ at time $t$, we analyze the QP feasibility in \eqref{QP_framework:follower} for follower agent $i$ in the following two main cases.

\subsection{Feasibility Case A: Both the CBF constraint $\mathcal{U}_{\text{cbf-i}}(t)$ and admissible control input constraint $\mathcal{U}_{\text{adm}}$ are active.}
This case indicates that the admissible control input  $\bm{u}_{\text{adm}}\in\mathcal{U}_{\text{adm}}$ cannot satisfy the CBF constraint with the nominal weight $\gamma_{\text{ref-ij}}$, i.e., 
\begin{equation}\label{cond:active_case_A}
L_{f_i}h_{ij}+L_{g_i}h_{ij}\bm{u}_{\text{adm}}+\gamma_{\text{ref-ij}}\alpha(h_{ij}) <0, \,\, j \in \mathcal{N}_{\text{cbf-i}}(t)
\end{equation}
Therefore, both $\mathcal{U}_{\text{cbf-i}}(t)$ and $\mathcal{U}_{\text{adm}}$ become active, imposing constraints to derive the feasible set for optimal solution $\bm{u}^*_i$. Consequently, the infeasibility problem may be caused by either \textit{Conflict Type I} or \textit{Conflict Type II} in this case.

For analysis, let us define a new control input error vector by $\bm{e}_{\bm{u}_i} = [
    e_{u_{i,1}}, e_{u_{i,2}} ]^\top = \bm{u}_i - \bm{u}_{\text{flock-i}}\in \mathbb{R}^2$,
and a weight error vector by
$\bm{e}_{\bm{\Upsilon}_i}= \mathop{\text{col}}\limits_{j\in\mathcal{N}_i}(e_{\gamma_{ij}})= \bm{\Upsilon}_i - \mathop{\text{col}}\limits_{j\in\mathcal{N}_i}( \gamma_{\text{ref-ij}})=\mathop{\text{col}}\limits_{j\in\mathcal{N}_i}( \gamma_{ij} - \gamma_{\text{ref-ij}})\in\mathbb{R}^{N_i} $.
Then, the QP in \eqref{QP:follower} can be rewritten as 
\begin{equation} \label{QP:new_error} 
\begin{aligned}
    \begin{bmatrix}\bm{e}^*_{\bm{u}_i} \\ \bm{e}^*_{\bm{\Upsilon}_i}  \end{bmatrix}=\mathop{\mathrm{argmin}}\limits_{\begin{smallmatrix}
    \bm{u}_i \in \mathcal{U}_{\text{adm}},\\ \bm{\Upsilon}_i\in \mathbb{R}^{Ni}_+ \end{smallmatrix}}
 \frac{1}{2} \left(\bm{e}^\top_{\bm{u}_i} \bm{e}_{\bm{u}_i} +  \bm{e}^\top_{\bm{\Upsilon}_i}\bm{e}_{\bm{\Upsilon}_i}  \right)    
 \end{aligned}
\end{equation}
\vspace{-5pt}
s.t.
\begin{equation*}
\begin{aligned}
 L_{f_i}h_{ij}+L_{g_i}h_{ij}(\bm{e}_{\bm{u}_i} + \bm{u}_{\text{flock-i}})+(e_{\gamma_{ij}} + \gamma_{\text{ref-ij}})\alpha(h_{ij}) \geq 0, & \\  \text{(Inter-agent CBF Constraint)} & 
\\ a_{\max} - e_{u_{i,1}} - {u}_{\text{flock-i,1}} \geq 0,\quad e_{u_{i,1}} + {u}_{\text{flock-i,1}} - a_{\min} \geq 0,  &
\\ \omega_{\max} - e_{u_{i,2}} - {u}_{\text{flock-i,2}} \geq 0, \quad e_{u_{i,2}} + {u}_{\text{flock-i,2}} - \omega_{\min} \geq 0, &
\\  \text{(Admissible Input Constraint)} & 
\end{aligned}
\end{equation*}
for all $j\in\mathcal{N}_{\text{cbf-i}}(t)$, following the non-negative weight constraint $\bm{\Upsilon}_i \in \mathbb{R}_+^{N_i} $.
 Denote the CBF constraint term by
\begin{equation}\label{CBF_cons_term}
   \mathcal{C}_{\text{CBF}} = L_{f_i}h_{ij}+L_{g_i}h_{ij}(\bm{e}_{\bm{u}_i} + \bm{u}_{\text{flock-i}})+(e_{\gamma_{ij}} + \gamma_{\text{ref-ij}})\alpha(h_{ij}),
\end{equation}
and it is used for the rest of the paper.
For the individual element of $\bm{u}_i$ (i.e., ${u}_{i,1}$ or ${u}_{i,2}$), it is obvious that it cannot violate both the lower and upper bound of the admissable control input set $\mathcal{U}_{\text{adm}}$ in \eqref{set:adm_conn} at the same time. Hence, we proceed to verify the feasibility in individual subcases where different saturation bounds are violated (i.e., $\mathcal{U}_{\text{adm}}$ is active).

\subsubsection{Feasibility Case A.1: Symmetric Saturation Bound Violations}
\begin{itemize}
    \item \textit{Case A.1.1}: Upper Bound Violation\\
We first consider the situation where the upper bounds of control inputs 
$\bm{u}_{\max} = [
    a_{\max},\,\omega_{\max}]^\top$ are violated, implying the constraint active condition as 
\begin{equation}\label{act:umax}
L_{f_i}h_{ij}+L_{g_i}h_{ij}\bm{u}_{\max}+\gamma_{\text{ref-ij}}\alpha(h_{ij}) <0.
\end{equation}
That is, the control input $\bm{u}_{\max}$ cannot satisfy the CBF constraint with nominal weight $\gamma_{\text{ref-ij}}$, resulting in both constraints becoming active in the derivation of $\bm{u}^*_i$.

Subsequently, the Lagrangian function $\mathcal{L}$ associated with the new QP \eqref{QP:new_error} can be given by a set of Lagrange multipliers $\bm{\lambda} = [\lambda_a, \lambda_\omega, \lambda_j,...]^\top$ with $j\in\mathcal{N}_{\text{cbf-i}}(t)$ as
$\mathcal{L} = \frac{1}{2}\left(\bm{e}^\top_{\bm{u}_i} \bm{e}_{\bm{u}_i} +  \bm{e}^\top_{\bm{\Upsilon}_i}\bm{e}_{\bm{\Upsilon}_i} \right) 
    -\sum_{j\in\mathcal{N}_i}\lambda_j \mathcal{C}_{\text{CBF}}
     -\lambda_a(a_{\max} - e_{u_{i,1}} - {u}_{\text{flock-i,1}})
     -\lambda_\omega(\omega_{\max} - e_{u_{i,2}} - {u}_{\text{flock-i,2}} )$,
where $\mathcal{C}_{\text{CBF}}$ is defined in \eqref{CBF_cons_term}.
The Karush-Kuhn-Tucker (KKT) conditions for the new QP problem \eqref{QP:new_error} are given by\\
\begin{subequations}
\text{(Stationarity:)} \quad
\begin{equation}\label{Cond:Stationarity}
\hspace{-0.35cm} 
\left.
\begin{aligned}
\frac{\partial \mathcal{L}}{\partial \bm{e}_{\bm{u}_i}} 
 = \bm{e}^\top_{\bm{u}_i} 
- \sum_{j \in \mathcal{N}_i} \lambda_j L_{g_i} h_{ij}
+ \begin{bmatrix} \lambda_a & \lambda_\omega \end{bmatrix} 
&= \begin{bmatrix} 0 & 0 \end{bmatrix} 
\\ \frac{\partial \mathcal{L}}{\partial e_{\gamma_{ij}}} 
= e_{\gamma_{ij}} - \lambda_j \alpha(h_{ij}) & = 0 
\end{aligned}
\right \},
\end{equation}
\text{(Complementary Slackness:)} \quad
\begin{equation}\label{Cond:Complementary Slackness}
\left.
\begin{aligned}
\lambda_j \mathcal{C}_{\text{CBF}}= 0 &  
\\ \lambda_a(a_{\max} - e_{u_{i,1}} - {u}_{\text{flock-i,1}}) = 0 &
\\ \lambda_\omega(\omega_{\max} - e_{u_{i,2}} - {u}_{\text{flock-i,2}}) = 0 &
\end{aligned}
\right \}, 
\end{equation}
\text{(Primal Feasibility:)}  \quad
\begin{equation}\label{Cond:Primal Feasibility}
\left.
\begin{aligned}
\mathcal{C}_{\text{CBF}} \geq 0& 
\\ a_{\max} - e_{u_{i,1}} - {u}_{\text{flock-i,1}} \geq 0&
\\ \omega_{\max} - e_{u_{i,2}} - {u}_{\text{flock-i,2}} \geq 0 &
\end{aligned}
\right \}, 
\end{equation}
\begin{flalign}\label{Cond:Dual Feasibility}
\text{(Dual Feasibility:)} \quad \lambda_a \geq 0, \,\,
\lambda_\omega \geq 0, \,\,
\lambda_j \geq 0. &&
\end{flalign}
\end{subequations}
In \textit{Feasibility Case A}, both the constraints $\mathcal{U}_{\text{cbf-i}}(t)$ and $\mathcal{U}_{\text{adm}}$ are active, $\lambda_j, \lambda_a, \lambda_\omega>0$ holds for all $j\in\mathcal{N}_{\text{cbf-i}}(t)$, and thus the complementary slackness condition \eqref{Cond:Complementary Slackness} becomes (i) $\mathcal{C}_{\text{CBF}} = 0$; (ii) $a_{\max} - e_{u_{i,1}} - {u}_{\text{flock-i,1}} = 0 $; (iii) $\omega_{\max} - e_{u_{i,2}} - {u}_{\text{flock-i,2}} = 0$.
Therefore, the resulting error $\bm{e}_{\bm{u}_i} = \bm{u}_{\max} - \bm{u}_{\text{flock-i}}$ renders the optimal control input $\bm{u}^*_i = \bm{u}_{\max} = [a_{\max},\, \omega_{\max}]^\top$.
By substituting $\bm{e}_{\bm{u}_i}$ into the first equality condition term (i), the weight error is computed by $ e_{\gamma_{ij}} = - \frac{L_{f_i}h_{ij} +L_{g_i}h_{ij}\bm{u}_{\max}+\gamma_{\text{ref-ij}}\alpha(h_{ij})}{\alpha(h_{ij}) }$.
Accordingly, given the stationarity condition $e_{\gamma_{ij}} = \lambda_j \alpha(h_{ij})$ in \eqref{Cond:Stationarity}, the Lagrange multipliers for each $j\in\mathcal{N}_{\text{cbf-i}}(t)$ can be obtained by
\begin{equation}\label{eq:lambda_j_final}
     \lambda_j = -\frac{L_{f_i}h_{ij} +L_{g_i}h_{ij}\bm{u}_{\max}+\gamma_{\text{ref-ij}}\alpha(h_{ij})}{\alpha^2(h_{ij}) }, 
\end{equation}
and it renders the final calculation of $\lambda_a, \lambda_\omega$ as
\begin{equation}\label{eq:lambda_a_w_final}
    \begin{bmatrix} \lambda_a \\ \lambda_\omega  \end{bmatrix} = \sum_{j\in\mathcal{N}_i}\lambda_j \Big(L_{g_i}h_{ij}\Big)^\top + \Big(\bm{u}_{\text{flock-i}}-\bm{u}_{\max}\Big).
\end{equation}
The updated weight for agent pair $(i,j)$ is obtained by
\begin{equation}\label{eq:gamma^*}
\begin{aligned}
    \gamma^*_{ij} &= \gamma_{\text{ref-ij}} +  e_{\gamma_{ij}}
     \\& = \gamma_{\text{ref-ij}} - \frac{L_{f_i}h_{ij} +L_{g_i}h_{ij}\bm{u}_{\max}+\gamma_{\text{ref-ij}}\alpha(h_{ij})}{\alpha(h_{ij}) }
    \\&=- \frac{L_{f_i}h_{ij} +L_{g_i}h_{ij}\bm{u}_{\max}}{\alpha(h_{ij}) }.
\end{aligned}
\end{equation}
Let us now discuss the QP feasibility.
As the primal problem \eqref{QP:new_error} is convex, the KKT conditions are sufficient for the pair of points $({\bm{u}}_i, {\gamma}_{ij},{\lambda}_a, {\lambda_\omega}, {\lambda}_j)$ to be primal optimal \cite{boyd2004convex}. Namely, if \eqref{eq:lambda_j_final} and \eqref{eq:lambda_a_w_final} are solvable, implying that there exists a solution of $({\bm{u}}_i, {\gamma}_{ij}, {\lambda}_a, {\lambda_\omega}, {\lambda}_j)$ satisfying the KKT conditions \eqref{Cond:Stationarity}-\eqref{Cond:Dual Feasibility}, then the constraints for the QP hold. 
 Hence, in order to analyze the existence of $\lambda_i$, we recall that the class $\mathcal{K}_\infty$ function $\alpha(h_{ij})\neq 0$ for the inter-agent pair $(\bm{\xi}_i,\bm{\xi}_j)\in\mathcal{S}_{ij}$ where $h_{ij}>0$. Furthermore, with the constraint active condition of \textit{Case A.1.1} in \eqref{act:umax}, $\lambda_j>0$ holds, as shown in \eqref{eq:lambda_j_final}.
Similarly, the updated weight $\gamma^*_{ij}$ exists as in \eqref{eq:gamma^*}, so as the $\lambda_a$ and $\lambda_\omega$. Therefore, the CBF constraint holds. For the existence proof, it is not surprising that the optimal solutions $\bm{u}^*_i$ and $\gamma^*_{ij}$ can be substituted into the feasible condition in \eqref{cond:feas} such that $\mathcal{F} =  \sup_{\bm{u}^*_i\in\mathcal{U}_{\text{adm}}} \left[ L_{f_i}h_{ij}+L_{g_i}h_{ij}\bm{u}^*_i+\gamma^*_{ij}\alpha(h_{ij})\right]=0$.
 Accordingly, the QP is feasible due to $\mathcal{U}_{\text{adm}}\bigcap\mathcal{U}_{\text{cbf-i}}(t)\neq \emptyset$. Specifically, the optimal weight $\gamma^*_{ij}$ obtained in \eqref{eq:gamma^*} updates the range of CBF constraint set $\mathcal{U}_{\text{cbf-i}}(t)$, for ensuring a non-empty intersection with the admissible input set $\mathcal{U}_{\text{adm}}$.
\end{itemize}
\begin{itemize}
    \item \textit{Case A.1.2}: Lower Bound Violation \\
Consider the scenario where both the lower bounds of control inputs, i.e.,  
    $\bm{u}_{\min} = [a_{\min},\, \omega_{\min}]^\top$ are violated. The resulting solution is given by $\bm{u}^*_i = \bm{u}_{\min}$, with the associated multiplier $ \lambda_j = -\frac{L_{f_i}h_{ij} +L_{g_i}h_{ij}\bm{u}_{\min}+\gamma_{\text{ref-ij}}\alpha(h_{ij})}{\alpha^2(h_{ij}) }$. It can be further used to compute $\gamma^*_{ij}, \lambda_a, \lambda_\omega$. A similar feasibility analysis (as in \textit{Case A.1.1}) is omitted here for brevity.
\end{itemize}

\subsubsection{Feasibility Case A.2: Asymmetric Saturation Bound Violation}
In the following, we analyze the feasibility when the individual control elements $u_{i,1}$ and $u_{i,2}$ violate asymmetric saturation bounds.
\begin{itemize}
    \item \textit{Case A.2.1}: The upper bound and lower bound are active for $u_{i,1}$ and $u_{i,2}$, respectively.\\
The corresponding Lagrangian function is rewritten as $ \mathcal{L} = \frac{1}{2}\left(\bm{e}^\top_{\bm{u}_i} \bm{e}_{\bm{u}_i} +  \bm{e}^\top_{\bm{\Upsilon}_i}\bm{e}_{\bm{\Upsilon}_i} \right) 
  - \sum_{j\in\mathcal{N}_i}\lambda_j \mathcal{C}_{\text{CBF}}
   -\lambda_a(a_{\max} - e_{u_{i,1}} - {u}_{\text{flock-i,1}})
     -\lambda_\omega(e_{u_{i,2}} + {u}_{\text{flock-i,2}} - \omega_{\min})$,
which follows a new construction of KKT conditions. Here we only highlight three main changes as: (i) $\frac{\partial \mathcal{L}}{\partial \bm{e}_{\bm{u}_i}} 
= \bm{e}^\top_{\bm{u}_i} 
- \sum_{j \in \mathcal{N}_i} \lambda_j L_{g_i} h_{ij}
+ [\lambda_a ,\, -\lambda_\omega]= [0,\, 0]$; (ii) $ \lambda_a(a_{\max} - e_{u_{i,1}} - {u}_{\text{flock-i,1}})=0 $; (iii) $\lambda_\omega(e_{u_{i,2}} + {u}_{\text{flock-i,2}} - \omega_{\min})=0$.
Following the similar analysis, it is accessible that the final solution $\bm{u}^*_i = [a_{\max},\, \omega_{\min}]^\top$ and $  \lambda_j = -\frac{L_{f_i}h_{ij} +L_{g_i}h_{ij}\bm{u}^*_i+\gamma_{\text{ref-ij}}\alpha(h_{ij})}{\alpha^2(h_{ij}) }$ exist, and the optimal weight parameter is calculated by
$\gamma^*_{ij} =  - \frac{L_{f_i}h_{ij} +L_{g_i}h_{ij}\bm{u}^*_i }{\alpha(h_{ij})}$.
\end{itemize}
\begin{itemize}
    \item \textit{Case A.2.2}: The lower bound and upper bound are active for $u_{i,1}$ and $u_{i,2}$, respectively.\\
A similar analysis is straightforward as \textit{Case A.2.1}.
\end{itemize} 

 As a short summary of \textit{Feasibility Case A}, the final expression of Lagrange multipliers $\lambda_j$ and the updated weight $\gamma^*_{ij}$ regarding each $j\in\mathcal{N}_\text{cbf-i}(t)$ can be written as
\begin{equation}\label{solution:Feas_A_final}
\left.
\begin{aligned}
 \lambda_j = -\frac{L_{f_i}h_{ij} +L_{g_i}h_{ij}\bm{u}^*_i+\gamma_{\text{ref-ij}}\alpha(h_{ij})}{\alpha^2(h_{ij})} &
 \\ \gamma^*_{ij} = \gamma_{\text{ref-ij}} + e_{\gamma_{ij}} = \gamma_{\text{ref-ij}} + \lambda_j \alpha(h_{ij})&
\end{aligned}
\right \}.
\end{equation}
Hence, with the constraint active condition and the property that $h_{ij}>0$ for the agent pair $(\bm{\xi}_i,\bm{\xi}_j)\in\mathcal{S}_{ij}$, the existence of $\lambda_j$ is guaranteed. Together with the calculated $\bm{u}^*_i, \gamma^*_{ij}$ and $(\lambda_j, \lambda_a, \lambda_\omega)$ that satisfy the KKT condition for the convex optimization  \eqref{QP:new_error}, it follows that $\bm{u}^*_i, \gamma^*_{ij}$ are optimal, ensuring the QP is feasible. Furthermore, as shown in \eqref{solution:Feas_A_final}, the role of $\gamma^*_{ij}> \gamma_{\text{ref-ij}}>0 $ is to relax the CBF constraint for guaranteeing $ \bm{u}^*_i \in \mathcal{U}_{\text{adm}}\bigcap\mathcal{U}_{\text{cbf-i}}(t)\neq \emptyset$.

\subsection{Feasibility Case B: Only CBF constraint $\mathcal{U}_{\text{cbf-i}}(t)$ is active.}
In this case, the admissible input constraint remains inactive, indicating that the optimization does not encounter saturation boundary restriction. Additionally, this case indicates that the reference control input $\bm{u}_{\text{flock-i}}$ cannot satisfy the nominal CBF constraint such that $\mathcal{U}_{\text{cbf-i}}(t)$ becomes active, i.e., 
\begin{equation}\label{eq:active_CBF}
     L_{f_i}h_{ij} +L_{g_i}h_{ij}\bm{u}_{\text{flock-i}} + \gamma_{\text{ref-ij}}\alpha(h_{ij}) < 0, \,\, j \in \mathcal{N}_{\text{cbf-i}}(t)
\end{equation}
Accordingly, the infeasibility concern may only come from \textit{Conflict Type I}: conflicts among individual CBF constraints with respect to each active neighboring agent $j\in\mathcal{N}_{\text{cbf-i}}(t)$.

Similar to the previous analysis, the KKT conditions for this case is given by (i) $\frac{\partial \mathcal{L}}{\partial \bm{e}_{\bm{u}_i}} =  \bm{e}^\top_{\bm{u}_i} - \sum_{j\in\mathcal{N}_i}\lambda_j L_{g_i}h_{ij}  = \bm{0}_{1\times 2} $; (ii) $ \frac{\partial \mathcal{L}}{\partial e_{\gamma_{ij}}} = e_{\gamma_{ij}} - \lambda_j \alpha(h_{ij}) = 0 $; (iii) $\mathcal{C}_{\text{CBF}}\geq 0 $; (iv) $ \lambda_j \geq 0$; (v) $\lambda_j \mathcal{C}_{\text{CBF}}= 0$ with $\mathcal{C}_{\text{CBF}}$ defined in \eqref{CBF_cons_term}.
Then the optimal control input $\bm{u}^*_i$ and weight $\gamma^*_{ij}$ are given by
\begin{equation}\label{eq:u_star_CBF}
   \left. \begin{matrix}
        \begin{aligned}
        \bm{u}^*_i &= e_{\bm{u}_i} + \bm{u}_{\text{flock-i}} = \sum_{j\in\mathcal{N}_i}\lambda_j \Big(L_{g_i}h_{ij}\Big)^\top + \bm{u}_{\text{flock-i}}\\
        \gamma^*_{ij} &= e_{\gamma_{ij}} + \gamma_{\text{ref-ij}} = \lambda_j \alpha(h_{ij})+ \gamma_{\text{ref-ij}}
    \end{aligned}
    \end{matrix} \right\}.
\end{equation}
To obtain the Lagrange multiplier vector $\bm{\lambda}=\left[ \lambda_j\right]^\top$ with respect to each active neighboring agent $j\in\mathcal{N}_{\text{cbf-i}}(t)$, we consider the sum of all CBF constraint terms in the case of $\lambda_j>0$, following $ \sum_{j\in\mathcal{N}_i}\mathcal{C}_{\text{CBF}}=0$. 
Then, substituting $e_{\bm{u}_i}$ and $e_{\gamma_{ij}}$ in \eqref{eq:u_star_CBF} into above, each element $\lambda_j$ is computed by
\begin{equation}\label{eq:lam}
    \lambda_{j} = -\frac{L_{f_i}h_{ij} +L_{g_i}h_{ij}\bm{u}_{\text{flock-i}}+\gamma_{\text{ref-ij}}\alpha(h_{ij})}{ \left\| L_{g_i}h_{ij}\right\|^2+\alpha^2(h_{ij}) }. 
\end{equation}
We recall that the constructed inter-agent CBF in \eqref{eq:CBF_hij} 
ensures that $\left \| L_{g_i}h_{ij} \right \|\neq 0$  for $(\bm{\xi}_i(t), \bm{\xi}_j(t))\in\mathcal{S}_{ij}$ (i.e., $h_{ij}>0$ and thus $\alpha(h_{ij})\neq 0$ follows). Therefore, under the constraint active condition given in \eqref{eq:active_CBF}, the Lagrange multiplier $\lambda_j$ in \eqref{eq:lam} exists and satisfies $\lambda_j>0$ for the set $\mathcal{S}_{ij}$. The corresponding $\bm{u}^*_{i}$ and $\gamma^*_{ij}$ can be obtained by substituting $\lambda_{j}$ into \eqref{eq:u_star_CBF}. Hence, the CBF constraints hold, and the supremum attains zero in \eqref{cond:feas}, indicating that the active CBF constraint is satisfied on the boundary, the QP remains feasible.

In summary, given the proofs of the \textit{Feasibility Case A} and \textit{Feasibility Case B},  we conclude that the QP problem for agent $i$ is feasible whenever $(\bm{\xi}_i(t), \bm{\xi}_j(t))\in\mathcal{S}_{ij}$ at time $t$.
\end{proof}

\section{Safety \& Connectivity Preservation }
\subsection{Proof of Theorem~\ref{thm:connectivity}: Inter-agent Safety and Connectivity}\label{Appendix:thm_connectivity}
\begin{proof} 
Given the condition that the leader is the neighbor of at least one follower in the initial connected graph $\mathcal{G}(t_0)$, and considering the identical CBF-QP framework for the leader and followers, we proceed with the proof by analyzing the QP derivation of the followers for clear presentation. 

Consider the follower agent $i$ and its neighbor $j\in\mathcal{N}_i(t)$, defined with states in \eqref{eq:ext_dynamics}. The corresponding inter-agent CBF $h_{ij}(\bm{\xi}_i,\bm{\xi}_j)$ relies on their relative information and its time derivative involves the control inputs $\bm{u}_i$ and $\bm{u}_j$ as 
\begin{equation}\label{eq:dot_hij}
   \begin{aligned}
       \dot{h}_{ij} 
       =\underbrace{ L_{f_i(\bm{\xi}_i)}h_{ij} + L_{g_i(\bm{\xi}_i)}h_{ij}\bm{u}_i }_{\frac{\partial h_{ij}}{\partial \bm{\xi}_i} \dot{\bm{\xi}}_i}+ \underbrace{L_{f_j(\bm{\xi}_j)}h_{ij} + L_{g_j(\bm{\xi}_j)}h_{ij}\bm{u}_j}_{\frac{\partial h_{ij}}{\partial \bm{\xi}_j}\dot{\bm{\xi}}_j}.
      \end{aligned}
   \end{equation} 
 By relaxing the ZCBF condition in \cite{Ames_TAC,Ames_TRO}, the growth rate of $h_{ij}$ is ideally constrained by an extended class $\mathcal{K}_{\infty}$ function $\alpha(\cdot)$ with a non-negative relaxation parameter $\gamma$, i.e., $\dot h_{ij} \geq -\gamma\alpha(h_{ij})$. With the connectivity property of the interconnection graph (i.e. $j \in \mathcal{N}_i(t) \Leftrightarrow  i \in \mathcal{N}_{j}(t)$), the pairwise CBF $h_{ij}$ defined in the specific form of \eqref{eq:CBF_hij} satisfies the relation of $h_{ij}(t) = h_{ji}(t)$ such that $\frac{\partial h_{ij}}{\partial \bm{\xi}_i} = \frac{\partial h_{ji}}{\partial \bm{\xi}_i}$ holds. 
 
 Hence, let us only consider the connected edge $(i,j)\in\mathcal{E}(t)$, following \eqref{eq:dot_hij}, the safety and connectivity constraint for agent $i$ regarding its neighbor $j\in \mathcal{N}_i(t)$ given in QP \eqref{QP:follower_QP_CBF} is  
\begin{align}\label{eq:cons_i}
  \underbrace{ L_{{f_i}(\bm{\xi}_i)}h_{ij} + L_{{g_i}(\bm{\xi}_i)}h_{ij}\bm{u}_i }_{\frac{\partial h_{ij}}{\partial \bm{\xi}_i}\dot {\bm{\xi}}_i} \geq -\gamma_{ij}\alpha(h_{ij}), 
\end{align}
and the corresponding constraint for this agent $j$ subject to this specific neighbor $i$ is given by
\begin{align}\label{eq:cons_j}
     \underbrace{ L_{{f_j}(\bm{\xi}_j)}h_{ji} + L_{{g_j}(\bm{\xi}_j)}h_{ji}\bm{u}_j }_{\frac{\partial h_{ji}}{\partial \bm{\xi}_j}\dot{\bm{\xi}}_j = \frac{\partial h_{ij}}{\partial \bm{\xi}_j}\dot{\bm{\xi}}_j } \geq -\gamma_{ji}\alpha(h_{ji}), 
\end{align}
It is noted that the relation $\frac{\partial h_{ji}}{\partial \bm{\xi}_j}\dot{\bm{\xi}}_j = \frac{\partial h_{ij}}{\partial \bm{\xi}_j}\dot{\bm{\xi}}_j$ holds in \eqref{eq:cons_j}, which can be substituted into the second term in \eqref{eq:dot_hij}. Correspondingly, the sum of the constraints \eqref{eq:cons_i} and \eqref{eq:cons_j} contributes to $\dot{h}_{ij} \geq -\gamma_{ij}\alpha(h_{ij}) - \gamma_{ji}\alpha(h_{ji}) $. Using the property of $h_{ij} = h_{ji}$, the inequality is derived as $\dot{h}_{ij} \geq -\gamma \alpha(h_{ij})$ with $\gamma = \gamma_{ij} + \gamma_{ji}$, which corresponds to the main constraint in \eqref{eq:dot_hij} for the agent $i$. Similarly, the constraint $\dot h_{ji} \geq -\gamma\alpha(h_{ji})$ is satisfied for its neighbor agent $j$. 
Note that the optimized weight parameter $\gamma$ is related to both control inputs $\bm{u}_i$ and $\bm{u}_j$ at time $t$, thus it can be distributed to each sub-constraints \eqref{eq:cons_i} and \eqref{eq:cons_j} and be optimized in the corresponding quadratic programming problems \eqref{QP:follower_QP_CBF} for agent $i$ and $j$, respectively.

Now consider a differentiable equation $\dot y = -\alpha(y)$ with $y(t_0)=h_{ij}(\bm{\xi}_i(t_0),\bm{\xi}_j(t_0))>0$ and $\alpha(\cdot)$ is a locally Lipschitz class $\mathcal{K}$ function. This ODE has a unique solution $y(t)=\sigma(h_{ij}(\bm{\xi}_i(t_0),\bm{\xi}_j(t_0)),t-t_0)$ where $\sigma(\cdot,\cdot)$ is a class $\mathcal{KL}$ function. According to Lemma 4.4 and the comparison Lemma in \cite{Khalil}, it is straightforward to get $h_{ij}(\bm{\xi}_i(t),\bm{\xi}_j(t))\geq \sigma(h_{ij}(\bm{\xi}_i(t_0),\bm{\xi}_j(t_0)),t-t_0)$ and $h_{ij}(\bm{\xi}_i(t),\bm{\xi}_j(t))\in\mathcal{S}_{ij}$ for all $t\geq t_0$ if $h_{ij}(\bm{\xi}_i(t_0),\bm{\xi}_j(t_0))>0$. As proved in Theorem~\ref{thm:feasibility} that QP is feasible if $(\bm{\xi}_i(t), \bm{\xi}_j(t))\in\mathcal{S}_{ij}$ at given time $t$, the set $\mathcal{S}_{ij}$ is forward invariant. This concludes the proof of {\bf P1}.

 With the fact that $h_{ij}(\bm{\xi}_i,\bm{\xi}_j)>0$ only corresponds to $d_r<\mu_{ij}(\bm{\xi}_i,\bm{\xi}_j)<r$, the connectivity of the initially connected edges (i.e., $\mu_{ij}(\bm{\xi}_i(t_0),\bm{\xi}_j(t_0))<r$) can be maintained and the dynamic communication graph $\mathcal{G}(t)$ is guaranteed to be connected at each switching time $t\geq t_0$. In addition, for the newly connected pairs during evolution where $(i,j)\notin \mathcal{E}(t_{0})$ and $(i,j)\in \mathcal{E}(t)$, the same set invariance argument can be applied such that $h_{ij}(\bm{\xi}_i(t), \bm{\xi}_j(t))\in\mathcal{S}_{ij}$ for $t\geq t_0$. Hence, the corresponding edge connectedness can be maintained once any new agent pairs are connected in the dynamic topology. In summary, if the initial states of the inter-agent pair start within the safe spatial communication set $\mathcal{S}_{ij}$ (i.e. $h_{ij}(\bm{\xi}_i(t_0), \bm{\xi}_j(t_0)) > 0$), then $h_{ij}(\bm{\xi}_i(t), \bm{\xi}_j(t)) > 0$ can be guaranteed for all $t\geq t_0$ by satisfying the distributed connectivity constraint \eqref{QP:follower_QP_CBF}. This concludes the proof of {\bf P2}.

On the other hand, the connectivity preservation condition in {\bf P2} also admits the collision avoidance between neighboring agents $(i,j)\in\mathcal{E}(t)$ with the guaranteed $\mu_{ij} > d_r$ for all time $t\geq t_0$. Given the signal flocking illustration in Assumption \ref{ass:J} and Section \ref{sec:flocking_pattern}, the estimated field gradient $\nabla \bm{J}(x,y)$ is related to the agent's position $(x,y)$, and the gradient measurement can be rewritten with the position argument as in \eqref{eq:mu_ij}.
Thus, given the forward invariant set $\mathcal{S}_{ij}$, the maintained $h_{ij}(t)>0$ (where $\mu_{ij}>d_r$) guarantees that the position of agent $i$ and $j$ never coincide. 
For the point-of-mass robot, the non-coincidence behavior ensures inter-agent collision avoidance. For the other agents with dimensions, the minimum safe margin $d_r$ can be designed based on the size of the robot and the Hessian of the field $J$, such that $\mu_{ij}>d_r$ still provides the collision-free guarantee. Here concludes the proof of {\bf P3} and the Theorem~\ref{thm:connectivity}.
\end{proof}

\subsection{Proof of Lemma~\ref{lemma:equil}: Steady-state Convergence}\label{Appendix:lemma_equil}
\begin{proof}
With the inter-agent CBF defined in \eqref{eq:CBF_hij}, let us define a differential function for each follower agent $i\in\mathcal{V}_f$ as $h_i = \prod_{j\in\mathcal{N}_i}h_{ij}$.
Then the followers are \textit{stationary} if $\dot h_i = \sum_{j\in\mathcal{N}_i}\dot h_{ij}\prod_{l\in\mathcal{N}_i\setminus \left\{ j\right\}} h_{il}=0$ and $ \bm{u}^*_i = [v^*_i,\, \omega^*_i]^\top = \bm{0}_{2,1}$.
Consider the feasible QPs and the forward invariance of set $\mathcal{S}_{ij}$ proved in Theorem~\ref{thm:feasibility} and \ref{thm:connectivity}, $h_{ij}(\bm{\xi}_i(t),\bm{\xi}_j(t))>0$ if $h_{ij}(\bm{\xi}_i(t_0),\bm{\xi}_j(t_0))>0$ and $t\geq t_0$. Hence, $\dot h_i(\bm{\xi}_i, \bm{\xi}_j) = 0$ iff $\dot h_{ij}(\bm{\xi}_i,\bm{\xi}_j)=0$ for all connected pairs $(i,j)\in\mathcal{E}(t)$. It implies that $h_{ij}$ is constant and time-invariant. Accordingly, all follower agents are stationary at a specific configuration with $v^*_i=0$, which can only happen when only the CBF constraint $\mathcal{U}_{\text{cbf-i}}$ is active as shown in the proof of Theorem~\ref{thm:feasibility} on the \textit{Feasibility Case B}, following the optimal control input $\bm{u}^*_i$ as in \eqref{eq:u_star_CBF}. 
Let us take the reference orientation-free flocking control input $\bm{u}_{\text{flock-i}}= [v_i,\, \omega_i]^\top$ as in \eqref{eq:ori_free_error_v}-\eqref{eq:ori_free_error_w} (where a similar argument can be used for the orientation-based flocking control input \eqref{eq:ori_con_error_v}-\eqref{eq:ori_con_error_w}), such that $\bm{u}_{\text{flock-i}}$ in \eqref{eq:u_star_CBF} is with
\begin{equation}\label{eq:reference}
     v_i = K_f e_{io} \left \langle {\bm{o}}_i, {\bm{p}}^\top \right \rangle, \;\;
    \omega_{i} =  - \frac{K_f e_{io}}{d} \left \langle {\bm{o}}^\perp_i, {\bm{p}}^\top \right \rangle, 
\end{equation}   
    where $\bm{o}_i, \bm{o}^\perp_i$ are as in \eqref{eq:orientation},
    and ${\bm{p}} =  \Big(\nabla^2 \bm{J}_{io}\Big)^{-1}  \begin{bmatrix}\cos(\beta_{i}) \\ \sin(\beta_{i}) \end{bmatrix}$. 
Then, the possible stationary cases for the followers are: 
\begin{itemize}
        \item {\textit{\textbf{Case I:} All followers are stationary with $\bm{u}^*_i = \bm{0}_{2,1}, e_{io} =0$.}} 
        This refers to the desired steady state where all followers have converged to the desired cohesion configuration. Particularly, the reference controller is $\bm{u}_{\text{flock-i}}=\bm{0}_{2,1}$ due to $e_{io}=0$. This case proves the first claim of the theorem. 
        \item {\textit{\textbf{Case II}: All followers are stationary with $\bm{u}^*_i = \bm{0}_{2,1}, e_{io} \neq 0$ and the reference controller $\bm{u}_{\text{flock-i}}$ is with $\omega_i \neq 0$. }} 
        In the following, we show that this stationary case is not possible by proving that $\bm{u}^*_i = \bm{0}_{2,1}$ can not be maintained at all time. Specifically, when the followers are stationary at the current configuration,  both $e_{io}$ and ${\bm{p}}$ are time-invariant in the reference controller \eqref{eq:reference} whereas $\omega_i \neq 0$. Consequently, the agent's orientation is time-varying, which implies that $v_i$ is time-varying (cf. \eqref{eq:reference}), thereby the reference controller $\bm{u}_{\text{flock-i}} = [v_i,\, \omega_i]^\top$ is not constant. By substituting the varying $\bm{u}_{\text{flock-i}}$ into the optimal control input solution in \eqref{eq:u_star_CBF}, it immediately shows that $\bm{u}^*_i = \bm{0}_{2,1} $ can not be maintained at all time. 
         \item {\textit{\textbf{Case III}: All followers are stationary with $\bm{u}^*_i = \bm{0}_{2,1}, e_{io} \neq 0$ and the reference controller $\bm{u}_{\text{flock-i}}$ is with $v_i\neq 0$ and $\omega_i = 0$.}}
        This case refers to the undesired steady states where the geometry conflict happens between inter-agent safety \& connectivity maintenance and the desired flocking configuration. 
        Specifically, the active CBF constraint $\mathcal{U}_{\text{cbf-i}}$ implies that the current reference flocking controller $\bm{u}_{\text{flock-i}}$ (aiming at flocking cohesion convergence) violates the safety and connectivity conditions. Similar to the analysis in Case II, given the time-invariant orientation of agent, the invariant optimal control input $\bm{u}^*_i = \bm{0}_{2,1}$ is maintained with a constant $\bm{u}_{\text{flock-i}}$ (where   $v_i\neq 0$ and $\omega_i=0 $ in \eqref{eq:reference}), such that the followers are stabilized at the undesired safe states where $e_{io}\neq 0$ and $h_{ij}>0$ for $(i,j)\in\mathcal{E}$. This case shows the last claim of the theorem. Corresponding examples are given in the simulation results in Figure~\ref{fig:ori_based_safe} in Section \ref{sec:simulation}.
\end{itemize}       

Similarly, for the leader, it can only be stationary with $\bm{u}^*_{\text{L}} = \bm{0}_{2\times 1}$, which either refers to the desired case that the leader stabilizes at the source location, or another case that the leader's motion is restricted for ensuring inter-agent safety and connectivity. For the second case, two similar scenarios are considered as in above \textit{Case II} and \textit{Case III} of the followers. Moreover, given the similar orientation principle of the source-seeking controller \eqref{eq:SS}, it is easy to analyze that the leader can be stationary with the constrained state $\bm{u}^*_{\text{L}} = \bm{0}_{2\times 1}$ while the reference $\bm{u}_{ss}$ is with constant $v_{\text{L}} \neq 0$ and $\omega_{\text{L}} = 0$.

In summary, driven by the optimized source-seeking (leader) and flocking-cohesion (followers) inputs from \eqref{QP_framework:leader}-\eqref{QP_framework:follower}, if there is no conflict between the final flocking-cohesion configuration and the inter-agent safety \& connectivity requirement in geometry, the multi-agent system will only be stabilized at the desired steady states (i.e., $e_{io} = 0$ in the orientation-free flocking method and $\widetilde{\bm{e}}_i = \bm{0}_{1,2}$ in the orientation-based method). Otherwise, the follower will stabilize at the states where $e_{io}\neq 0$ (or $ \widetilde{\bm{e}}_i\neq \bm{0}_{1,2}$) to ensure $h_{ij}>0$ for $(i,j)\in\mathcal{E}$.
\end{proof}

\bibliographystyle{IEEEtran}
\bibliography{bibtex/bib/IEEEabrv.bib, bibtex/bib/reference.bib}{}

\end{document}